\documentclass[review]{elsarticle}

\usepackage{lineno,hyperref}
\modulolinenumbers[5]

\journal{International Journal of Multiphase Flow}

\usepackage{color}
\usepackage{multirow}
\usepackage{dcolumn}
\usepackage{multicol}
\usepackage{graphicx}
\usepackage{caption}
\usepackage{amssymb}
\usepackage{amsmath}
\usepackage{epstopdf}
\usepackage{mathtools}
\usepackage{float}
\usepackage{bm}

\usepackage{subcaption}
\usepackage{graphicx}
\usepackage{soul}

\usepackage{color}

\biboptions{authoryear}

\biboptions{sort&compress}

\usepackage{color}

\def\We{{\it We}}
\def\Oh{{\it Oh}}
\def\Bo{{\it Bo}}

\newcommand{\ks}{\textcolor{black}} 

\begin{document}

\begin{frontmatter}

\title{\ks{Coalescence of non-spherical drops with a liquid surface}}
\author{Nagula Venkata Anirudh$^a$, Sachidananda Behera$^a$\footnote{sbehera@mae.iith.ac.in} and Kirti Chandra Sahu$^{b}$\footnote{ksahu@che.iith.ac.in} }
\address{$^a$Department of Mechanical and Aerospace Engineering, Indian Institute of Technology Hyderabad, Kandi - 502 284, Sangareddy, Telangana, India  \\ 
$^b$Department of Chemical Engineering, Indian Institute of Technology Hyderabad, Kandi - 502 284, Sangareddy, Telangana, India}

\begin{abstract}
We employ three-dimensional numerical simulations to explore the impact dynamics of non-spherical drops in a deep liquid pool by varying the aspect ratios $(A_r)$ and Weber numbers $(\We)$. We observe that when a non-spherical drop is gently placed on a liquid pool, it exhibits a partial coalescence phenomenon and the emergence of a daughter droplet for $A_r>0.67$. In contrast to the prolate $(A_r<1)$ and spherical drops $(A_r=1)$, an oblate $(A_r>1)$ drop with a high aspect ratio encapsulates air in a ring-like bubble within the pool and emerges a liquid column that undergoes Rayleigh-Plateau capillary instability, leading to the formation of two daughter droplets with complex shapes. When the parent drop is impacted with finite velocity, our observations indicate that increasing the Weber number leads to elevated crater heights on the free surface for all aspect ratios. A prolate drop produces a less pronounced wave swell and exhibits a prolonged impact duration owing to its negligible impact area. Conversely, an oblate drop generates a much wider wave swell than spherical and prolate drops. We analyze the relationship between rim formation dynamics and the kinetic and surface energies of the system. Finally, we establish an analogy by comparing the dynamics of a freely falling non-spherical drop, undergoing topological oscillations during its descent from a height, with the impact dynamics of parent drops of various shapes striking the liquid surface with an equivalent velocity. Our investigation involving non-spherical drops contrasts the extensive studies conducted by various researchers on the impact of a parent spherical drop just above the free surface of a liquid pool.
\end{abstract}

\end{frontmatter}

\noindent Keywords: Drop, coalescence, partial coalescence, interfacial flow, liquid-air interface

\section{Introduction} \label{sec:intro}

The dynamics of a drop impacting solid and liquid surfaces have been explored for centuries and remain a topic of great interest even today. Leonardo Da Vinci documented this phenomenon in the 1500s in Codex Leicester, as later reported in \cite{worthington1877xxviii}. Over the past few decades, there have been significant advances in understanding the interaction of two interfaces \citep{josserand2016drop,deka2019coalescence,sprittles2023gas}. This progress has contributed to diverse fields, such as microfluidics, combustion, coatings, inkjet printing, soil erosion, air trapping on the sea surface, separation of the water phase from the oil phase in oil recovery \citep{stone2004engineering,thoroddsen2008high,kavehpour2015coalescence}, and natural phenomena, such as raindrops falling on a liquid surface and ejection of marine microplastics from lakes and oceans by raindrops \citep{low1982collision,thomson1886v,pumphrey1989underwater,veron2015ocean,liu2018experimental}, to name a few. Apart from the wide range of practical applications, this subject involves complex physics, such as the generation of capillary waves leading to the formation of daughter droplets (partial coalescence) at low impact velocities, as well as the formation of craters and splashing at high impact velocities \citep{ray2015regimes,deka2019coalescence,kirar2022coalescence}.
 
The impact velocity of the drop is characterised by the Weber number, defined as $\We=\rho_l v_0^2 R_{eq}/\sigma$, where $\rho_l$ denotes the density of the liquid, $v_0$ is the impact velocity, $R_{eq}$ represents the radius of the parent/primary drop and $\sigma$ is the interfacial tension between the drop and the surrounding medium. At a low Weber number, when a primary droplet contacts the free surface of a liquid pool, it hovers until the trapped surrounding fluid (air cushion between the drop and the free surface) drains out. Subsequently, coalescence results in a rapidly expanding neck at the contact point due to high capillary pressure. This triggers upward-moving capillary waves, forming a liquid column and initiating necking. Over time, the neck diameter decreases, causing a daughter droplet to pinch off \citep{blanchette2006partial}. The competition between vertical and horizontal collapse rates influences the pinch-off process. If capillary waves sufficiently delay vertical collapse, horizontal collapse merges the neck, producing a daughter/secondary droplet over the free surface. This process is commonly known as the partial coalescence phenomenon \citep{charles1960mechanism}. In contrast to previous beliefs attributing partial coalescence to an inviscid instability mechanism in the liquid column, later research indicates that the dynamics of partial coalescence are predominantly influenced by gravity, viscosity, and interfacial tension \citep{thoroddsen2000coalescence,chen2006partial}. \cite{Chen2006b} found gravity, inertial-capillary, and viscosity-dominated regimes and described that gravity drives coalescence at high Bond numbers $(\Bo = \rho_l g {R_{eq}}^2/\sigma)$, while viscosity dominates at lower values of $\Bo$. Here, $g$ represents the acceleration due to gravity. At intermediate $\Bo$ values, coalescence is driven by inertia and capillary forces, with partial coalescence observed in the inertial-capillary regime. Recently, \cite{ivanova2023self} demonstrated that the levitation of millimetre-sized droplets above a liquid pool is caused by capillary convection within the liquid resulting from a surface tension gradient created by the non-uniform distribution of vapour molecules from the droplet at the pool surface.  \cite{paul2023investigation} numerically investigated the impact of two vertically aligned drops on a liquid surface, exploring coalescence dynamics through variations in the ratio of diameters between the lower and upper drops.

When a drop impacts a liquid pool at high velocity, it gives rise to several distinct phenomena, including crater formation, air bubble entrainment inside the pool, the release of a jet and splashing dynamics. In this situation, immediately following impact, inertia triggers the creation of an ejecta sheet, forming an edge. This edge, in turn, develops into a cylindrical-shaped rim, evolving through a combination of Rayleigh–Taylor and Rayleigh–Plateau instabilities. Nonlinearities then emerge, resulting in breakup through end pinching and forming a cascade of droplets \citep{josserand2016drop,constante2023impact}. Such impacts trigger the expulsion of a sheet above the pool surface, potentially resulting in a subsequent splash \citep{weiss1999single,thoroddsen2002ejecta,agbaglah2015drop,josserand2003droplet,ray2015regimes,gielen2017oblique}. The impact dynamics are contingent on whether the interaction takes place with a thin liquid layer or a deep pool. It was found that a droplet descending onto a thin liquid layer manifests the generation of an upward Worthington jet \citep{gekle2010,gordillo2010}, while in a deep liquid pool, it culminates in the formation of the well-known milkdrop coronet \citep{josserand2016drop,constante2023impact}. A few researchers have also examined the oblique impact of a drop on a liquid surface \citep{gielen2017oblique,kirar2022coalescence}. It has been observed that oblique impacts, unlike normal impacts, suppress splashing and secondary droplet formation \citep{gielen2017oblique,ray2012oblique,okawa2008effect}. \cite{leneweit2005regimes,liu2016numerical} found that at large Weber numbers, capillary waves are absent, and instead, a lamella is ejected at the droplet front, comprising partially of the droplet fluid. The interplay between surface tension and inertial forces in the contact-line region determines whether this region of high curvature is influenced by surface tension, leading to the creation of capillary waves, or if the stagnation pressure due to the oblique impact results in the extrusion of a lamella-shaped surface. By performing axisymmetric simulations, \cite{behera2023investigation} presented a regime map delineating the coalescence and splashing regimes within the Weber and Ohnesorge numbers space. \cite{deka2017regime} employed axisymmetric simulations to elucidate the mechanism of large bubble entrapment within the pool during the coalescence process and compared their findings with the experimental observations of \cite{wang2013we}. \cite{constante2023impact} investigated the influence of surfactants during the impact of a spherical droplet on thin liquid films by conducting three-dimensional numerical simulations. Recently, \cite{liu2021role} experimentally investigated the shape of the drop on the impact dynamics, albeit on a solid substrate, and showed that altering the shape of a droplet can regulate its spreading and splashing behaviour on a substrate.

Research on the impact of spherical droplets into liquid pools relies on a combination of experimental techniques involving high-speed imaging \citep{thoroddsen2008high,blanchette2006partial,bisighini2010crater, kirar2022coalescence} and axisymmetric numerical simulations \citep{ray2010generation,ray2012oblique,ray2015regimes,gielen2017oblique,josserand2003droplet}. However, many real-world scenarios, such as raindrops \citep{kostinski2009raindrops}, charged drops \citep{ristenpart2009non}, and oscillating drops \ks{\citep{zhang2019short,deka2019dynamics,agrawal2017nonspherical,agrawal2020experimental,balla2019shape}} exhibit a variety of non-spherical shapes. Apart from the axisymmetric study by \cite{deka2017regime}, all other investigations focused on the impact of spherical droplets into liquid pools. Moreover, the emphasis in the study of \cite{deka2017regime} was specifically on the large bubble entrapment observed in prolate-shaped drops, attributed to their deeper penetration of the resultant vortex ring into the liquid pool. Consequently, their study can not capture the three-dimensional behaviour,  as demonstrated experimentally by \cite{liu2021role} in the context of impacts involving non-spherical drops on solid surfaces. Thus, the present investigation aims to explore the impact dynamics of a non-spherical drop in a deep liquid pool for different Weber numbers via three-dimensional numerical simulations. Our findings reveal asymmetrical dynamics and the formation of complex-shape daughter drops due to partial coalescence. Additionally, an analogy is established by comparing the dynamics of a freely falling non-spherical drop from height $L$, undergoing topological oscillations, with parent drops of different shapes impacting the liquid surface with an equivalent velocity ($\sqrt{2gL}$). This analysis enables us to explore the influence of shape oscillations during the free fall of a drop, resulting from the interplay between inertia and surface tension \citep{agrawal2017nonspherical}, on the coalescence phenomenon. This is in contrast to the impact of a parent drop just above the free surface, a phenomenon extensively studied by various researchers \citep{thoroddsen2008high,blanchette2006partial,ray2015regimes,deka2017regime}.

The rest of the manuscript is organised as follows. The problem is formulated in \S\ref{sec:method}, where we elaborate the numerical method for three-dimensional simulations and extensively validate the flow solver. The results are presented in \S\ref{sec:dis}. In this section, we first explore phenomena related to zero impact velocity $(\We=0)$. Then, we examine the effect of the Weber number on the dynamics of a drop impacting a deep liquid pool. Finally, we present the equivalence between the dynamics of a freely falling drop from a height and that of a drop impacting near the liquid pool. The conclusion remarks are given in \S\ref{sec:conc}.

\section{Formulation} \label{sec:method}

\begin{figure}
\centering
\hspace{0.1cm} {\large ($a$)} \hspace{4.2cm} {\large ($b$)} \\
\includegraphics[scale=0.25]{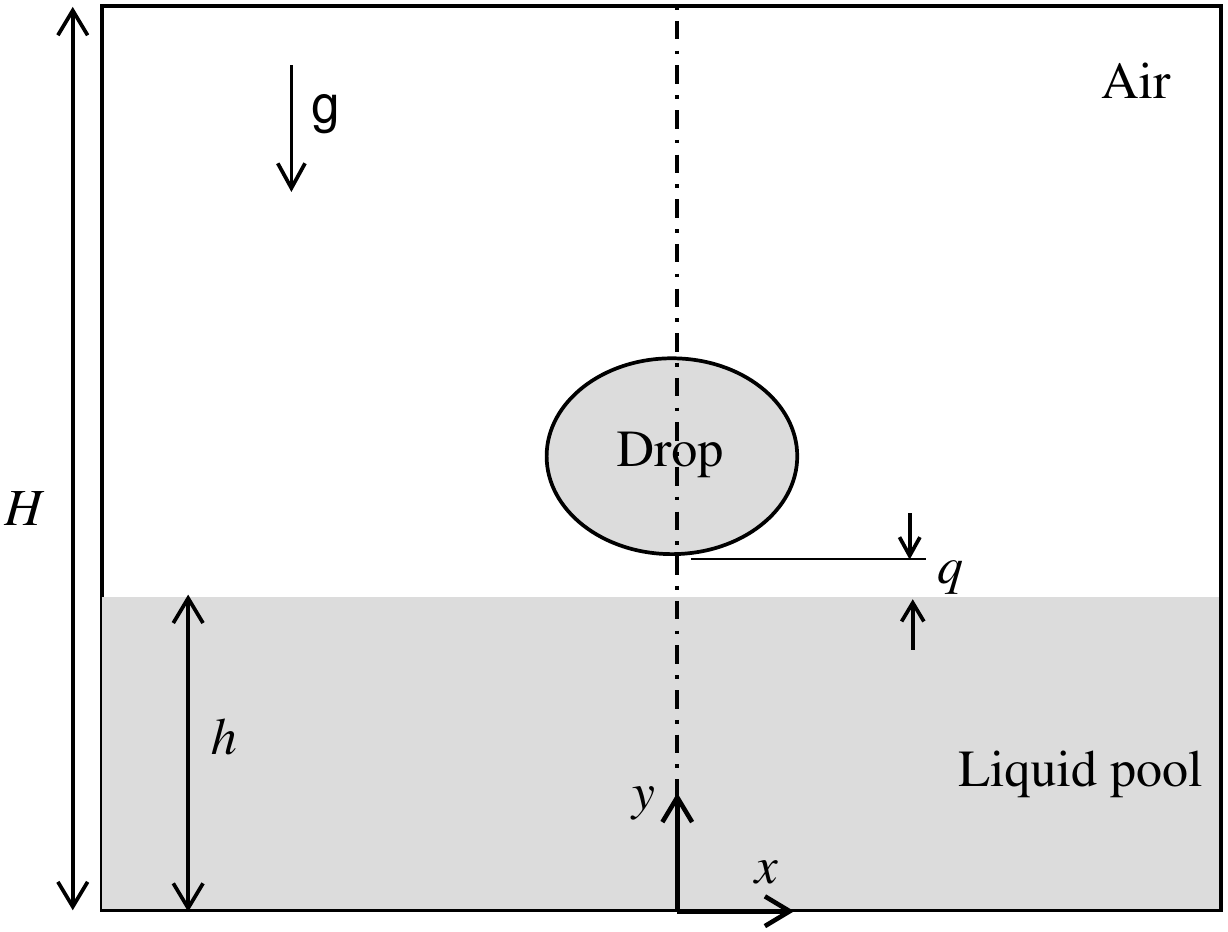} \hspace{2mm} \includegraphics[scale=0.25]{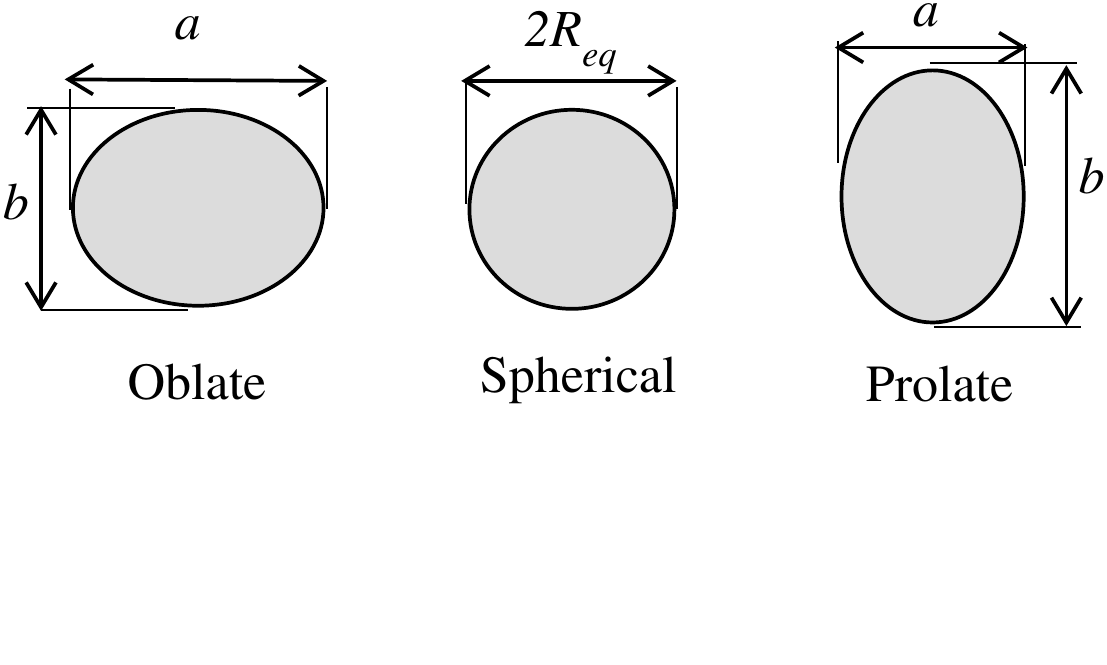}
\caption{($a$) Schematic diagram of the cross-sectional $(xy)$ view at $z=0$ of the cubic computational domain of dimension $H^3$ showing the initial configuration of a non-spherical drop impacting a liquid pool under gravity $(g)$ acting in the negative $y$ direction. ($b$) Three shapes of drops (oblate/spherical/prolate) have been considered in our study. The diameter of the non-spherical drop in the horizontal $(x)$ and vertical $(y)$ directions are $a$ and $b$, respectively. The diameter of the drop in the spanwise direction $(z)$ is the same as that of the horizontal direction, such that the equivalent spherical radius of the drop, $R_{eq}=(a^2b/8)^{1/3}$. The aspect ratio of the drop, $A_r=a/b$, such that $A_r>1$, $A_r=1$ and $A_r<1$ represent oblate, spherical and prolate drops, respectively.}
    \label{fig:schematic}
\end{figure}

We investigate the collision dynamics of a non-spherical primary drop impacting a deep liquid pool by performing three-dimensional numerical simulations of the Navier-Stokes and continuity equations. The drop and the pool constitute the same liquid, and the surrounding medium is air. The dynamic viscosities and the densities of liquid and air are denoted by $(\mu_l, \rho_l)$ and $(\mu_a, \rho_a)$, respectively. The fluids are assumed to be Newtonian and incompressible. The schematic diagram showing the initial configuration of the drop is depicted in Figure \ref{fig:schematic}$a$. To minimize the boundary effect, we consider sufficiently large cubic computational domains with dimensions of $16 R_{eq} \times 16 R_{eq} \times 16 R_{eq}$ for $\We=0$ and $30R_{eq} \times 30R_{eq} \times 30R_{eq}$ for the finite Weber numbers examined in the current study. Here, $R_{eq}$ denotes the volume equivalent spherical radius of the droplet, and $h$ represents the height of the liquid pool, which is maintained at $5.5R_{eq}$ and $16 R_{eq}$ for $\We=0$ and finite Weber numbers, respectively. Except for the case associated with the freely falling droplet from a height $(L)$, the separation distance between the bottom of the drop and the free surface of the liquid pool $(q)$ is maintained at a constant value of $0.09R_{eq}$..

We consider three types of initial droplet shapes: oblate $(A_r > 1)$, spherical $(A_r=1)$, and prolate $(A_r < 1)$, as depicted in Figure \ref{fig:schematic}$b$. We vary the aspect ratio of the primary non-spherical droplet $(A_r =a/b)$  while keeping its volume constant. Here, $a$ and $b$ represent the initial diameters of the primary droplet along its initial major and minor axes, respectively, such that $R_{eq}=(a^2b/8)^{1/3}$. Our study employs a Cartesian coordinate system $(x,y,z)$, with gravity $(g)$ acting in the negative $y$ direction. Initially, the surrounding air is at rest. 

\subsection{Governing equations}

The three-dimensional Navier-Stokes and continuity equations and the associated boundary conditions are non-dimensionalized using $R_{eq}$ as the length scale, $t_s = \sqrt{\rho_l R_{eq}^3/\sigma}$ (the capillary time) as the time scale and $V_s=R_{eq}/t_s$ as the velocity scale, such that
\begin{eqnarray} 
(x,y,z) = R_{eq} (\widetilde x,\widetilde y,\widetilde z), ~ t = t_s \widetilde t, ~ \bm{U} = V_s \bm{\widetilde U}, ~ P = (\sigma/R_{eq}) \widetilde P, \nonumber \\ \mu (\alpha) = {\widetilde \mu (\alpha)} \mu_l,  ~ \rho (\alpha) = {\widetilde \rho (\alpha)} \rho_l, ~ {\rm and} ~~ \delta_s = {\widetilde \delta_s} R_{eq}.
\label{scale}
\end{eqnarray}
Here, the tildes designate dimensionless quantities; $\bm{U} = (u,v,w)$ is the velocity field, wherein $u$, $v$ and $w$ denote the components of $\bm{U}$ in the $x$, $y$ and $z$ directions, respectively; $P$ represents the pressure field; $\sigma$ represents surface tension, and $\delta_{s}(\bm{x}-\bm{x}_f)$ is the delta distribution function which is zero elsewhere except on the interface, where $\bm{x}=\bm{x}_f$. 

After dropping tildes from all non-dimensional variables, the dimensionless continuity and Navier–Stokes equations governing the coalescence dynamics of the drop into the liquid pool are given by
\begin{eqnarray}
\nabla \cdot \bm{U} &=& 0, \label{eqn:eq1} \\
\rho (\alpha) \left(\frac{\partial \bm{U}}{\partial t} + \bm{U}. \nabla \bm{U} \right) &=& -\nabla P + Oh \nabla \cdot \left [\mu(\alpha)(\nabla \bm{U} + \nabla \bm{U}^T) \right]  \nonumber \\ 
&+& \kappa \bm{n} \delta_s + Bo \rho(\alpha)\bm{g} \bm{j}.  ~~~~  ~~\label{eqn:eq2}
\end{eqnarray}
Here, $\bm{j}$ denotes the unit vector along the vertical (negative $y$) direction, $\kappa~(\equiv - \nabla \cdot n)$ is the interfacial curvature, wherein $n$ is the outward-pointing unit normal to the interface. Note that the surface tension force is included as a body force term in Eq. (\ref{eqn:eq2}) by employing the continuum surface force (CSF) formulation proposed by \cite{brackbill1992continuum}. The Ohnesorge number, $Oh = \mu_l / {\sqrt{\rho_l \sigma R_{eq}}}$ represents the competition between the viscous and surface tension forces, which can also be expressed as $Oh=\sqrt{We}/Re$, wherein $Re=v_0 \rho_l R_{eq}/ \mu_l$ represents the Reynolds number.

The interface between the air and liquid phases is monitored by solving an advection equation for the volume fraction of the liquid phase, denoted as $\alpha$ (where $\alpha=0$ and $1$ correspond to the air and liquid phases, respectively). This is expressed as:
\begin{eqnarray}
{\partial \alpha \over \partial t} + \bm{U} \cdot \nabla \alpha  = 0. \label{eqn:VOF}
\end{eqnarray}
The density field, $\rho(\alpha)$, and the viscosity field, $\mu(\alpha)$, are given by
\begin{eqnarray} 
\rho(\alpha) = \rho_r (1-\alpha) + \alpha , \label{eq_rho} \\ 
\mu(\alpha) = \mu_r (1-\alpha) + \alpha , \label{eq_mu}
\end{eqnarray}
where $\rho_r (=\rho_a/\rho_l)$ and $\mu_r (=\mu_a/\mu_l)$ represent the density and viscosity ratios, respectively. In the present study, as we consider a water-air system, $\rho_r=10^{-3}$ and $\mu_r=0.018$, unless otherwise specified. In subsequent sections, we normalize the time $(t)$, such that the rescaled time, $\tau=0$, signifies the initiation of coalescence between the drop and the free surface of the liquid pool.

\subsection{Numerical method} \label{sec:numeric}

We adopt a two-phase flow solver within the OpenFOAM framework to simulate the impact dynamics of a drop into a liquid pool governed by eqs. (\ref{eqn:eq1}) and (\ref{eqn:eq2}) in a collocated grid arrangement, where the velocity vector, pressure and volume fraction are defined at the cell centers. A non-uniform mesh that assigns a large number of grids near the interfacial regions and slightly coarser mesh away from the interface is employed.  The numerical schemes used are second-order accurate in both space and time. The time integration is performed using the Crank-Nicolson scheme. In the momentum equation (\ref{eqn:eq2}), the diffusion and gradient terms are discretized using Gauss linear, and the convective terms are discretized employing a van Leer scheme. The pressure-velocity coupling is handled using the PIMPLE algorithm. This method is a combination of PISO (Pressure Implicit with Splitting of Operator) and SIMPLE (Semi-Implicit Method for Pressure-Linked Equations) algorithms. 
 
The simulation employs a volume-of-fluid (VoF) method incorporating a numerical interface compression method based on the multidimensional universal limiter with explicit solution (MULES) to track the interface separating the liquid and air phases. Ideally, the interface between the two fluids must represent a sharp discontinuity. However, a standard VOF method diffuses the interface, resulting in the volume fraction ($\alpha$) varying between 0 and 1 in a thin region across the interface. To address this issue, we use the MULES technique that enables the numerical compression of the interface by introducing the extra compression term into the standard VOF equation. This is given by
\begin{equation} \label{alpha_decomposed_transport}
 \frac{\partial\alpha}{\partial t}+\nabla\cdot(\bm{U}\alpha) \underbrace{-\nabla\cdot\left \{U_{r,f} (1-\alpha)\alpha \right \}}_\text{compression term} =0.
\end{equation}
It is to be noted that the compression term acts exclusively at the interface and is computed in the normal direction of the interface. The term $U_{r,f}$ signifies the relative velocity between the two fluids, which is modeled as
\begin{equation}\label{URF}
    U_{r,f} = \min \left (  C_{\rm compr.} \frac{|\phi_f|}{|S_f|} , \max \left [ \frac{|\phi_f|}{|S_f|} \right ]\right ) n_f.
\end{equation}
Here, $C_{\rm compr.}$ indicates a user-defined interface compression factor, and its value is set to 1 in the present study as suggested by \cite{deshpande2012evaluating} to avoid errors in the calculation of interfacial curvatures and interface smearing. The term $\phi_f$ represents the volumetric flux through the cell face, $S_f$ is the outward pointing face area vector, and $n_f$ is the face-centered interface normal vector. For a more comprehensive discussion on the numerical implementation of the MULES method, refer \cite{deshpande2012evaluating}.

We employ the following boundary conditions in order to solve the governing equations. The free-slip boundary condition is imposed on all side boundaries of the computational domain. The no-slip and no-penetration boundary conditions are used at the bottom wall, and the Neumann boundary condition is used at the top of the computational domain.

\subsection{Validation}

\begin{figure}
\centering
\includegraphics[scale=0.5]{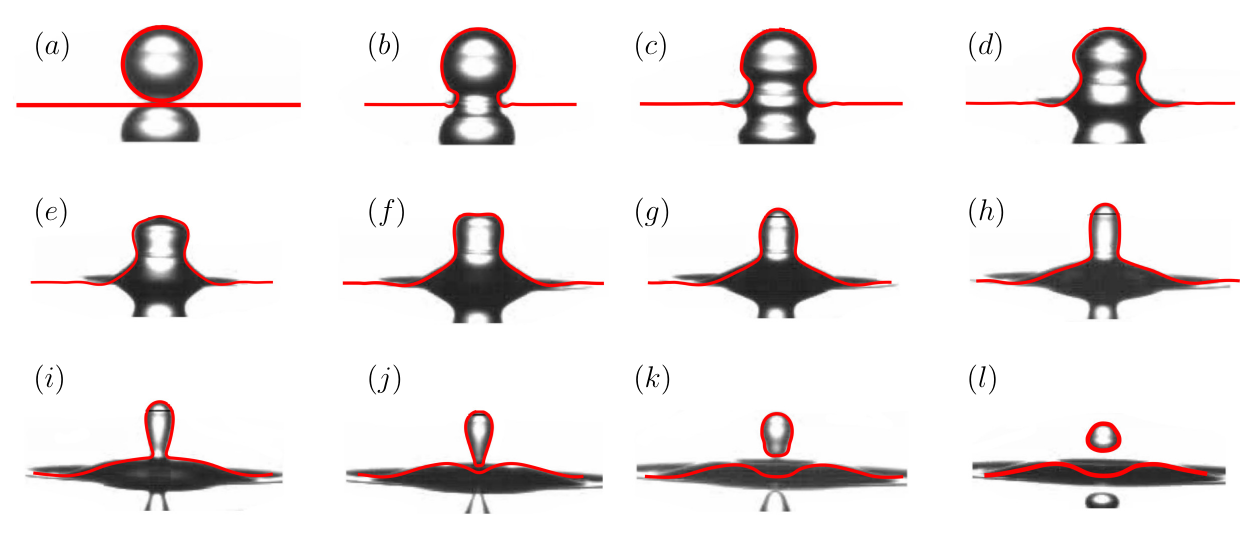}
\caption{Comparison of the results obtained from our numerical simulation (depicted by the red line) with the experimental data from \cite{chen2006partial} (shown in the background). Panel ($a$) corresponds to the instant $(\tau=0)$ when the drop comes into contact with the free surface of the liquid pool. Subsequent panels follow at intervals of 0.542 ms. The values of the non-dimensional numbers considered in our simulations are $\rho_r = 0.763$, $\mu_r = 2$, $\Oh = 0.008$, $\Bo = 0.10$ and $\We = 0$.}
\label{fig:chen_Validation}
\end{figure}

In order to validate the present solver, we compare the results obtained from our numerical simulations with the experimental observations of \cite{chen2006partial} and \cite{wang2000splashing, cossali1997impact}. In the experiment by \cite{chen2006partial}, a water droplet of diameter 1.1 mm was gently deposited onto the free surface of a liquid pool containing 20\% polybutene in decane and water, and the dynamics were recorded using a high-speed camera. The associated dimensionless numbers in our simulation are $\rho_r = 0.763$, $\mu_r = 2$, $\Oh = 0.008$, $\Bo = 0.10$ and $\We = 0$. It can be seen in Figure \ref{fig:chen_Validation} that the coalescence initiates at the re-scaled dimensionless time, $\tau = 0$, with the drop draining into the pool for $\tau > 0$. This process gives rise to a cylindrical column due to the upward-moving capillary wave. The neck of this column constricts due to surface tension, leading to the pinching off of a daughter drop as discussed in \S\ref{sec:intro}. Figure  \ref{fig:chen_Validation} also reveals that the dynamics obtained from our numerical simulation at various instants agree with the experimental findings of \cite{chen2006partial}. 

\begin{figure}
\centering
\includegraphics[scale=0.5]{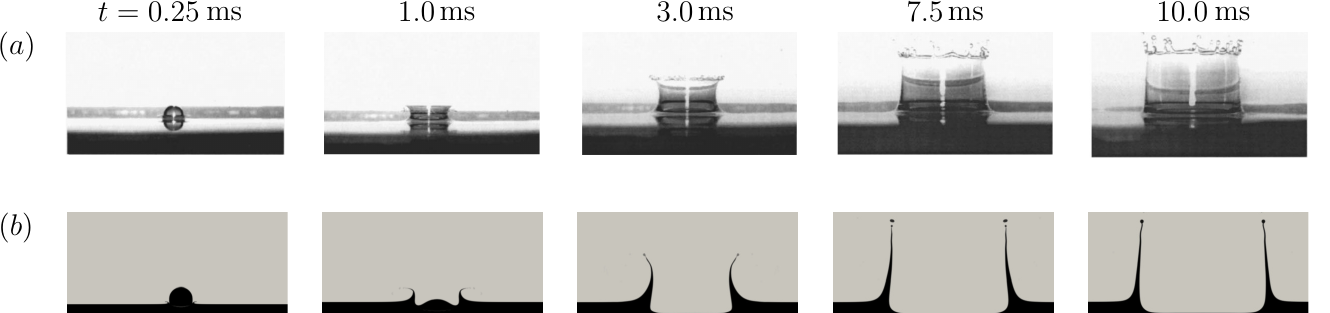}
\caption{Comparison of the results obtained from our numerical simulation (panel $b$) with the experimental results of \cite{wang2000splashing} (panel $a$). The values of the non-dimensional numbers considered in our simulations are $\rho_r = 10^{-3}$, $\mu_r = 8.3 \times 10^{-4}$, $\Oh = 0.054$, $\Bo = 0.79$ and $\We = 1001.35$.}
\label{fig:wang_chen_Validation}
\end{figure}

\begin{figure}
\centering
\hspace{-0.2cm} {$(a)$} \hspace{5.5cm} {$(b)$} \\
\begin{subfigure}[h]{0.4\textwidth}
\includegraphics[width=\textwidth]{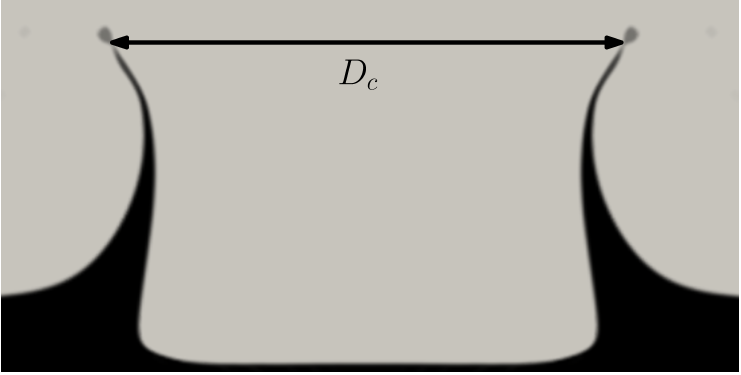}
\end{subfigure}
\hspace{2mm}
\begin{subfigure}[h]{0.45\textwidth}
\includegraphics[width=\textwidth]{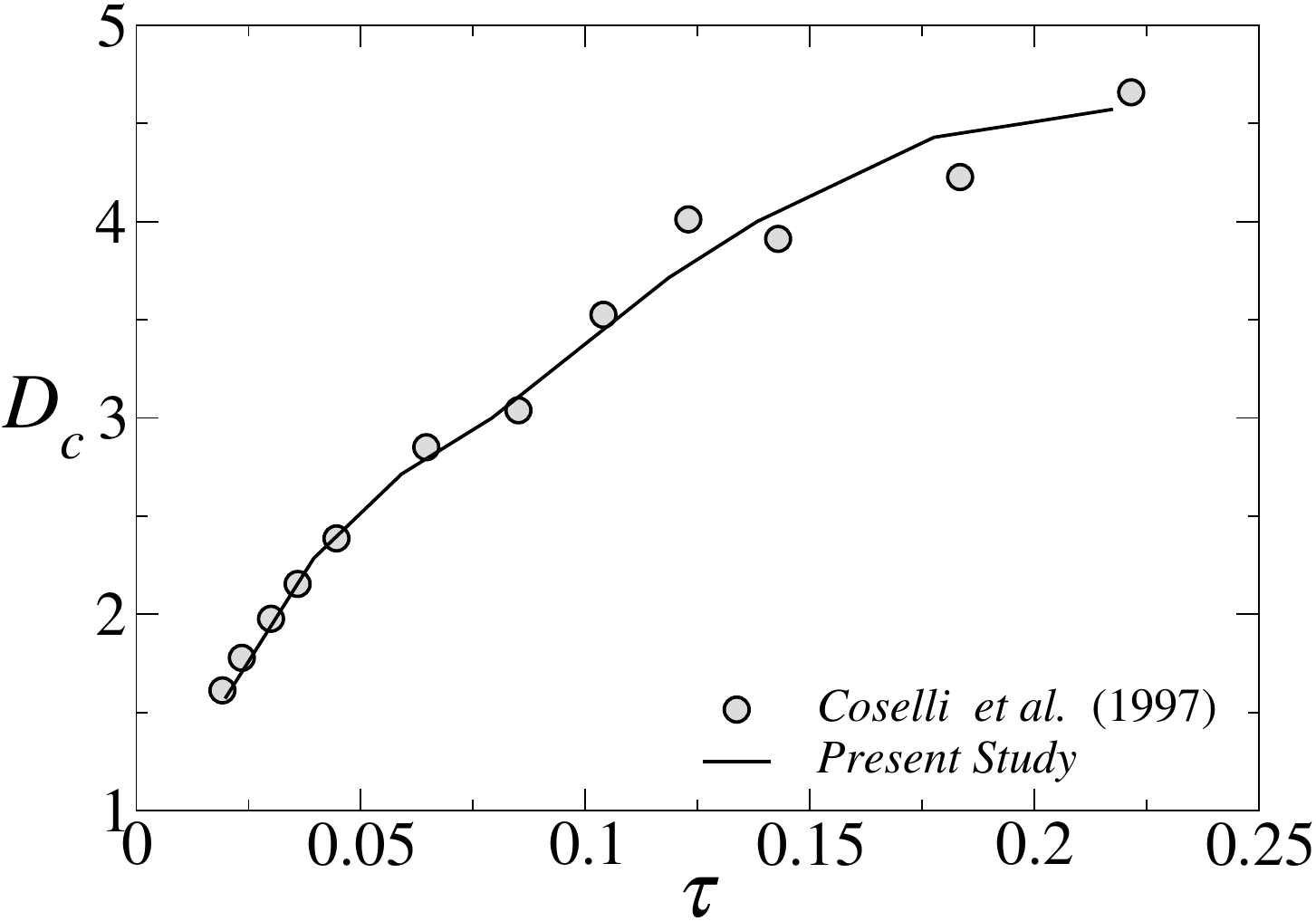}
\end{subfigure}
\caption{($a$) A typical crown formation during the impact of a drop on a liquid pool. ($b$) Comparison of the temporal evolution of the dimensionless crown diameter ($D_c$) obtained from our numerical simulation with the experimental results of \cite{cossali1997impact}. The values of the non-dimensional numbers considered in our simulations are $\rho_r = 0.0012$, $\mu_r = 0.014$, $\Oh = 0.003$, $\Bo = 0.87$ and $\We = 160.15$.}
\label{fig:cossali}
\end{figure}

Figures \ref{fig:wang_chen_Validation} and \ref{fig:cossali} illustrate the comparison of our numerical results with the experimental observations of \cite{wang2000splashing,cossali1997impact} in the context of finite impact velocities of a drop colliding with a deep liquid pool. In the investigation conducted by \cite{wang2000splashing}, the drop and the pool liquid is a mixture of 70\% glycerol in water by weight. Consequently, the values of the various dimensionless numbers in our formulation become $\rho_r = 10^{-3}$, $\mu_r = 8.3 \times 10^{-4}$, $\Oh = 0.054$, $\Bo = 0.79$ and $\We = 1001.35$. It can be seen that, upon impact, the drop exhibits splashing behavior, and the free surface manifests crown structures and a milkdrop coronet. Our results qualitatively match with the experimental observations quite well. For a quantitative validation of our solver, in Figure \ref{fig:cossali}$b$, we compare the temporal evolution of the crown diameter, $D_c$ (as illustrated in Figure \ref{fig:cossali}$a$), with the experimental findings of \cite{cossali1997impact}. The close agreement observed further strengthens the validation of our solver. \ks{We further validate our solver by comparing the shape oscillations of a freely falling oblate drop with an initial aspect ratio ($A_r=1.95$) with the results from \cite{agrawal2017nonspherical}. The comparison is presented in the appendix (Figure \ref{fig:Agrawal_validation}).}

\begin{figure}[H]
\centering
\includegraphics[scale=0.55]{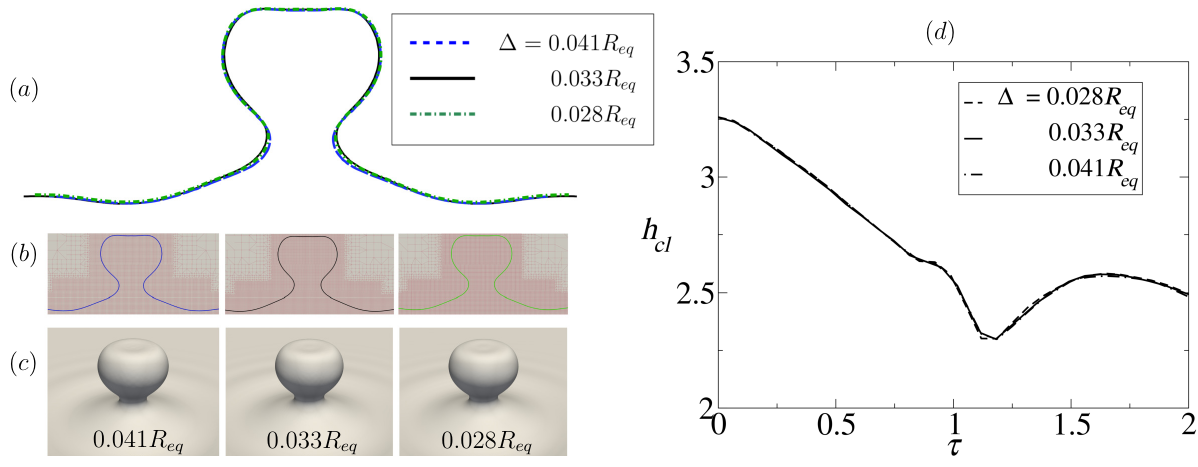}
\caption{The topology of a non-spherical primary droplet with $A_r=0.5$ coalescing with the liquid pool at $\tau=1.22$ obtained using three different grids with the smallest (dimensionless) cell sizes of $\Delta=0.041 R_{eq}$, $0.033 R_{eq}$, and $0.028 R_{eq}$. The rest of the dimensionless numbers are $\rho_r=10^{-3}$, $\mu_r=0.018$, $\Oh=7.82 \times 10^{-3}$, $\Bo=0.04$ and $\We=0$. Panel $(a)$ shows the overlap profiles, and panels $(b,c)$ depict the cross-sectional views and the iso-surfaces of the drop obtained using different grids, respectively. \ks{Panel $(d)$ shows the temporal evolutions of the maximum height of the column ($h_{cl}$) for $A_r = 0.5$ obtained using different grid sizes.}}
\label{fig:Grid_convergence}
\end{figure}

\begin{figure}[H]
\centering
\includegraphics[scale=0.45]{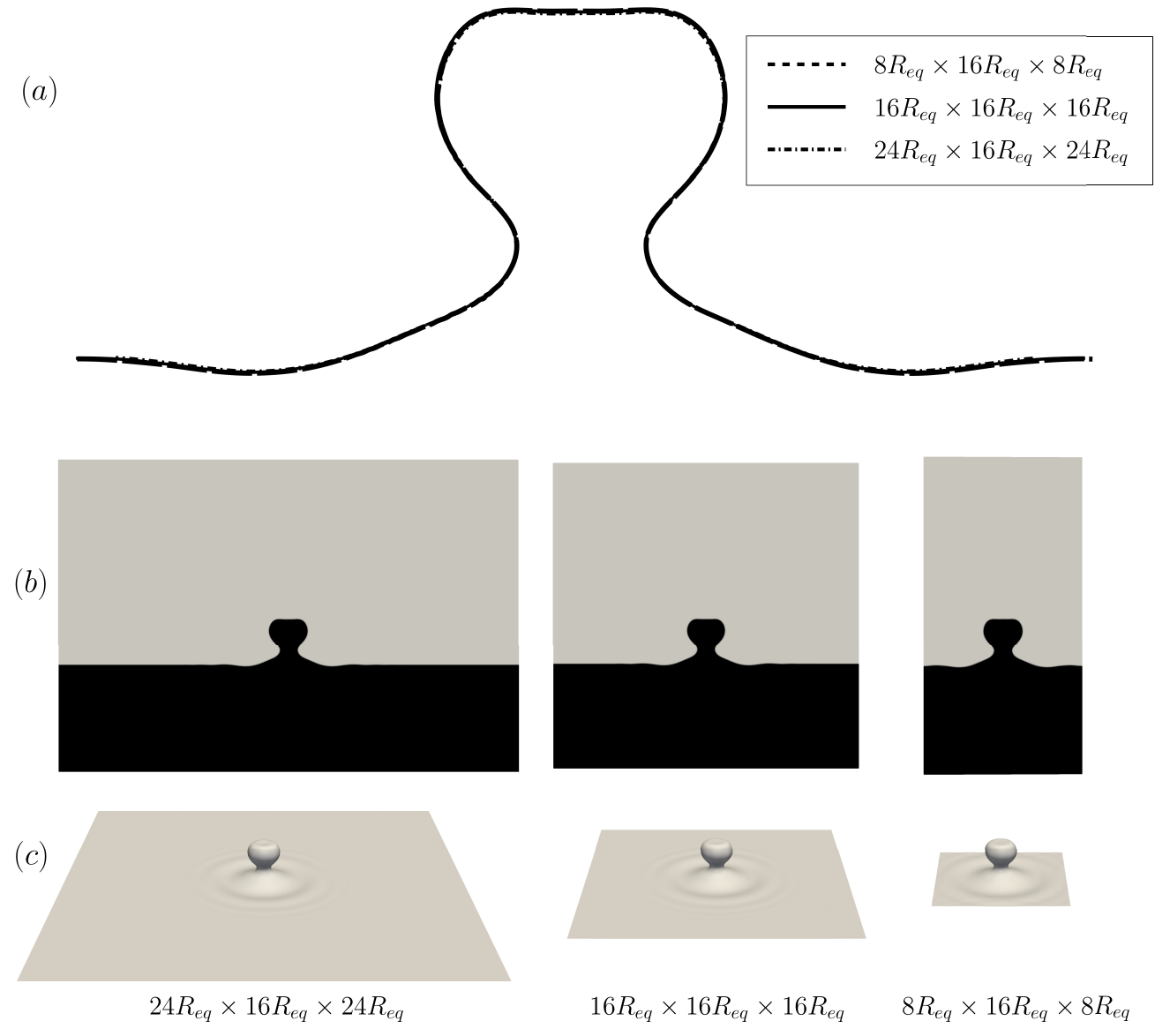}
\caption{The topology of a non-spherical primary droplet with $A_r=0.5$ coalescing into the liquid pool at $\tau=1.22$ obtained using three different computational domains. The rest of the dimensionless numbers are $\rho_r=10^{-3}$, $\mu_r=0.018$, $\Oh=7.82 \times 10^{-3}$, $\Bo=0.04$ and $\We=0$. Panel $(a)$ shows the overlap profiles, and panels $(b,c)$ depict the cross-sectional views and the isosurfaces of the drop obtained using different computational domains, respectively.}
\label{fig:Domain_convergence}
\end{figure}

Furthermore, in Figures \ref{fig:Grid_convergence} and \ref{fig:Domain_convergence}, we conduct grid convergence and domain dependence tests for a typical set of parameters considered in our study ($\rho_r=10^{-3}$, $\mu_r=0.018$, $\Oh=7.82 \times 10^{-3}$, $\Bo=0.04$, and $\We=0$). \ks{An octree mesh, generated using SnappyHexMesh, an OpenFOAM tool specifically designed for octree mesh creation, is employed to discretize the computational domain. \ks{A detailed discussion on octree-based hexahedral mesh generation can be found in \cite{schneiders2000octree}. The octree mesh facilitates a flexible and efficient representation of complex geometries, adapting the mesh resolution to specific regions based on the problem requirement, and is particularly advantageous in simulating two-phase flows where intricate physics dominates near the interface.} Consequently, a finer mesh zone is established in the proximity of both the droplet and the free surface of the liquid pool by refining the grid close to the interface, while larger-sized grids are employed in the outer regions. This strategic approach effectively minimizes computational costs.} In Figure \ref{fig:Grid_convergence}(a-c), the topology of a non-spherical primary droplet with $A_r=0.5$ coalescing with the liquid pool at $\tau=1.22$ is depicted using three different grids. These grids are characterized by the smallest (dimensionless) cell sizes of $\Delta =0.041 R_{eq}$, $0.033 R_{eq}$, and $0.028 R_{eq}$, respectively. The corresponding far-field maximum grid sizes of $\Delta_{max}=0.64 R_{eq}$, $0.53 R_{eq}$, and $0.46 R_{eq}$, respectively. Notably, the results obtained with $\Delta =0.033 R_{eq}$ and $\Delta =0.028 R_{eq}$ are practically indistinguishable. \ks{Additionally, an inspection of the temporal evolution of the maximum column height ($h_{cl}$), as illustrated in Figure \ref{fig:Grid_convergence}$(d)$ for $A_r = 0.5$ across three different grid sizes, reveals minimal disparities.} Similarly, Figure \ref{fig:Domain_convergence} demonstrates that results obtained with different sizes of the computational domain considered for this test are also indistinguishable. Given these observations, we adopt the grid with $\Delta=0.033 R_{eq}$, and a computational domain size of $16R_{eq} \times 16R_{eq} \times 16R_{eq}$ for generating the rest of the results presented in the following section.

\section{Results and discussion} \label{sec:dis}

\subsection{Coalescence of non-spherical drops gently placed on a pool} \label{sec:pc}

\begin{figure}
\centering
\includegraphics[scale=0.5]{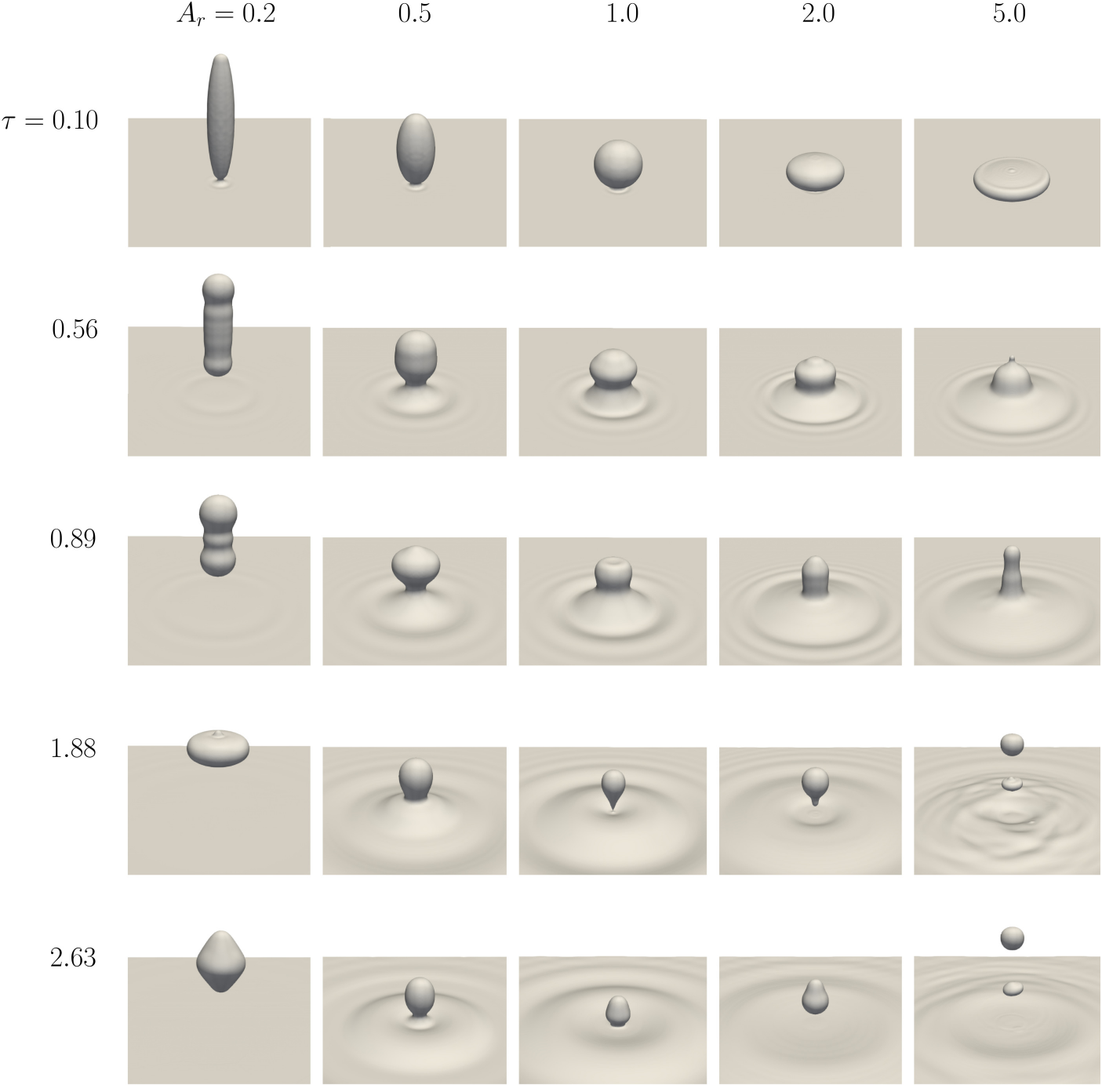}
\caption{Temporal evolution of the coalescence of the parent drop of different aspect ratios $(A_r)$ gently placed on a deep liquid pool $(We=0)$. The values of the rest of the dimensionless parameters are $\Oh = 7.82 \times 10^{-3}$ and $\Bo = 0.04$. \ks{The coalescence dynamics for $A_r=0.2$, 0.5, 1.0, 2.0 and 5.0 are included as supplementary video 1-5, respectively.} }
\label{fig:3_D}
\end{figure}

\begin{figure}
\centering
\includegraphics[scale=0.42]{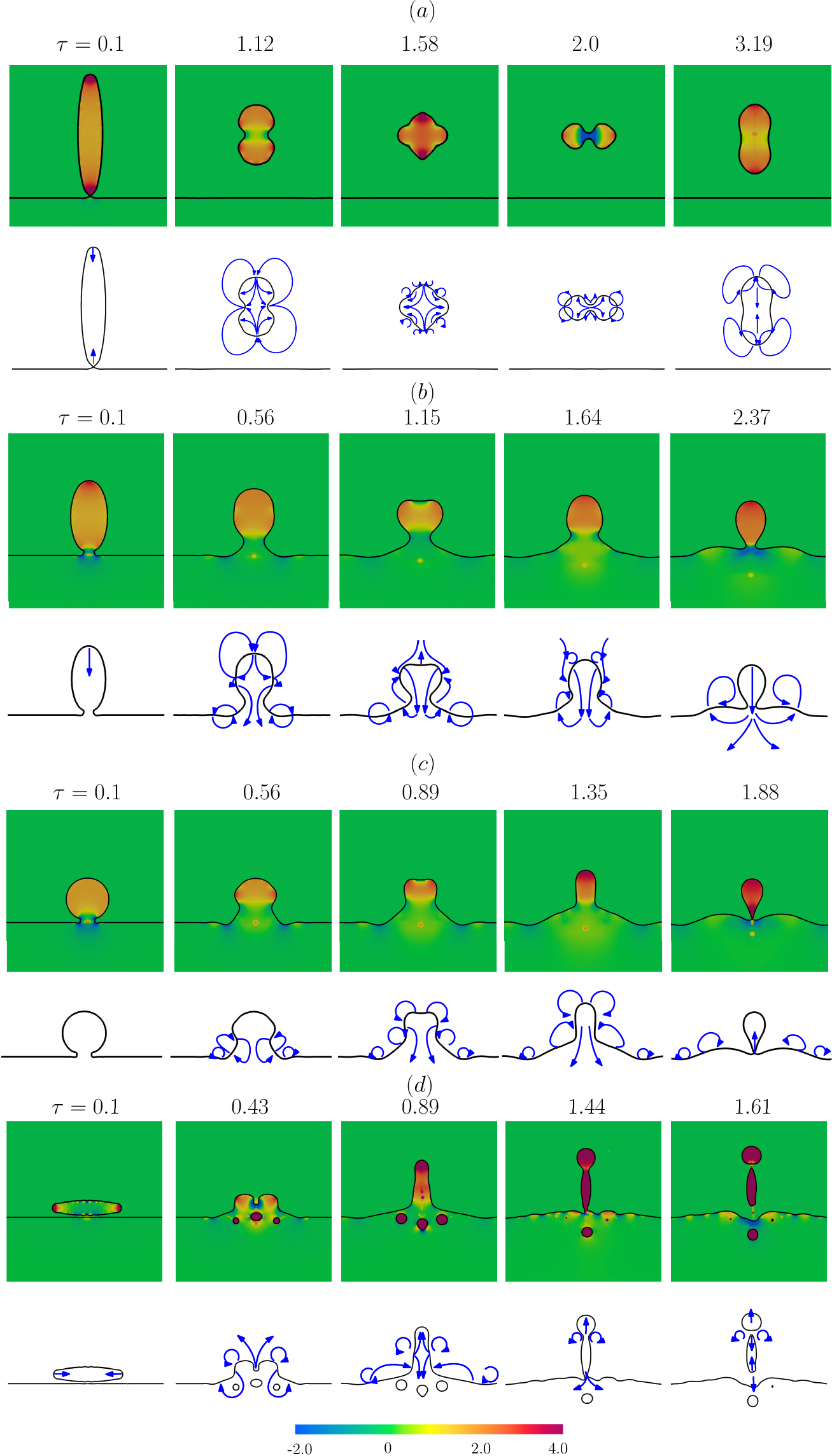}
\caption{Evolution of pressure field and representation of flow field in the $x-y$ plane crossing the center of the drop for different values of the aspect ratio $(A_r)$ of the parent drop. (a) $A_r = 0.2$, (b) $A_r = 0.5$, (c) $A_r = 1.0$ and $A_r = 5.0$. The color bar represents the pressure counters. The values of the rest of the dimensionless parameters are $\Oh = 7.82 \times 10^{-3}$, $\Bo = 0.04$ and $\We=0$.}
\label{fig:flowfield}
\end{figure}

The shape of the parent droplet is expected to play a crucial role in the coalescence process and subsequent dynamics. This aligns with the observations made by \cite{liu2021role} in the context of the collision of non-spherical drops on a solid substrate. Thus, we begin the presentation of our results by examining the effect of the aspect ratio $(A_r)$ of the non-spherical parent droplet gently placed on a liquid pool $(We=0)$. Figure \ref{fig:3_D} illustrates the temporal evolution of the coalescence phenomenon for the parent non-spherical drops with $A_r=0.2$, 0.5, 1, 2, and 5. The temporal evaluations of the pressure and the flow field (cross-sectional views in the $x-y$ plane crossing the center of the drop) for different aspect ratios are presented in Figure \ref{fig:flowfield}. It is important to note that $A_r<1$, $A_r=1$, and $A_r>1$ represent parent drops with prolate, spherical, and oblate shapes, respectively. To exclusively examine the impact of the aspect ratio of the parent droplet on the coalescence phenomenon, we maintain the other dimensionless numbers constant at $\rho_r=10^{-3}$, $\mu_r=0.018$, $\Oh = 7.82 \times 10^{-3}$ and $\Bo = 0.04$, while varying the aspect ratio of the parent drop in Figures \ref{fig:3_D} and \ref{fig:flowfield}.

\begin{figure}
\centering
\hspace{0.7cm}{($a$)}\\
\hspace{0.7cm}
\includegraphics[scale=0.3]{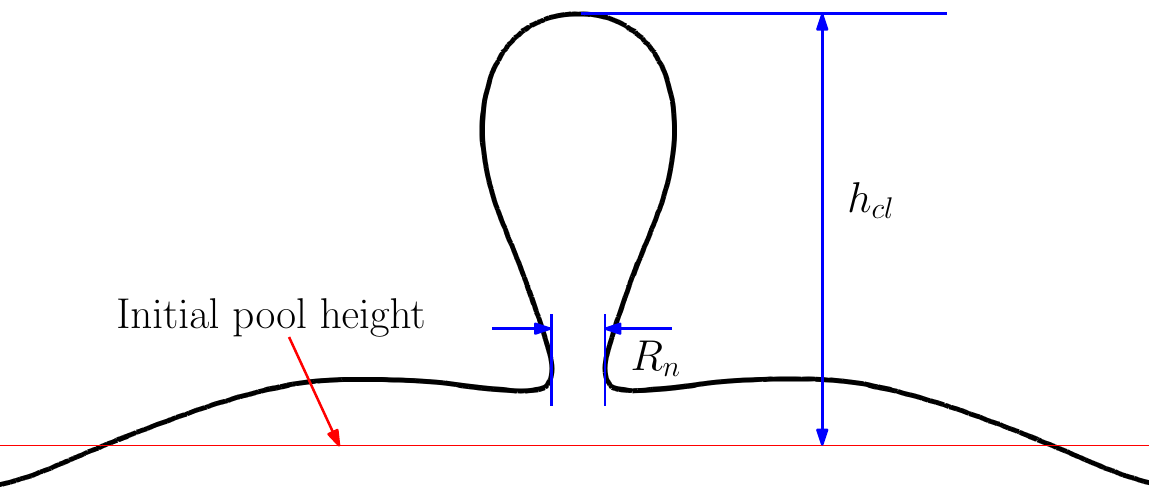} \\
\hspace{0.25cm} {($b$)} \hspace{4.4cm} {($c$)} \\
\includegraphics[scale=0.24]{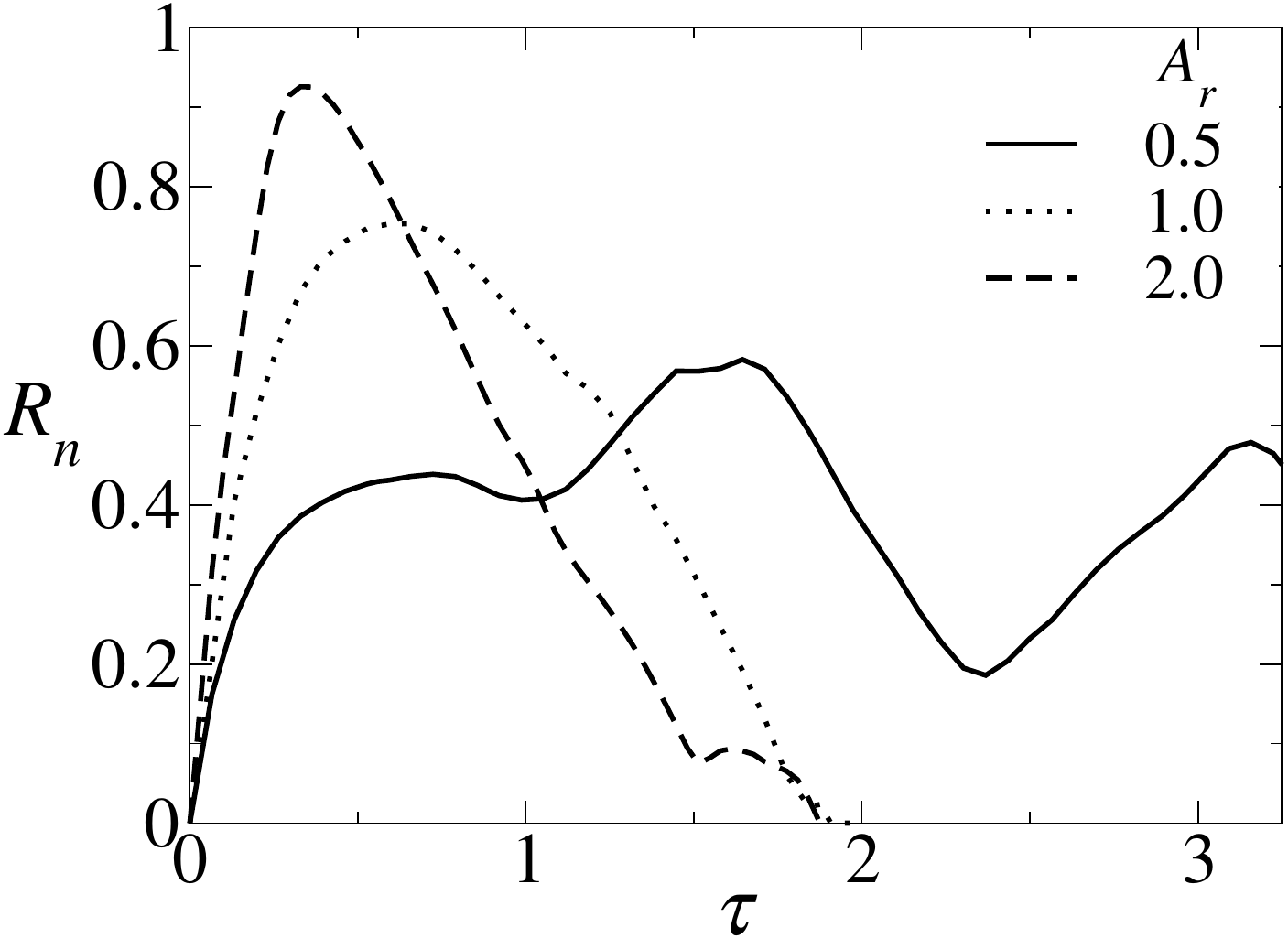}
\includegraphics[scale=0.24]{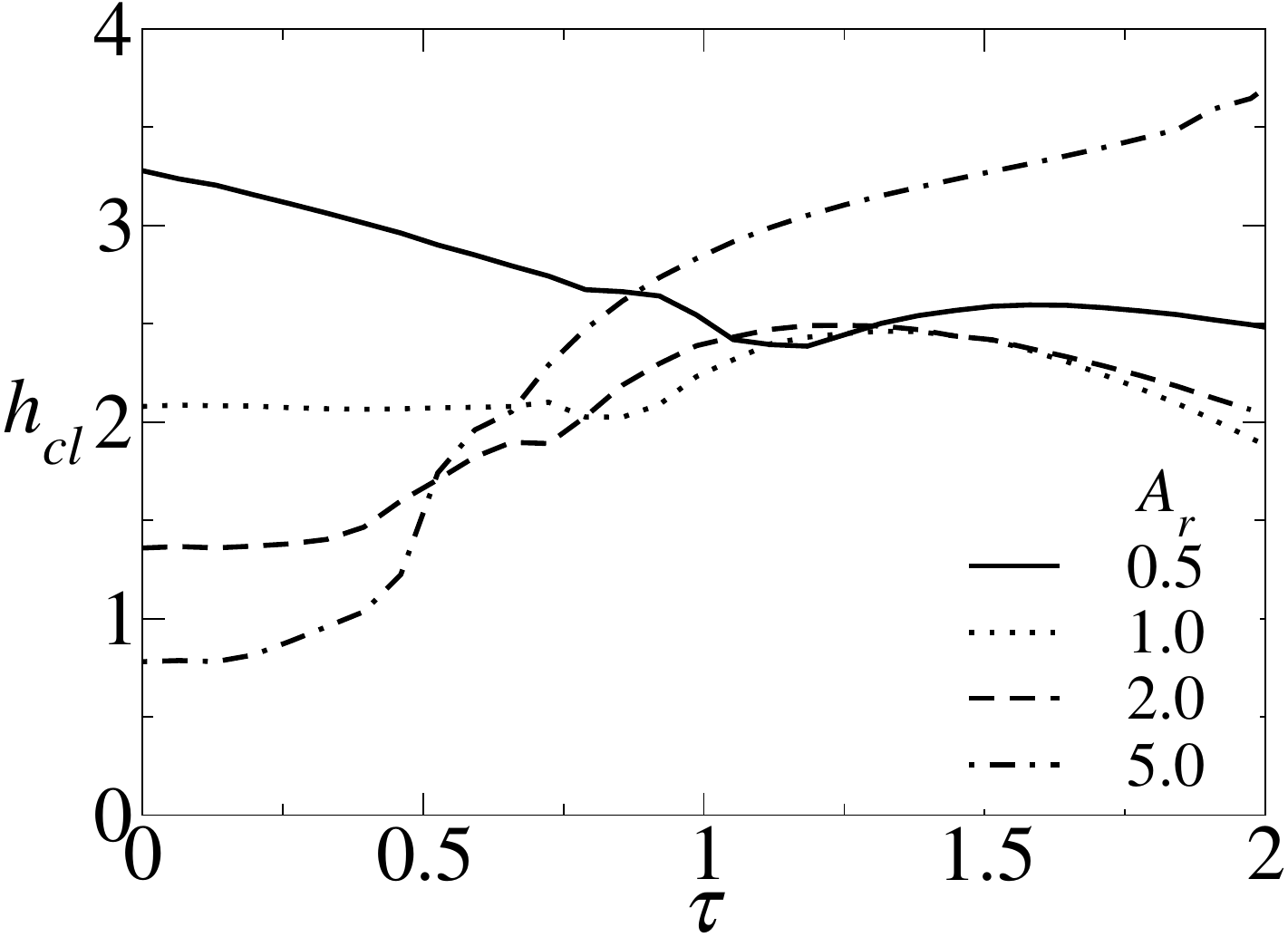}
\caption{\ks{$(a)$ A schematic representation of the column configuration, depicting the neck radius $(R_{n})$ and column height $(h_{cl})$ with respect to the undisturbed free surface of the liquid pool. $(b, c)$ Temporal evolutions of the neck radius $(R_{n})$ and height of the column $(h_{cl})$ for different values of $A_r$.}}
 \label{fig:neck_and_col}
\end{figure}

\ks{For a spherical drop ($A_r=1$), as depicted in Figure \ref{fig:3_D}, the drop initially contacts the liquid pool at $\tau=0$, following the evacuation of the air cushion between the drop and the free surface. As time progresses ($\tau > 0$), the drop undergoes drainage into the pool due to the pressure difference created by the negative curvature of the interface (Figure \ref{fig:3_D} at $\tau = 0.56$). The drop exhibits high capillary pressure near the contact region, generating two capillary waves, one upwards and the other into the pool. Once the wave within the droplet regime converges at the top, it induces the formation of a dent-like structure, observed at $\tau = 0.89$, followed by a whipping action of the tip at the free surface, initiating the formation of a cylindrical column. This whipping action leads to a sudden increase in the column height ($h_{cl}$) just after $\tau$ = 0.89, then increases slowly till it reaches a maximum height of $h_{cl} = 2.45R_{eq}$ at $\tau = 1.35$, which is also seen in Figure \ref{fig:neck_and_col}$(c)$. The illustration defining the column height ($h_{cl}$) as well as the neck radius ($R_{n}$) is presented in Figure \ref{fig:neck_and_col}$(a)$. The capillary wave, after rebounding, causes the droplet to neck, which, if dominated by the horizontal surface tension force, prevents secondary droplet formation. In the case of $A_r = 1.0$, the momentum induced by the capillary wave rebound dominates the horizontal surface tension force, resulting in the pinching off of a daughter drop (partial coalescence) with an inverted-tear drop shape at $\tau = 1.88$. This phenomenon is further highlighted in Figure \ref{fig:neck_and_col}$(b)$, where the neck radius diminishes at $\tau = 1.88$, signifying the occurrence of partial coalescence. Subsequently, the daughter drop deforms back towards a spherical shape, coalesces again into the pool, and undergoes a coalescence cascade until fully submerged. As discussed in \S\ref{sec:intro}, this phenomenon has been extensively studied experimentally and numerically by previous researchers \citep{thoroddsen2000coalescence,chen2006partial,blanchette2006partial,ray2010generation}.}

\ks{In contrast, a significantly prolate primary drop ($A_r=0.2$), touching the liquid pool at almost a single point at $\tau=0$, reverts to a spherical shape due to exceptionally high-pressure gradients within the droplet, a characteristic unique to $A_r=0.2$. Upon contact with the liquid pool, competition arises between the capillary pressure due to negative curvature near the contact point and the Laplace pressure inside the parent drop. The prevailing Laplace pressure induces a low-pressure zone at the center, with increased pressure at the top and bottom ends (Figure \ref{fig:flowfield}$a$). Consequently, a velocity field is initiated inside the droplet, directed towards the center, where the flow remains nearly stationary. This central stationary flow obstructs the flow from the polar ends, leading to bulges forming away from the center ($\tau = 1.12$ in Figure \ref{fig:flowfield}$a$). The resulting crater exhibits low pressure attributed to negative curvature. The droplet morphology exhibits two high-pressure regions on each bulge, prompting fluid flow towards the center, causing the crater to overshoot ($\tau = 1.58$). This sequence results in the droplet adopting a dumbbell shape at $\tau = 2.0$ and eventually transforming into a shape resembling a bicone at $\tau = 2.63$ and later, reverting to a shape similar to the initial form of the drop, albeit less predominantly prolate (observe $\tau = 3.19$), before undergoing oscillations and returning to a spherical shape or making contact with the free surface.}

\ks{For $A_r=0.5$, the capillary pressure near the contact point and the Laplace pressure inside the parent drop become comparable, unlike a detachment condition observed for $A_r=0.2$. In the initial stages of coalescence, the parent drop experiences high pressure at its uppermost region, leading to liquid migration towards the center and subsequent horizontal bulging. Simultaneously, liquid continuously drains into the pool due to the negative curvature caused by the neck ($\tau = 0.10$ in Figure \ref{fig:flowfield}$b$). A capillary wave manifests above and below the neck of the droplet, displaying dynamics similar to the $A_r=1.0$ case, albeit with delayed whipping action and column formation observed at $\tau = 1.15$ (Figure \ref{fig:flowfield}$b$) and in Figure \ref{fig:neck_and_col}$(c)$. Unlike other $A_r$ values, the increase in column height persists within $\tau = 1.15 - 1.5$, indicating a lesser induced upward momentum by the capillary wave—i.e., the change in $h_{cl}$ is less for an extended period, resulting in significantly slower column height increase. Consequently, the droplet does not undergo pinch-off but instead undergoes a cycle of expansion and contraction, as depicted in Figure \ref{fig:neck_and_col}$(b)$, where $R_n$ oscillates until all the liquid is drained into the pool. This phenomenon is known as total coalescence. Parent drops with $A_r > 0.67$ exhibit partial coalescence behavior, displaying dynamics similar to that of a spherical drop.}

\begin{figure}
\centering
\includegraphics[scale=0.6]{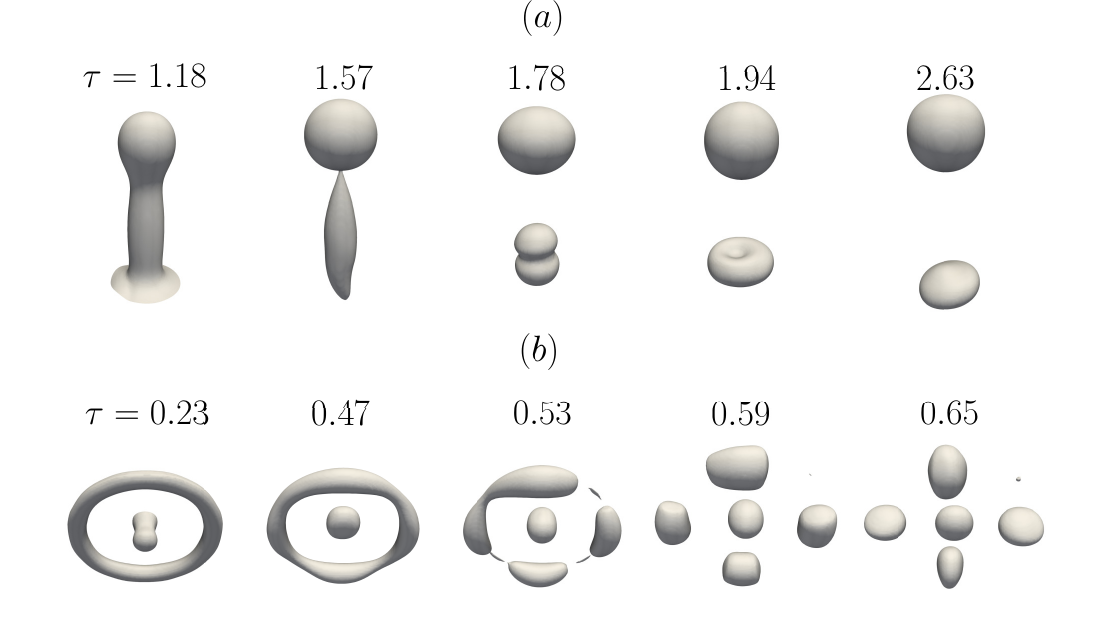}
\caption{$(a)$ Temporal evolution of the splitting of the liquid column due to Rayleigh-Plateau capillary instability above the free surface of the liquid pool. $(b)$ Evolution of the entrapped air bubble inside the liquid pool for $A_r = 5.0$. The rest of the dimensionless parameters are $We=0$, $\Oh = 7.82 \times 10^{-3}$ and $\Bo = 0.04$.}
\label{fig:RP_instability}
\end{figure}

\ks{An intriguing phenomenon unfolds with $A_r=5.0$, where the emergence of two daughter droplets is observed. In this scenario, a significant volume of air between the parent oblate drop and the pool is displaced outward upon contact with the free surface. Consequently, the drop establishes contact across a large area, introducing inherent asymmetry into the coalescence process and trapping multiple air bubbles inside the liquid pool. The parent drop undergoes a horizontal Laplace pressure gradient, inducing horizontal collapse and liquid migration from the ends toward the center, reminiscent of the dynamics observed in $A_r = 0.2$. However, this process leads to faster bulging near the horizontal periphery, forming an annular contact region with the pool and trapping air in a ring-like bubble, as depicted in Figure \ref{fig:RP_instability}$(b)$. The instability of the ring-shaped bubble causes fragmentation into smaller air bubbles, which move towards the free surface asymmetrically. The free surface develops a three-dimensional morphology, and surface-wave structures become chaotic at $\tau=1.88$, eventually settling over time. Furthermore, a significant capillary wave above the free surface contributes to the formation of a pronounced depth in the crater ($\tau = 0.43$ in Figure \ref{fig:flowfield}$d$) just before the capillary waves converge. This leads to an intense whipping action at the topmost tip of the droplet regime, evident in Figure \ref{fig:neck_and_col}$(c)$ during the time period of $\tau = 0.43 - 0.89$. The formation of the column is remarkably rapid compared to any other case discussed, resulting in a very high upward momentum within the column, i.e., $h_{cl}$ varies from 1.1$R_{eq}$ to 2.7$R_{eq}$. This heightened upward momentum is the primary factor behind the formation of the tall column structure at $\tau = 0.89$. The liquid column undergoes a Rayleigh-Plateau capillary instability, eventually pinching off at $\tau = 1.61$ to form one daughter droplet. The remaining liquid in the column also forms another non-spherical blob of liquid due to surface tension, whose evolution is depicted in Figure \ref{fig:RP_instability}$(a)$. The phenomenon of multiple daughter droplets due to Rayleigh-Plateau instability has been observed by several researchers \citep{ray2010generation,fedorchenko2004some,charles1960mechanism}. As the aspect ratio of the parent drop increases, the width of the column structure decreases while the maximum height of migration for the daughter drop increases. The detachment time $(\tau_d)$ of the daughter drop and its volume $(V_d)$ appear to be strongly influenced by the aspect ratio of the parent drop.}

\ks{The temporal evolution of the neck radius ($R_n$) and the column height ($h_{cl}$) for different $A_r$ values is presented in Figure \ref{fig:neck_and_col}$(b, c)$. In the case of $A_r = 0.5$, as shown in Figure \ref{fig:neck_and_col}$(b)$, the oscillation of the neck radius suggests no formation of a secondary droplet, whereas $(R_n)$ approaches zero for $A_r$ = 1.0 and 2.0, indicating the formation of a secondary droplet in these instances. Figure \ref{fig:neck_and_col}$(c)$ illustrates the variation of the column height $(h_{cl}$) with time for different $A_r$. It is evident that the rate of increase in $h_{cl}$ is higher with an increase in $A_r$ for an oblate droplet. This heightened rate is attributed to the higher capillary pressure gradient for larger $A_r$, which generates a horizontal velocity field directed towards the center. This induces a rush of fluid in the vertical direction, causing an increase in $h_{cl}$. Conversely, for $A_r = 0.5$, the maximum capillary pressure is concentrated at the vertical poles of the droplet, resulting in a vertical flow towards the center and a consistent decrease in $h_{cl}$. For a spherical droplet ($A_r = 1.0$), the change in $h_{cl}$ appears considerably less compared to a non-spherical droplet. This is because there is no internal capillary pressure gradient independently caused by the droplet. A sudden but slight change in $h_{cl}$ observed around $\tau = 0.5 \sim 1.0$ indicates the capillary wave meeting at the topmost end of the droplet, causing a sudden catapulting action. The rate of this sudden increase in $h_{cl}$ determines the occurrence of either the phenomenon of total coalescence or partial coalescence. Further insights into these findings are provided in \S\ref{sec:pc}, where the dynamics of coalescence with zero impact velocity of the droplet are described in detail.}

\begin{figure}
\centering
\hspace{0.8cm} {\large ($a$)} \hspace{5.5cm} {\large ($b$)} \\
\includegraphics[scale=0.24]{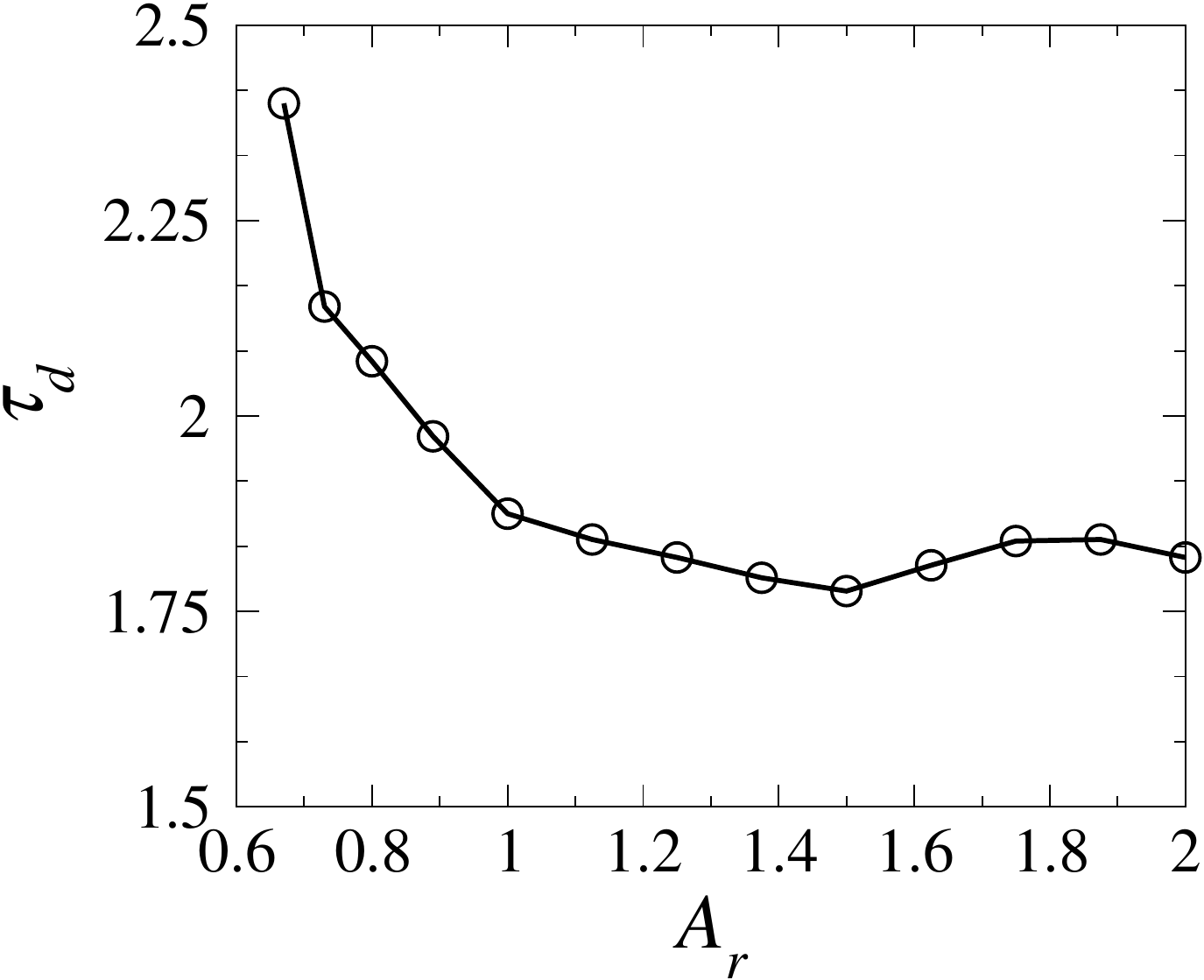} \hspace{2mm} 
\label{fig: detach time}
\includegraphics[scale=0.24]{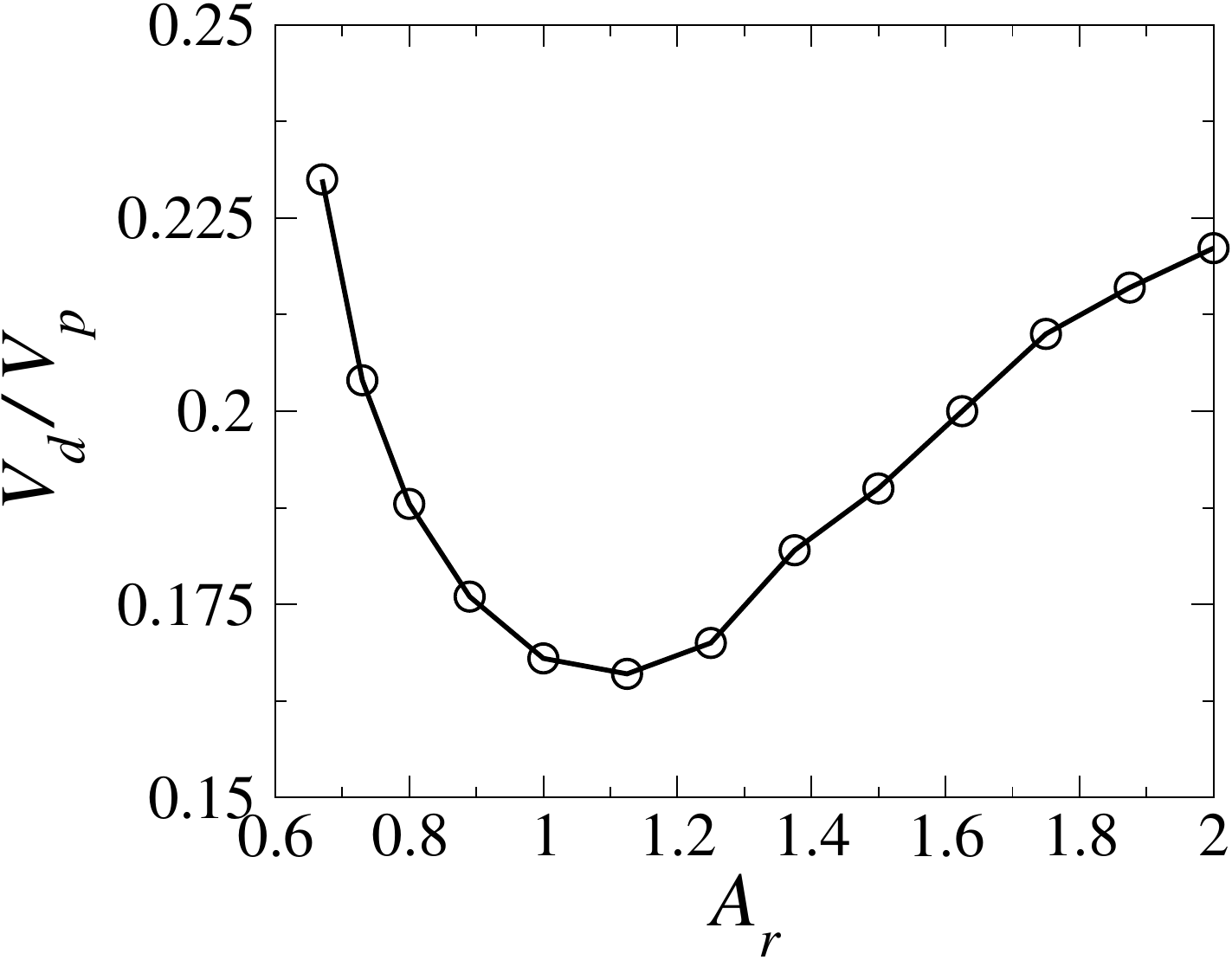}
\caption{The variation of the ($a$) detachment time ($\tau_d$) and ($b$) normalized volume of the daughter droplet $(V_d/V_p)$ with the aspect ratio of the parent drop $(A_r)$. The rest of the dimensionless parameters are $\Oh = 7.82 \times 10^{-3}$, $\Bo = 0.04$ and $\We=0$.}
\label{fig:2nd_drop_plots}
\end{figure}

Figure \ref{fig:2nd_drop_plots}($a$) and ($b$) depict the variations of the detachment time ($\tau_d$) and the normalised volume of the daughter drop $(V_d/V_p)$ emerging due to the partial coalescence phenomenon with the aspect ratio of the parent drop $(A_r)$. As discussed above, the parent drops with  $A_r<0.67$ do not experience the partial coalescence phenomenon; thus, no daughter droplet is generated.  For $A_r > 0.67$, the presence of upward-moving capillary waves is significant enough to delay the vertical collapse of the liquid column with the dominating horizontal collapse that promotes the neck formation near the free surface and the detachment of liquid mass to form a daughter/secondary drop. Inspection of Figure \ref{fig:flowfield} reveals that as the aspect ratio ($A_r$) increases, the lateral velocity also increases. This lateral velocity then extends across the longitudinal axis of the liquid column, resulting in a higher upward momentum at the top. This, in turn, leads to an earlier pinch-off of the droplet, as depicted in Figure \ref{fig:2nd_drop_plots}($a$). In this situation, the parent droplet will not have sufficient time to release the fluid into the pool, forming a larger daughter droplet. For $A_r < 1$, the initial curvature difference in the droplet results in a notable pressure gradient. In other words, the droplet exhibits more pressure at the longitudinal tips and less pressure at the lateral ends. This creates a net fluid flow towards the lateral direction, causing the parent droplet to bulge sideways. Despite having a longer detachment time, this bulge prevents the fluid from draining into the pool due to the bulk momentum being in the lateral direction. Consequently, the daughter droplet attains a minimum size near $A_r = 1$, as depicted in Figure \ref{fig:2nd_drop_plots}$(b)$. It is essential to highlight that as the capillary wave moves across the surface of the droplet in an upward direction, the variation of the detachment time with aspect ratio follows a similar trend as that of the circumference of the droplet. The detachment requires the capillary wave to traverse the entire droplet surface twice, covering a total travel length of $\pi (a + b)$, with $a$ and $b$ representing the major and minor diameters of the parent drop.

\begin{figure}
\centering
\includegraphics[scale=0.55]{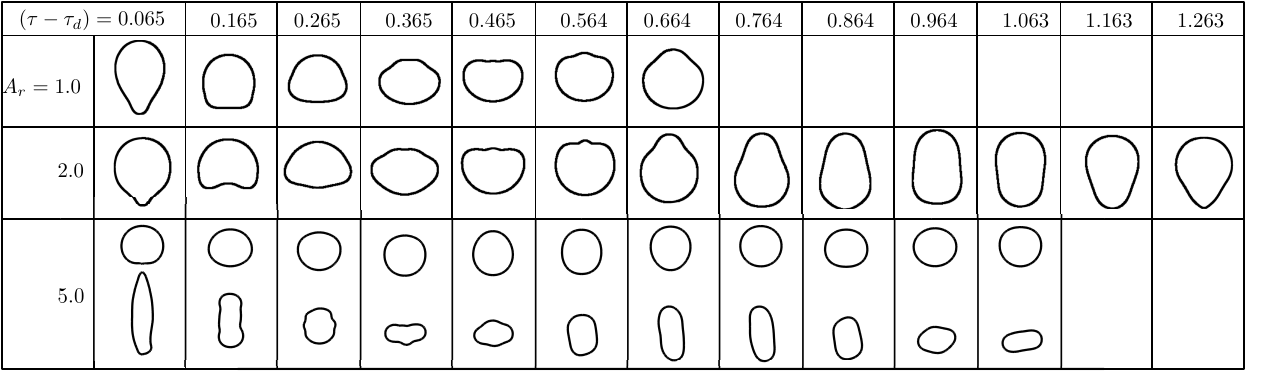}
\caption{Temporal evolution of the shapes of the secondary drop after detachment for different aspect ratios $(A_r)$. The rest of the dimensionless parameters are $\Oh = 7.82 \times 10^{-3}$, $\Bo = 0.04$ and $\We=0$.}
\label{fig:secondarydrop}
\end{figure}

Figure \ref{fig:secondarydrop} shows the temporal evolution of the morphology of the daughter droplet (cross-sectional view) associated with the spherical ($A_r=1$) and oblate parent drops with $A_r=2$ and 5. The observation spans from just after the detachment instant ($\tau-\tau_d=0$) until the daughter droplet reconnects with the pool. As depicted in Figure \ref{fig:3_D}, drops with $A_r=1$ and 2 exhibit nearly identical behaviour, oscillating significantly from inverted-teardrop to teardrop shapes. However, for $A_r=2$, the daughter droplet ascends to a higher height than $A_r=1$. Consequently, the daughter droplet for $A_r=2$ requires significantly more time to return to the liquid pool while falling under the action of gravity compared to the spherical drop ($A_r=1$). In contrast, $A_r=5$ exhibits two daughter drops. The liquid column, generated by the upward-moving capillary wave, experiences a detachment process near the free surface, primarily driven by the prevailing horizontal collapse over the vertical collapse. This process is similar to that observed for lower aspect ratios. Concurrently, the top part of the liquid column undergoes fragmentation due to Rayleigh-Plateau instability, forming a slightly oblate-shaped daughter drop that subsequently undergoes oblate-prolate shape oscillations. The remaining liquid in the column amalgamates into a blob of complex shape (as also depicted in Figure \ref{fig:RP_instability}$a$. This drop also exhibits shape oscillations, albeit asymmetrical, that result from the interplay between inertia and surface tension. These observations also highlight the importance of the three-dimensional numerical simulations conducted in the present study. 

\subsection{Splashing}\label{sec:splashing}

Next, we investigate the collision of non-spherical parent drops with varying aspect ratios into the liquid pool, taking into account different impact velocities characterized by the Weber number ($\We=\rho_l v_0^2 R_{eq}/\sigma$). Here, we consider the impact velocities, $v_0 = 3.5$ m/s and 5 m/s, while maintaining the remaining parameters consistent with those considered in \S\ref{sec:pc}. They are associated with $We=93.4$ and 190.6, respectively. In order to minimize the boundary effects, we employ a larger computational domain of size $30R_{eq} \times 30R_{eq} \times 30R_{eq}$ in our simulations. 

\begin{figure}
\centering
\includegraphics[scale=0.43]{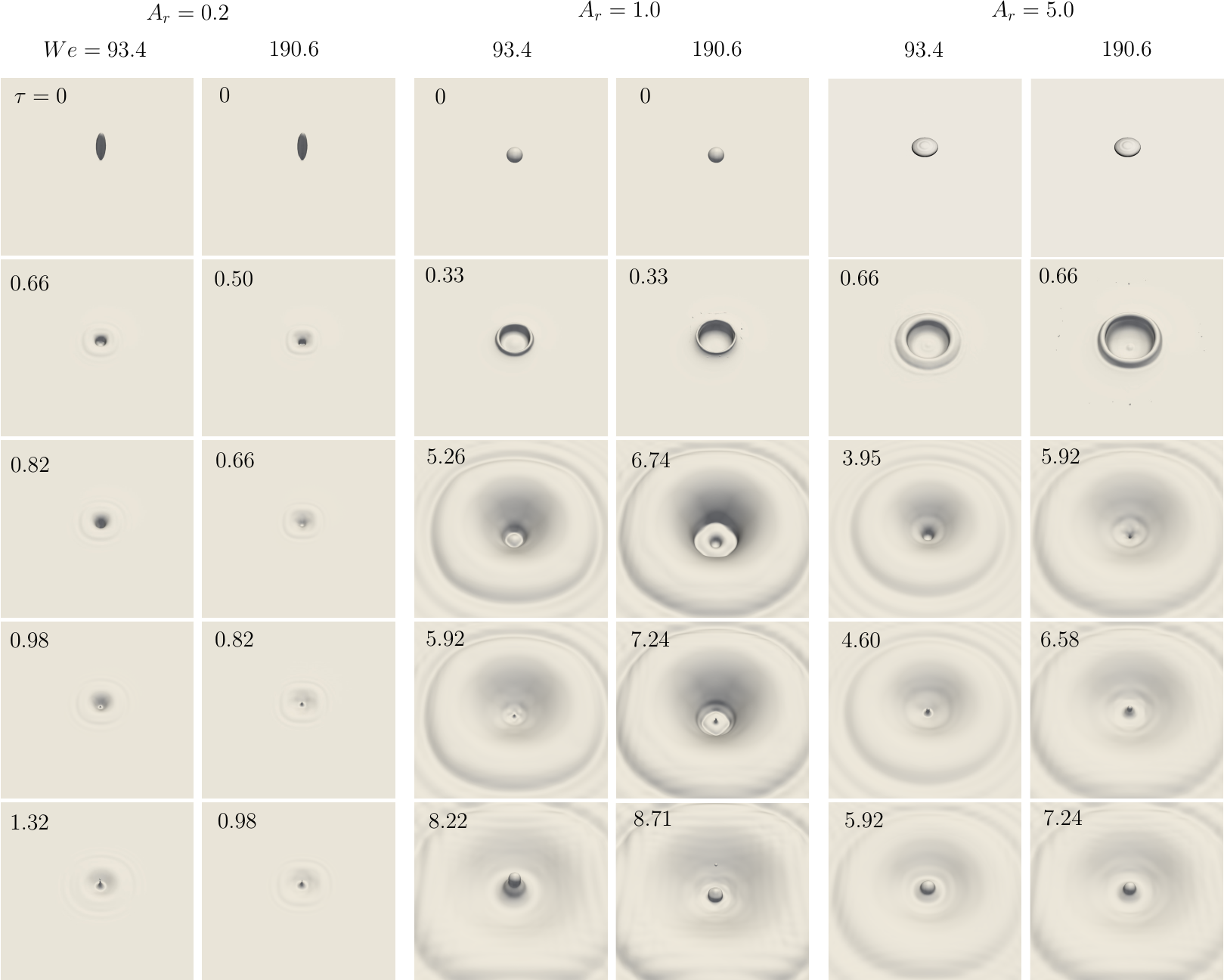}
\caption{Temporal evolution of the interface morphology resulting from the impact of a parent drop with different aspect ratios $(A_r)$ and different impact velocities. The rest of the dimensionless parameters are $Oh = 7.82 \times 10^{-3}$ and $Bo = 0.04$.}
\label{fig:Splash_3d}
\end{figure}

\begin{figure}
\centering
\includegraphics[scale=0.45]{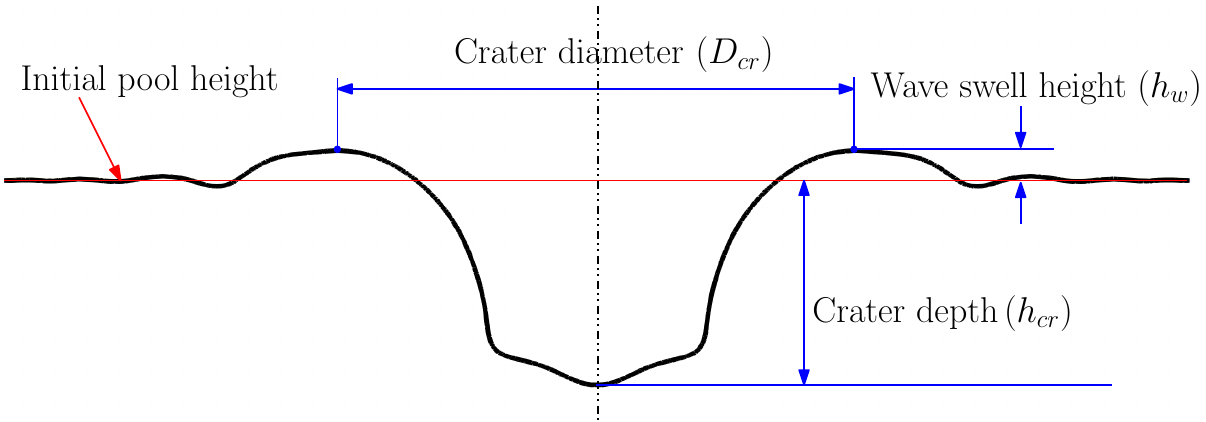}
\caption{A schematic representation illustrating the crater configuration, depicting the crater depth $(h_{cr})$, diameter $(D_{cr})$, and wave swell size $(h_w)$ with respect to the undisturbed free surface of the liquid pool.}
\label{fig:Notation_cd_wh}
\end{figure}

\begin{figure}
\centering
\hspace{0.3cm} {($a$)} \hspace{4.8cm} {($b$)} \\
\includegraphics[scale=0.2]{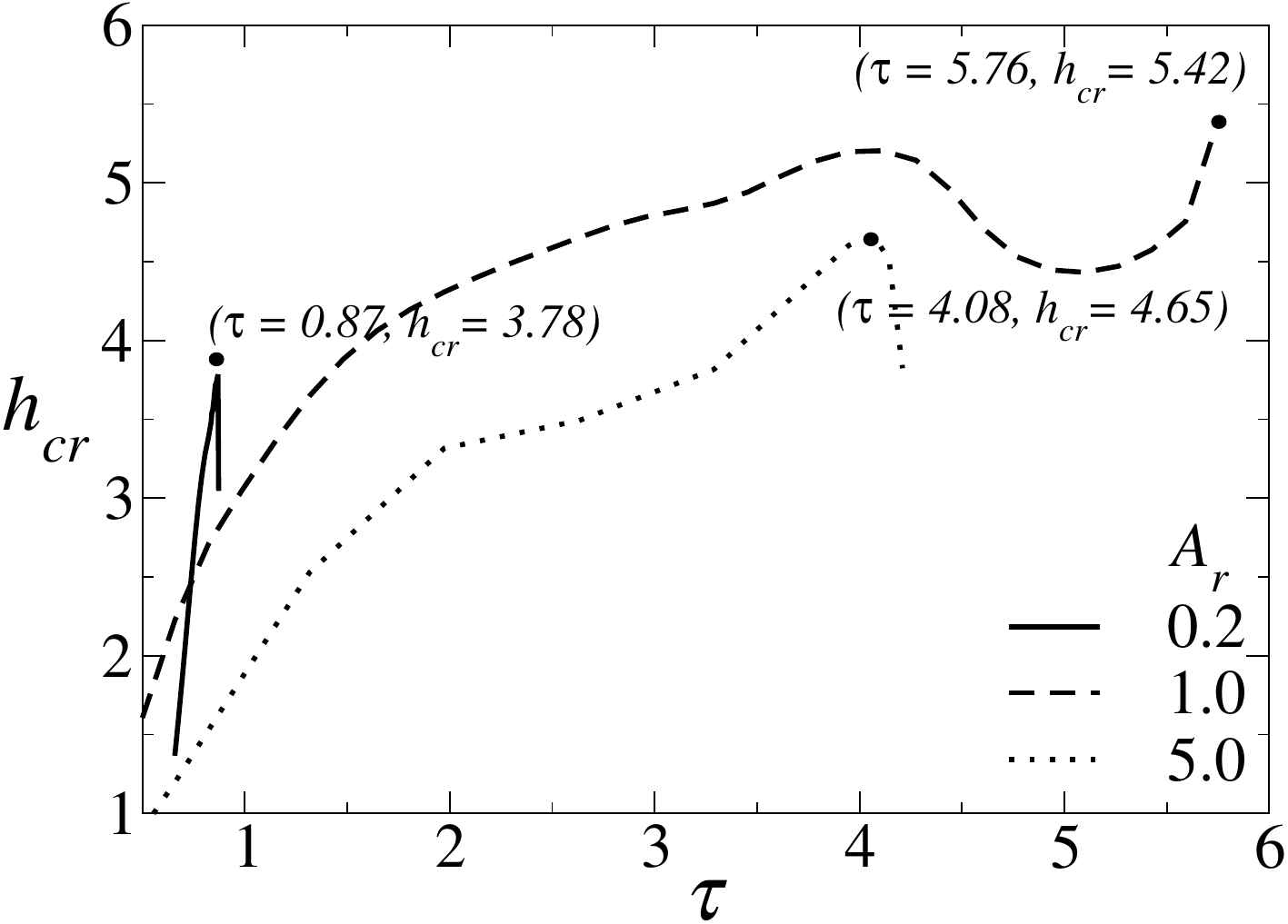} \hspace{0.2cm} \includegraphics[scale=0.2]{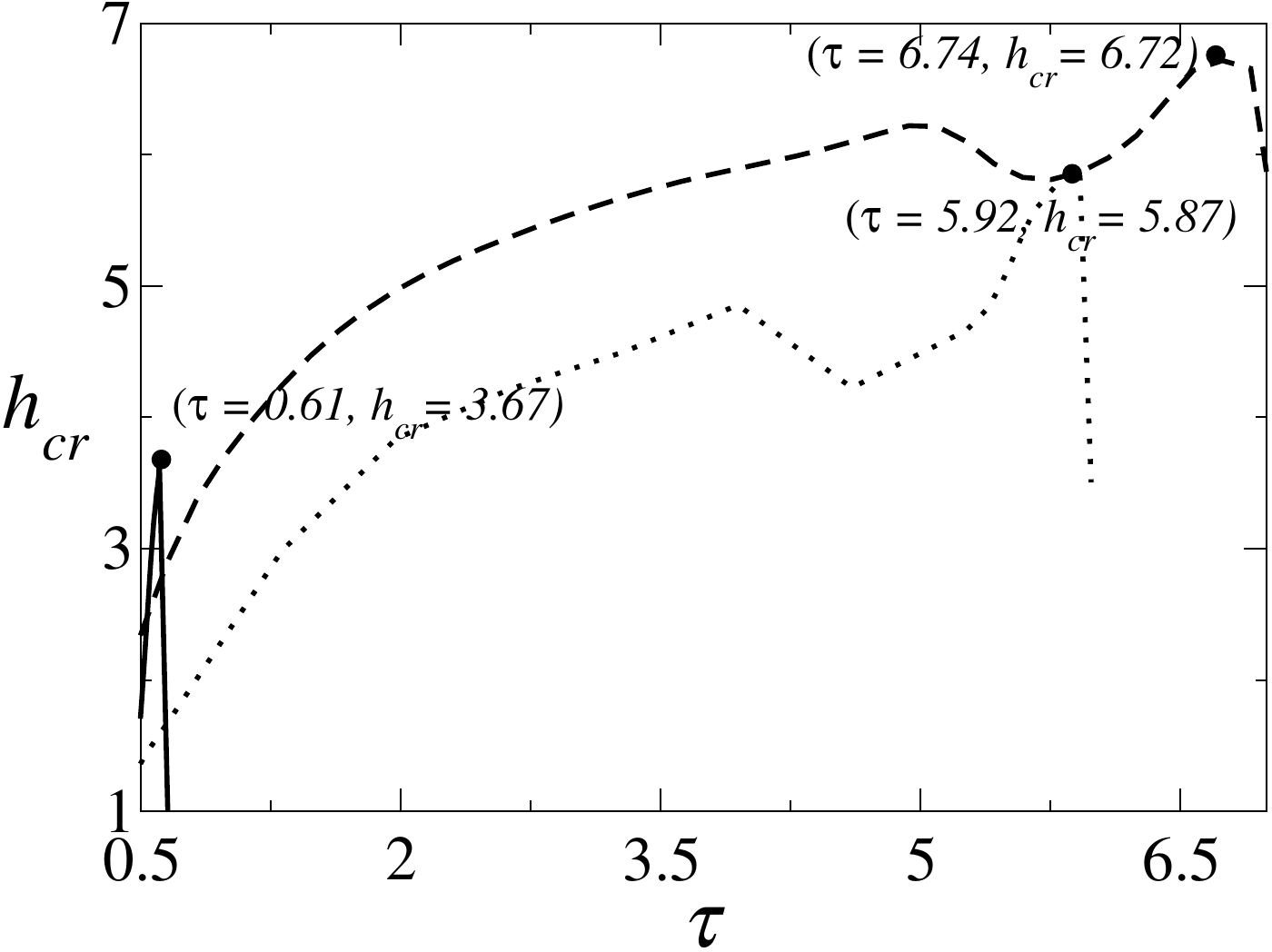} \\
\hspace{0.3cm} {($c$)} \hspace{4.8cm} {($d$)} \\
\includegraphics[scale=0.2]{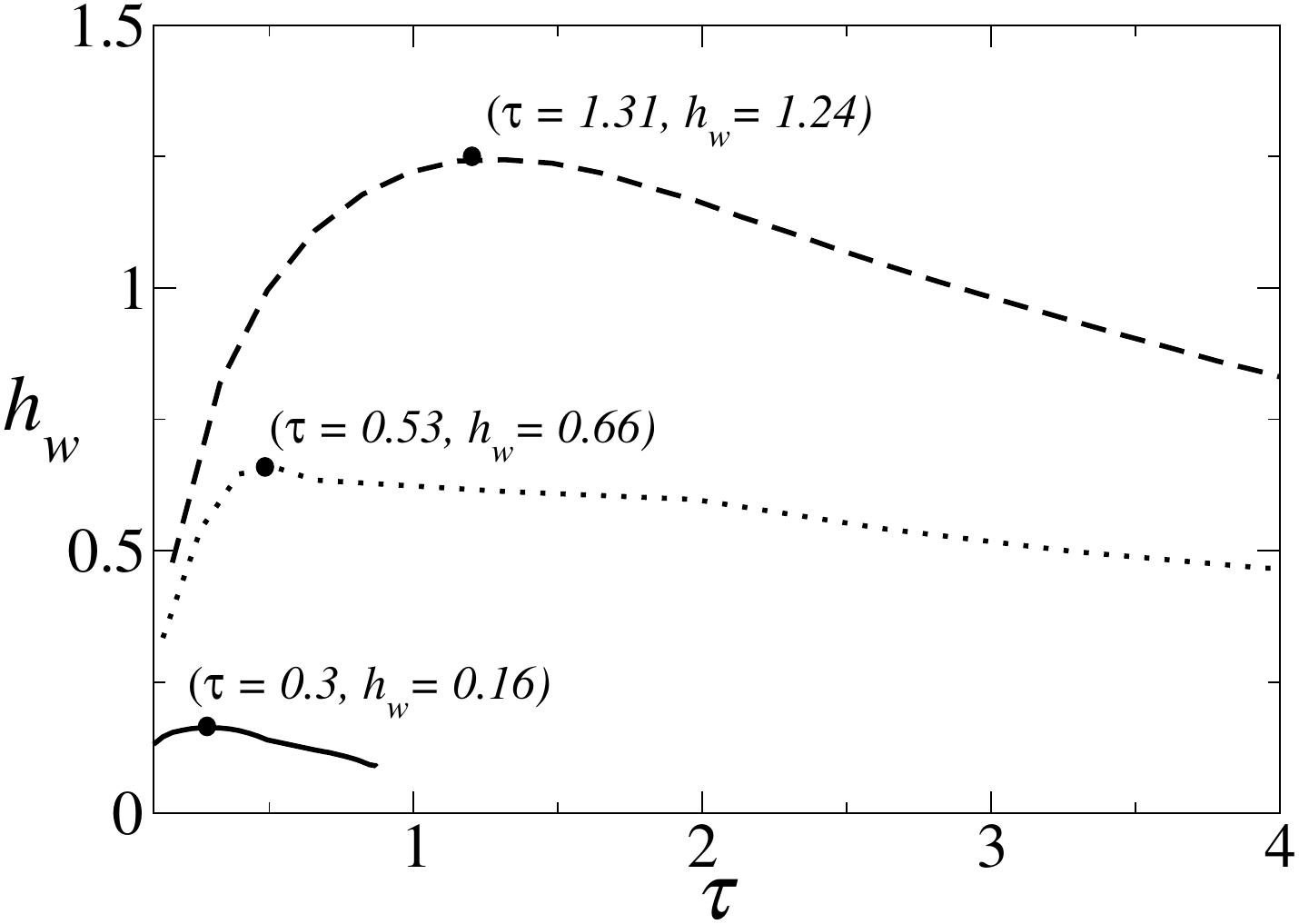} \hspace{0.25cm} \includegraphics[scale=0.2]{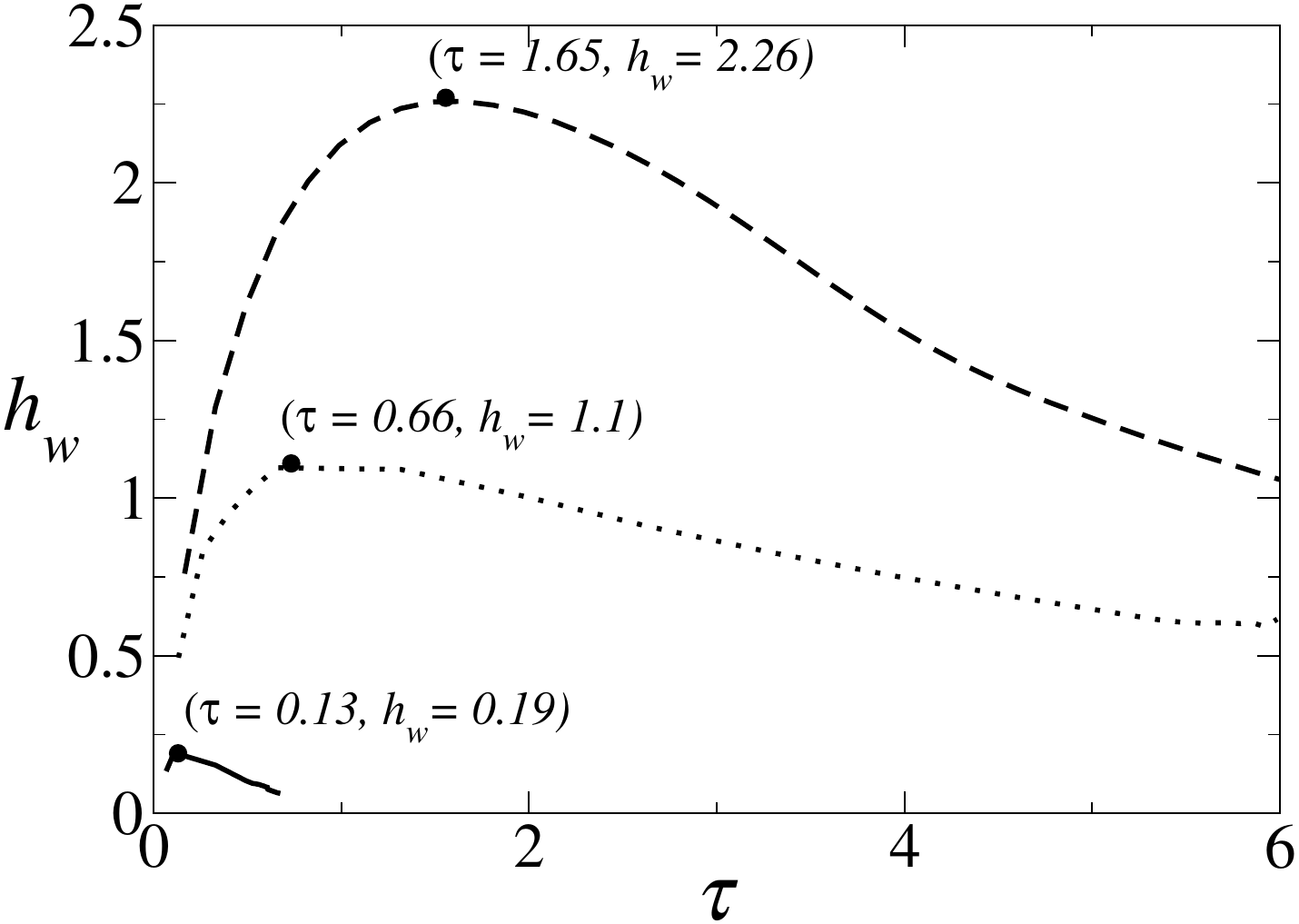} \\
\caption{\ks{Temporal variation of non-dimensional crater depth $(h_{cr})$ and wave swell height $(h_{w})$ for different aspect ratios and Weber numbers. Panels $(a, b)$ and $(c, d)$ are associated with $h_{cr}$ and $h_{w}$, respectively. Panels $(a,c)$ and $(b,d)$ correspond to $We=93.4$ ($v_0 =  3.5$ m/s) and $We=190.6$ ($v_0 = 5$ m/s), respectively. The rest of the parameters are $Oh = 7.82 \times 10^{-3}$ and $Bo = 0.04$.}}
\label{fig:CD_WH}
\end{figure}

In Figure \ref{fig:Splash_3d}, two effects are investigated: (i) the effect of the impact velocity $(We)$ of a water droplet with $A_r=0.2$, 1, and 5.0, and (ii) the influence of $A_r$ for $We=93.4$ and 190.6 on the temporal evolution of the morphology of the air-water interface due to the impact of the drops. The remaining dimensionless parameters are $\Oh = 7.82 \times 10^{-3}$ and $\Bo = 0.04$. We first discuss the effect of the Weber number for $A_r=1$. Figure \ref{fig:Splash_3d} (for $A_r=1.0$) depicts the manifestation of three distinct phenomena in the collision process, namely, (i) the expansion of the crater and the rim of the wave (wave swell), (ii) the retraction of the wave swell leading to the retraction of the crater sides, and (iii) the retraction of the crater base. It can be seen that at $\tau = 0$, the drop comes in contact with the liquid pool and forms a hemispherical crater surrounded by a rim-like structure. As expected, the rim appears sharper at higher Weber numbers (see, at $\tau = 0.33$). Over time, the rim expands, transforming into a wave swell. In Figure \ref{fig:Notation_cd_wh}, a comparison between craters at $\We=93.4$ and $190.6$ illustrates that increasing the impact velocity increases crater depth $(h_{cr})$, diameter $(D_{cr})$ and wave swell size $(h_w)$. Figure \ref{fig:CD_WH}$(a,b)$ and $(c,d)$ depict the temporal variation of non-dimensional crater depth and wave swell height for various aspect ratios and Weber numbers. For $A_r = 1.0$ i.e., the spherical droplet, they clearly show that the crater depth is greater for $\We = 190.6$ in comparison to $\We = 93.4$. Consequently, the wave swell retracts earlier for $\We=93.4$ compared to $\We=190.6$. Thus, in the case of $\We = 93.4$, the rapid retraction of the crater prevents it from rebounding fully, resulting in an air pocket before collapse. The wave swell height attains its peak much sooner for $\We = 93.4$ in comparison to $\We = 190.6$, indicating that the wave retraction occurs earlier for $\We = 93.4$ than for $\We = 190.6$. Conversely, at $\We = 190.6$, the larger crater amplitude delays the retraction, allowing the crater to rebound sufficiently, preventing air entrapment. This leads to a high pressure gradient at the crater base in both cases, with $\We = 190.6$ exhibiting a stronger gradient. The pressure gradient prompts jet formation at the crater base. The strength of the jet plays an important role in the development of the Rayleigh-Plateau instability, which triggers the breakup of the jet, giving rise to smaller daughter droplets. In the case of $\We = 93.4$, the jet is comparatively weaker due to air bubble entrapment, resulting in a less pronounced pressure gradient than $\We = 190.6$, where no air bubble entrapment occurs. The strength of the jet is also influenced by the maximum height reached by the wave swell. Despite achieving an earlier maximum in wave swell height, the actual height is not as substantial for $\We = 93.4$ compared to $\We = 190.6$. This height plays a crucial role in determining the intensity of the crater collapse, consequently leading to a stronger catapulting effect of the crater base. The dissimilarity in jet strength becomes apparent at $\tau = 8.71$ for $\We = 190.6$ in Figure \ref{fig:Splash_3d}, where the Rayleigh-Plateau instability induces jet breakup and droplet formation. In contrast, at $\tau = 8.22$ for $\We = 93.4$ in Figure \ref{fig:Splash_3d}, no droplet formation takes place. Subsequently, the jet thickens and transforms into a flat interface due to Laplace pressures and viscous dissipation. These phenomena also resemble those observed by \cite{ray2015regimes} for the impact of a spherical drop ($A_r=1.0$) on a liquid pool.

\begin{figure}
\centering
\hspace{0.3cm} {\large ($a$)} \hspace{5.5cm} {\large ($b$)} \\
\includegraphics[scale=0.23]{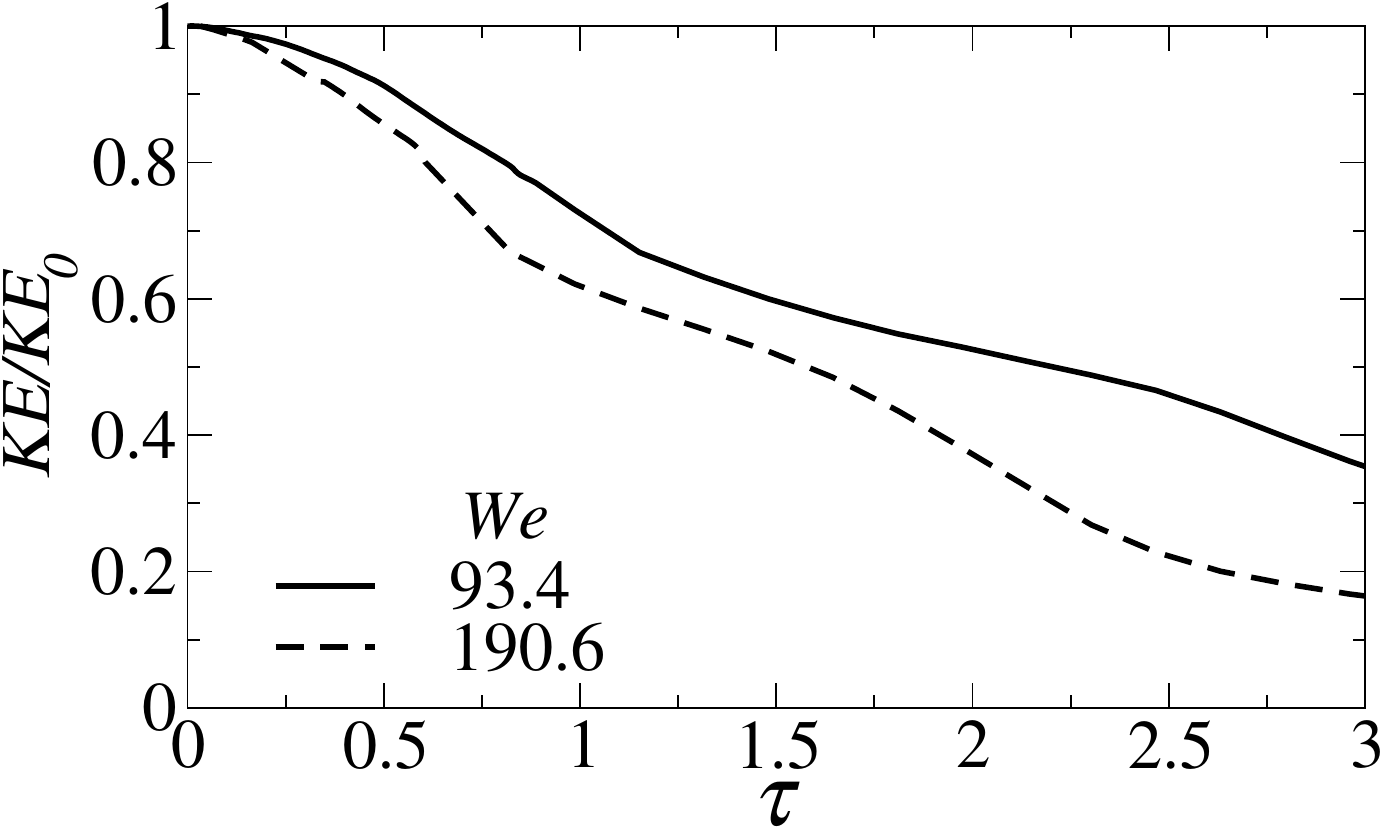} \hspace{0.2cm} \includegraphics[scale=0.23]{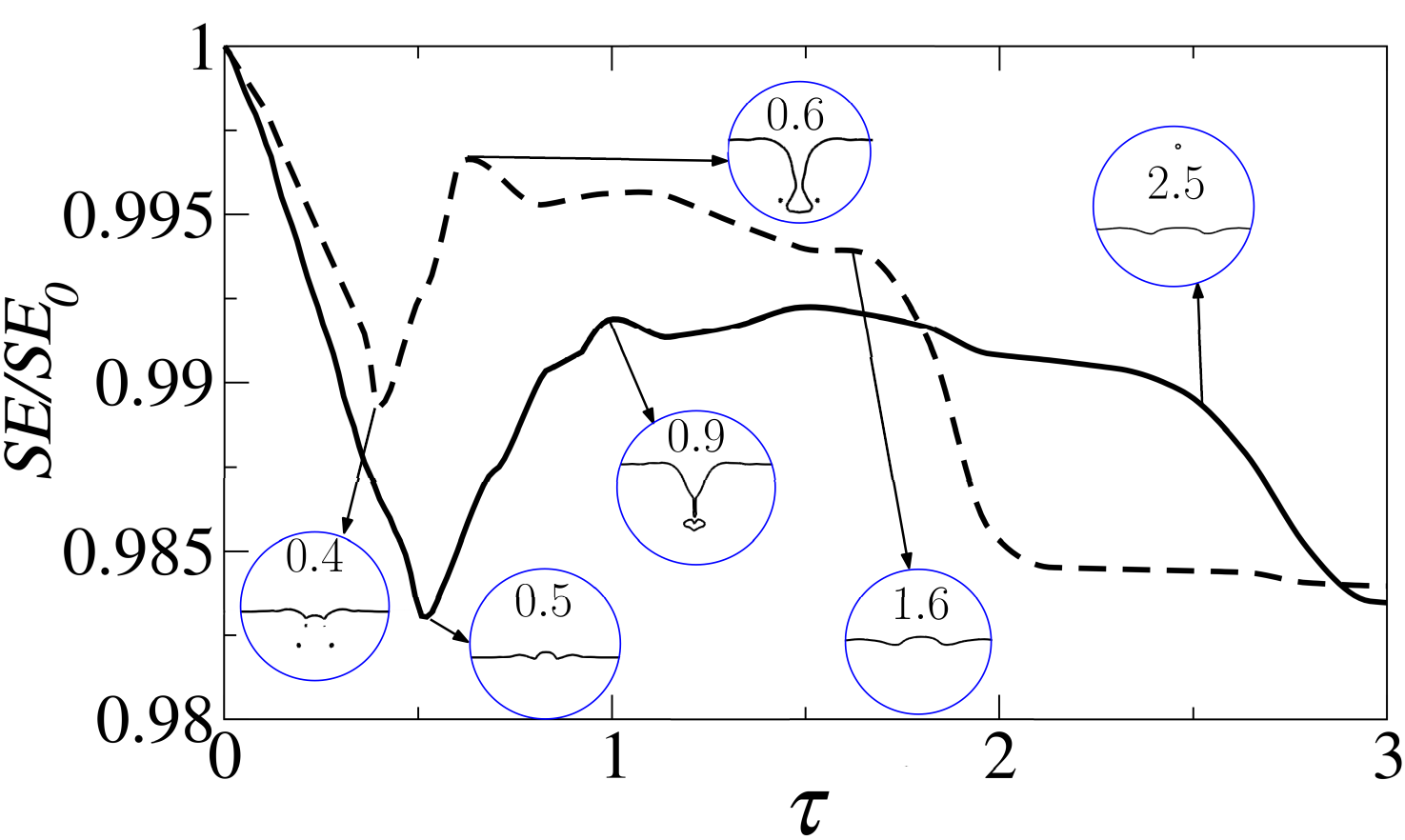} \\
\vspace{0.2cm}
\hspace{0.3cm} {\large ($c$)} \hspace{5.5cm} {\large ($d$)} \\
\includegraphics[scale=0.23]{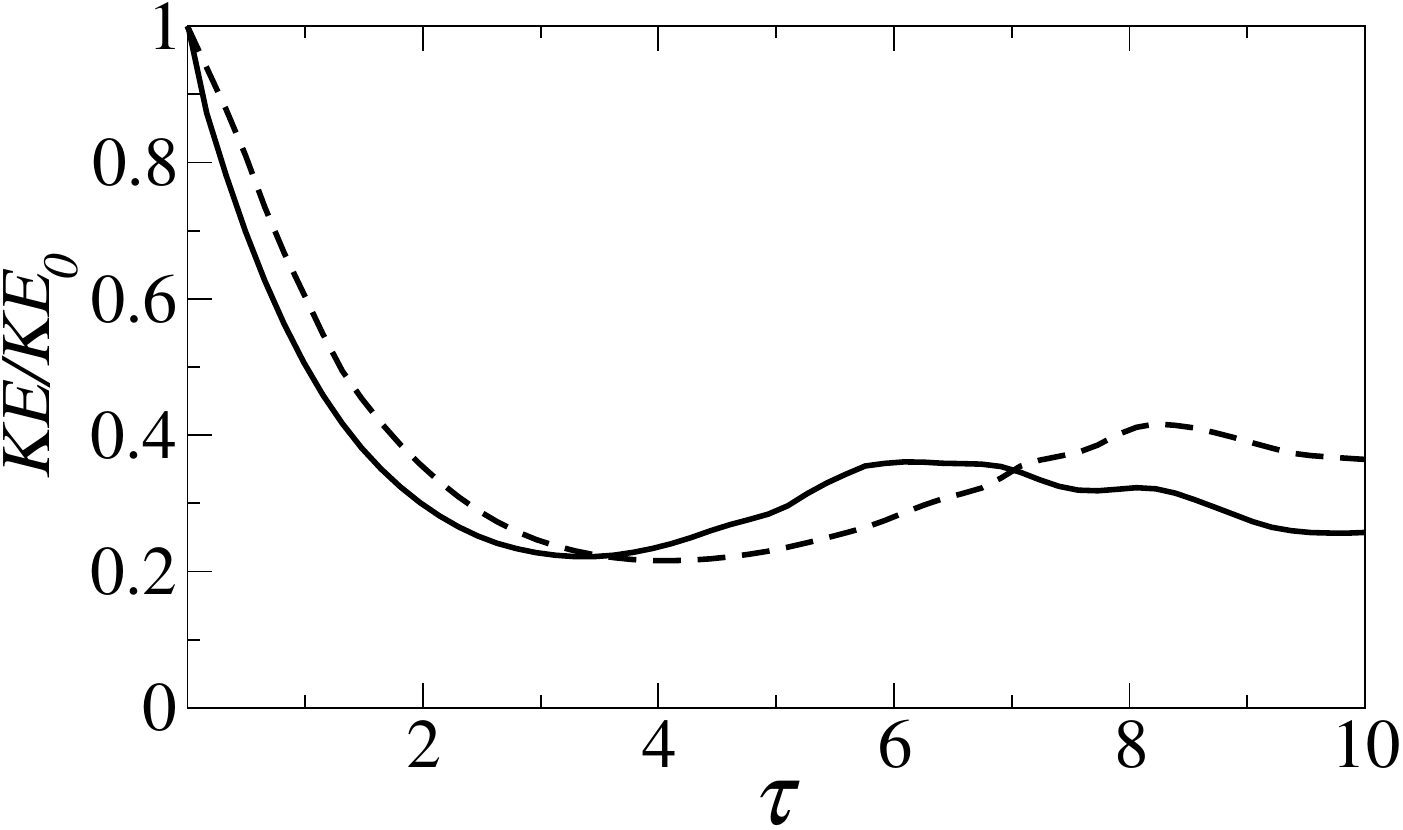} \hspace{0.25cm} \includegraphics[scale=0.23]{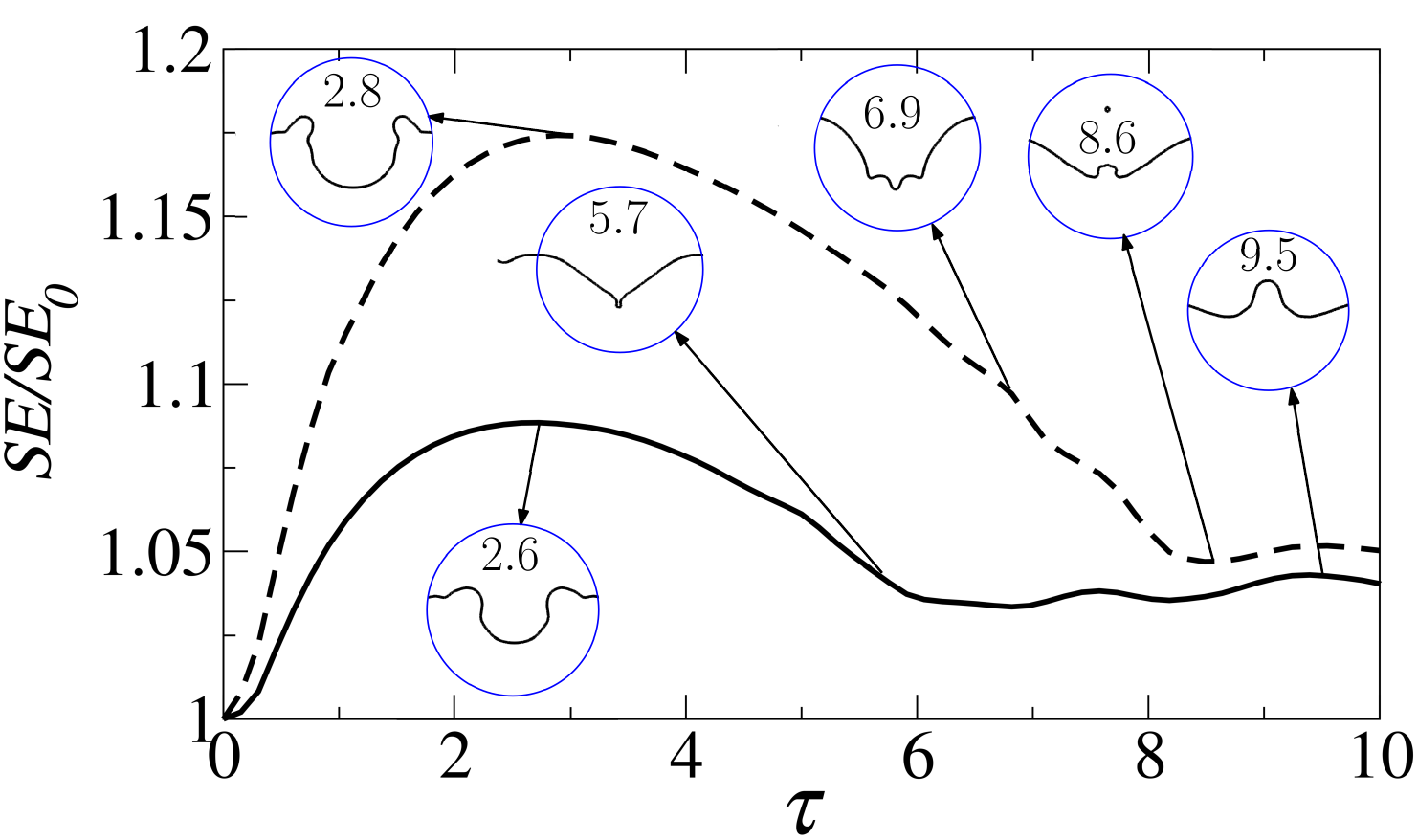} \\
\vspace{0.2cm}
\hspace{0.3cm} {\large ($e$)} \hspace{5.5cm} {\large ($f$)} \\
\includegraphics[scale=0.23]{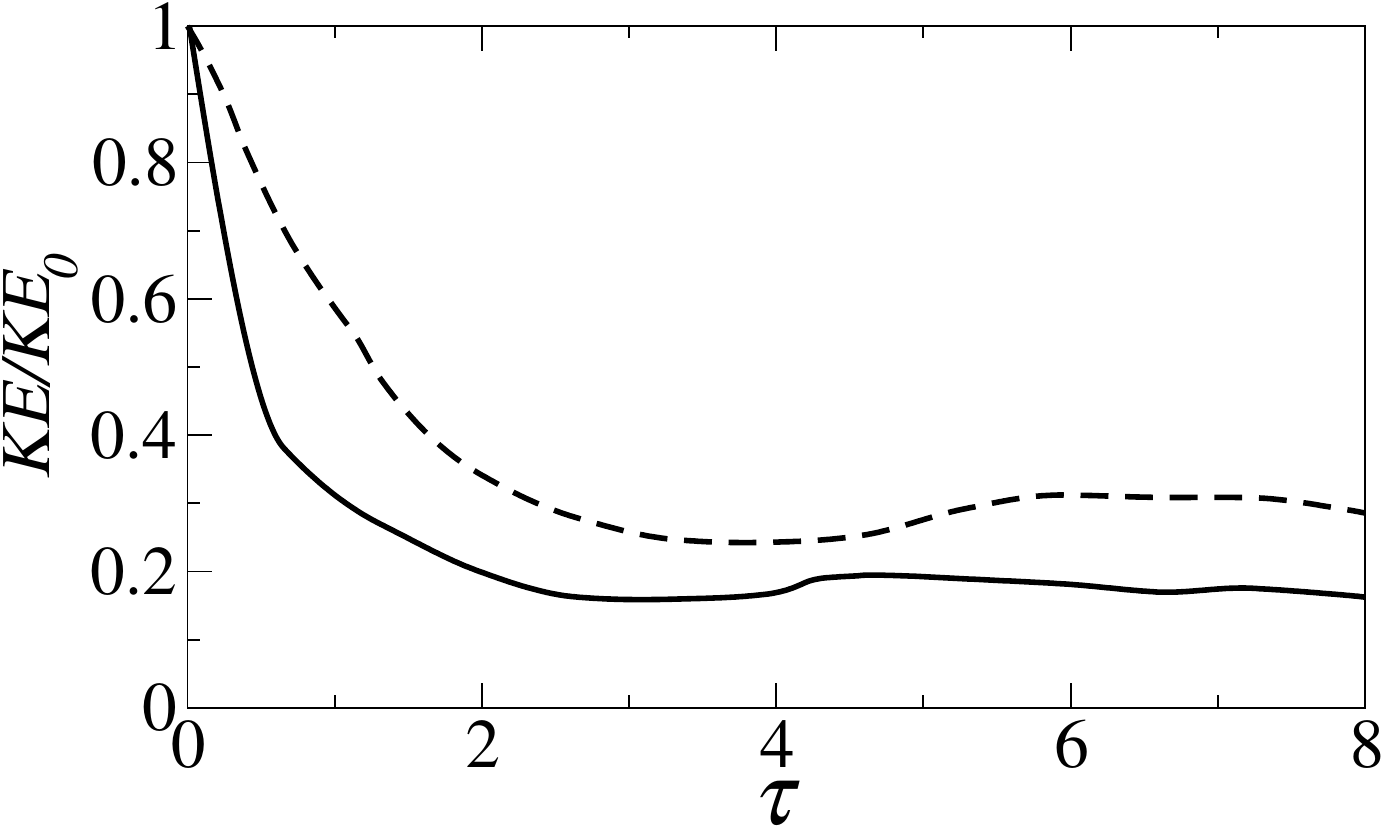} \hspace{0.25cm} \includegraphics[scale=0.23]{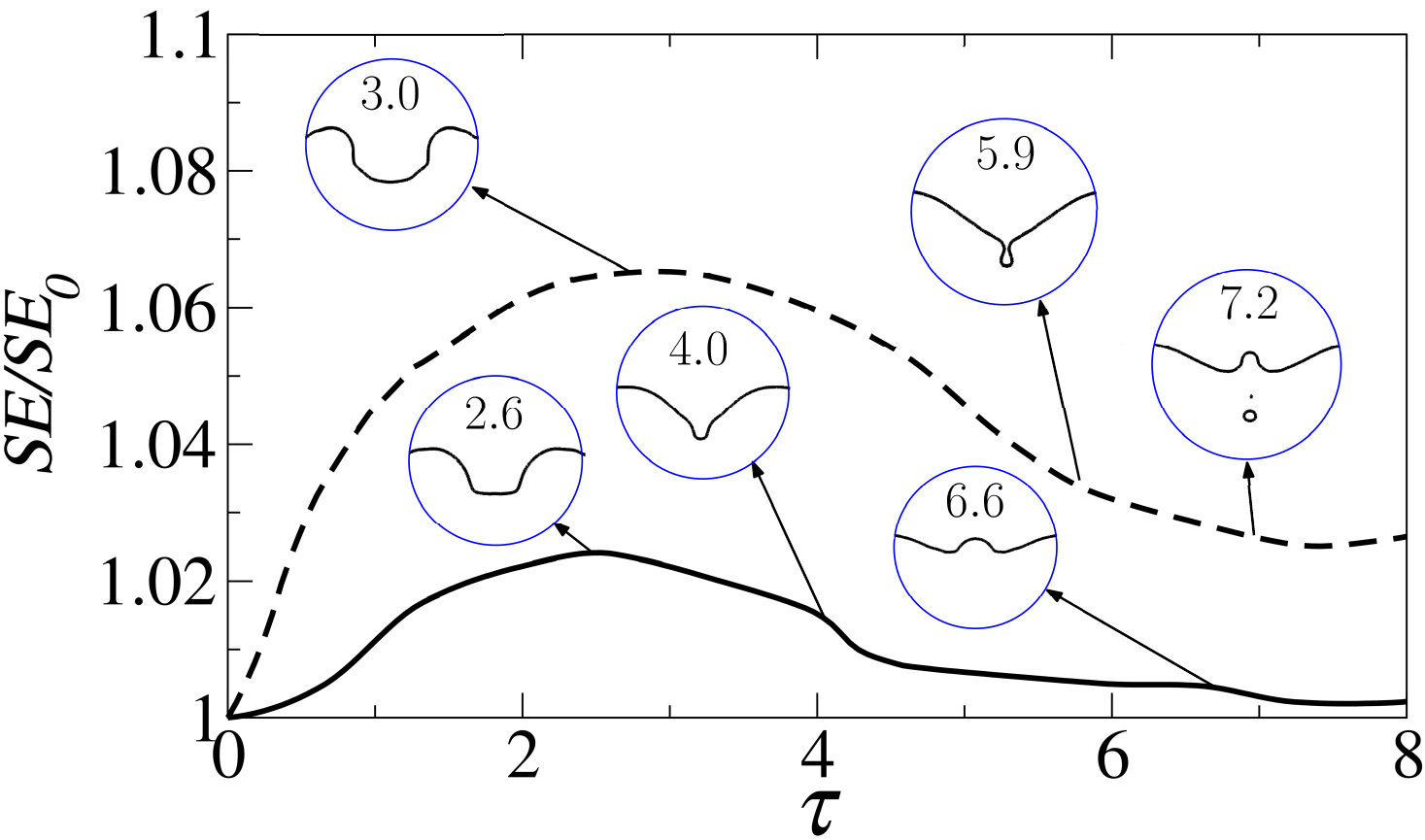}
\caption{Temporal variations of $KE/KE_0$ and $SE/SE_0$ for different values of the aspect ratio and Weber number. The panels $(a, b)$, $(c, d)$ and $(e,f)$ are for $A_r=0.2$, $A_r=1.0$ and $A_r=5.0$, respectively. The solid and dashed lines represent the results for $We = 93.4$ and $We = 190.6$, respectively. The rest of the parameters are $\Oh = 7.82 \times 10^{-3}$ and $\Bo = 0.04$. The interface shapes at different instants (mentioned inside the circle) are inserted in panels $b$, $d$ and $f$.}
\label{fig:Energy_splashing}
\end{figure}

The impact area of a prolate drop i.e., $A_r = 0.2$ is significantly smaller than that of a spherical drop due to the lower radius of curvature in the prolate shape. Consequently, in the case of the impact of a prolate drop ($A_r = 0.2$), Figure \ref{fig:Splash_3d} reveals a less pronounced wave swell during impact. The wave height resulting from the droplet impact reaches its maximum earlier for $\We = 190.6$ compared to $\We = 93.4$ (Figure \ref{fig:CD_WH}). Moreover, due to its bullet-like shape, the impact duration is prolonged, resulting in the persistence of a shearing zone within the liquid pool and the generation of more intense vortices. The impact zone forms a crater that delves into the pool until the crater collapses inward, driven by the wave swell retraction phenomenon. This process creates an air pocket that gets smaller with time as the crater walls retract within the pool. It can be seen that the size of this air pocket is proportional to the impact velocity (Weber number) at time $\tau = 0.6$ and $\tau = 0.9$ for $We = 190.6$ and 93.4 respectively (Figure \ref{fig:Energy_splashing}$b$). The collapse of the wave height prevents the crater from expelling all the air from the crater base, forming larger air bubbles within the liquid. A similar phenomenon was also observed in the axisymmetric simulations of \cite{deka2017regime}. As the crater collapses entirely, the air pocket transforms into a bubble, fully submerged by the collapsing crater. For $\We = 190.6$, the maximum crater height is achieved sooner compared to $\We = 93.4$, leading to the entrapment of larger-volume air pockets in the pool. As the crater collapses, a jet is formed, shooting upwards (see, $\tau = 0.98$ for $\We = 93.4$ and $\tau = 0.82$ for $\We = 190.6$ in Figure \ref{fig:Splash_3d}). This jet entrains the surrounding air, causing it to recirculate (see, $\tau = 1.32$ for $\We = 93.4$ and $\tau = 0.98$ for $\We = 190.6$).

In contrast, an oblate drop with $A_r = 5.0$, despite having an earlier maximum wave height than the spherical drop (Figure \ref{fig:CD_WH}) has a significantly higher impact area, which in turn produces a much wider wave swell (see, at $\tau = 0.66$ for $\We = 93.4$ and $\We = 190.6$ in Figure \ref{fig:Splash_3d}). Despite this, the droplet lacks sufficient polar length to sustain the wave swell for an extended period, causing it to retract quickly. Meanwhile, the crater depth continues to increase, with $\We = 190.6$ exhibiting a greater depth compared to $\We = 93.4$.

This results in the entrapment of a substantial air pocket when the crater walls collapse, a phenomenon observed in $\We = 190.6$, unlike in $\We = 93.4$, where the crater has not descended as far. This distinction is evident at $\tau = 3.95$ for $\We = 93.4$ and $\tau = 5.92$ for $\We = 190.6$. A jet forms upward once the crater fully collapses, as indicated at $\tau = 4.6$ for $\We = 93.4$ and $\tau = 6.58$ for $\We = 190.6$. The subsequent stages are similar to that observed for $A_r = 1.0$, except in this case, the jet lacks the strength to generate a secondary droplet due to the Rayleigh-Plateau instability.

\begin{figure}
\centering
\hspace{0.4cm} {($a$)} \hspace{3.2cm} {($b$)} \hspace{3.2cm} {($c$)} \\
\includegraphics[scale=0.15]{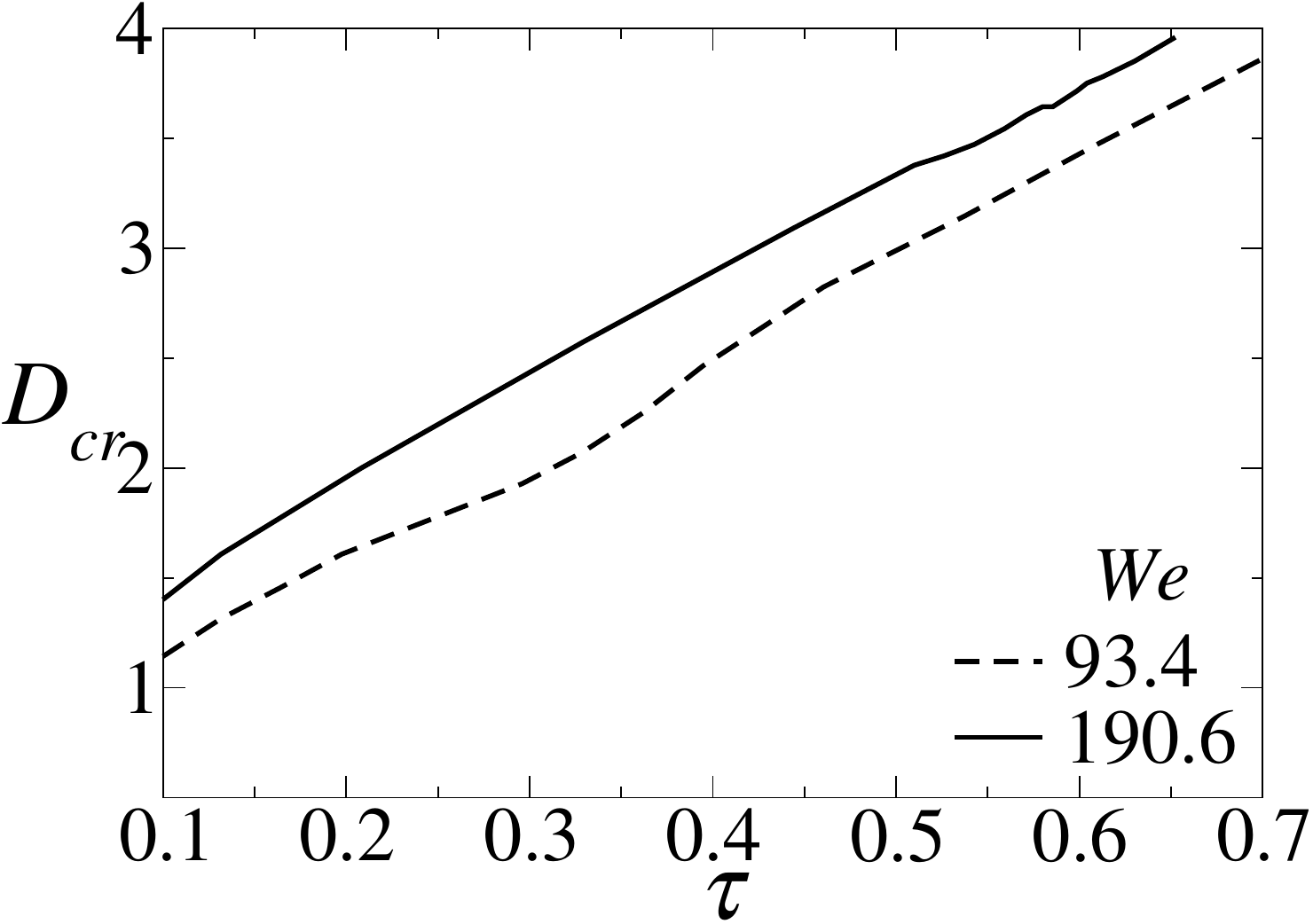}
\includegraphics[scale=0.15]{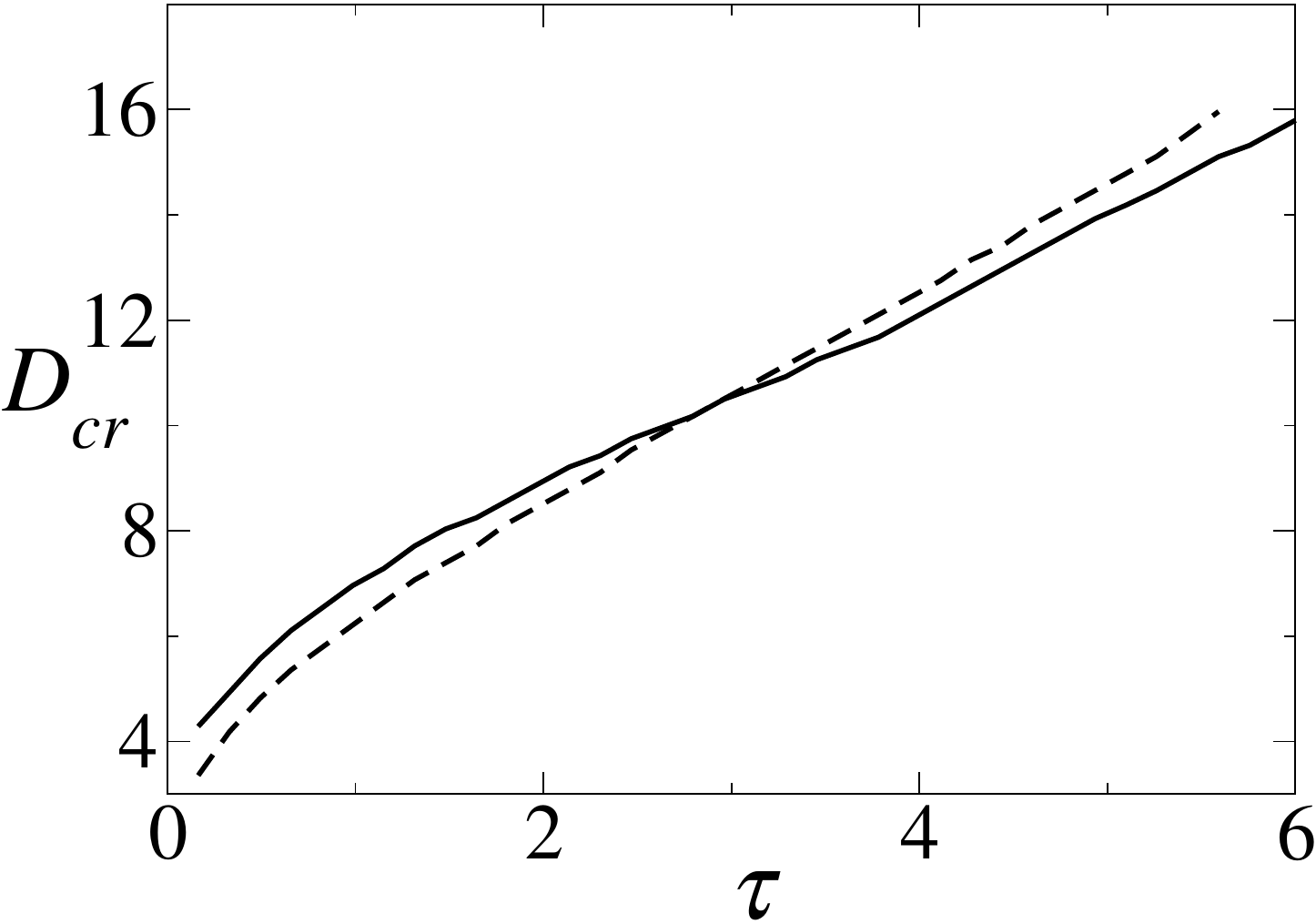}
\includegraphics[scale=0.15]{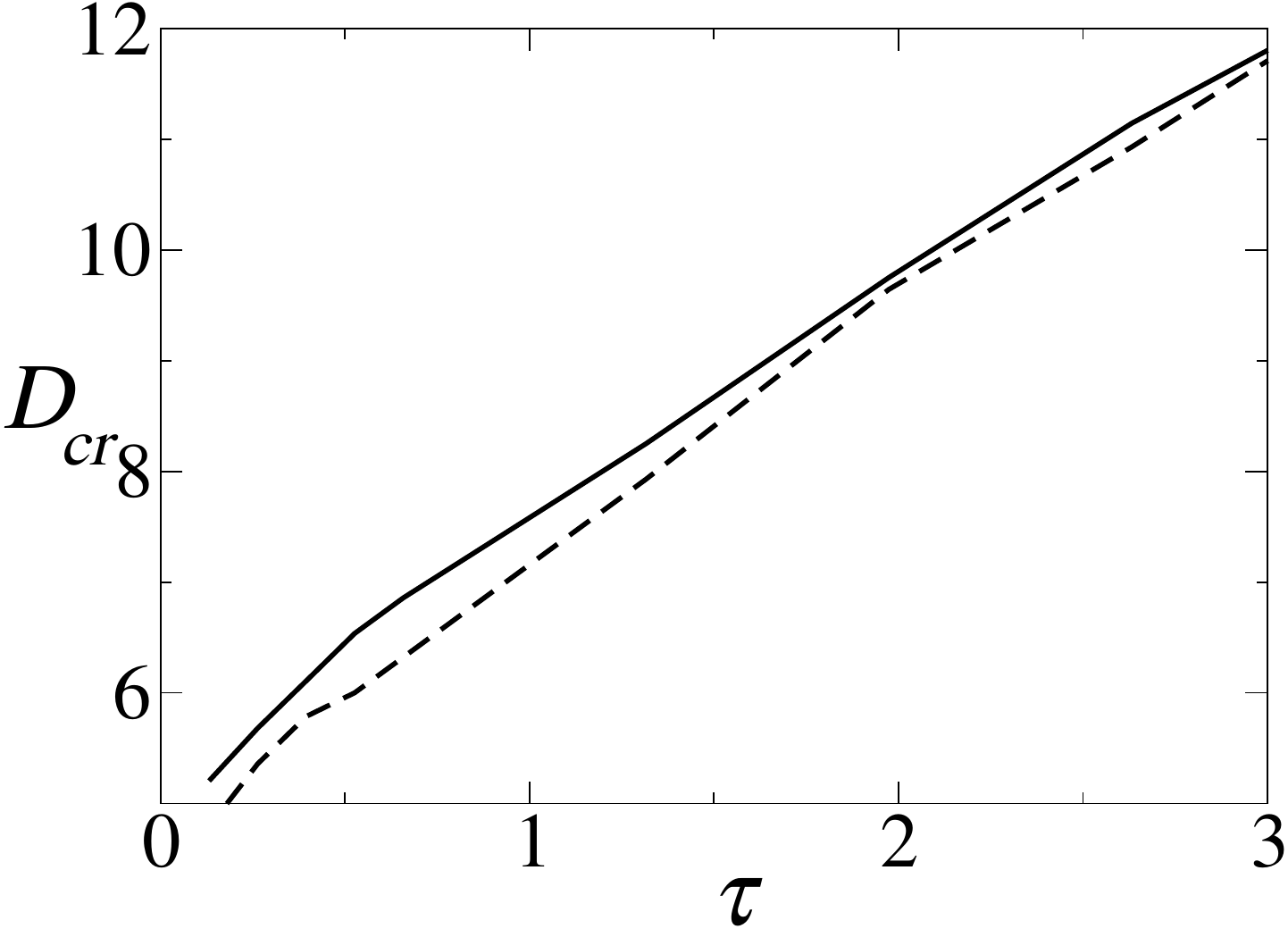}
\caption{Temporal variation of non-dimensional crater diameter $(D_{cr})$ for different values of the aspect ratio and Weber number. $(a)$ $A_r=0.2$, $(b)$ $A_r=1.0$ and $(c)$ $A_r=5.0$. The rest of the parameters are $\Oh = 7.82 \times 10^{-3}$ and $\Bo = 0.04$.}
\label{fig:crater_diameter}
\end{figure}

The temporal evolutions of the crater diameter (spreading of rim) are illustrated in Figure \ref{fig:crater_diameter}$(a-c)$ for $A_r = 0.2$, 1.0 and 5.0, respectively. Upon contact with the free surface of the liquid pool, the droplet imparts velocity to the pool, forming a rim around the droplet. This rim steadily expands until its energy is entirely dissipated. For $A_r = 0.2$, the rim exhibits a constant spread rate due to its small impact area, as observed for both $\We = 93.4$ and $190.6$. In contrast, for $A_r = 1.0$ and 5.0, the initial spreading rate is higher for higher $\We$, but it eventually decays due to the elevated vertical component of the velocity. This trend is evident in Figure \ref{fig:crater_diameter}$(b,c)$, where the solid line decays more rapidly than the dashed line. In Figure \ref{fig:crater_diameter}$b$, it can be observed that the diameter of the rim for $\We=93.4$ and 190.6 in the case of the spherical drop clearly undergoes a crossover around $\tau \approx 3$. Below, to understand the behaviour, we examine the kinetic $(KE)$ and surface energy $(SE)$ of the system due to the impact of the drop on the free surface of the pool, which are defined as 
\begin{equation}
KE= \frac{1}{2}\sum \alpha \rho_l (u^2+v^2+w^2) dV_{cell}, ~ {\rm and} ~ SE=\sum \sigma  A_s,
\end{equation}
where $A_s$ denotes the surface area of the interface and $dV_{cell}$  represents the volume of the computational cell.

\begin{figure}
\centering
\hspace{0.35cm} {($a$)} \hspace{5.5cm} {($b$)} \\
\includegraphics[scale=0.25]{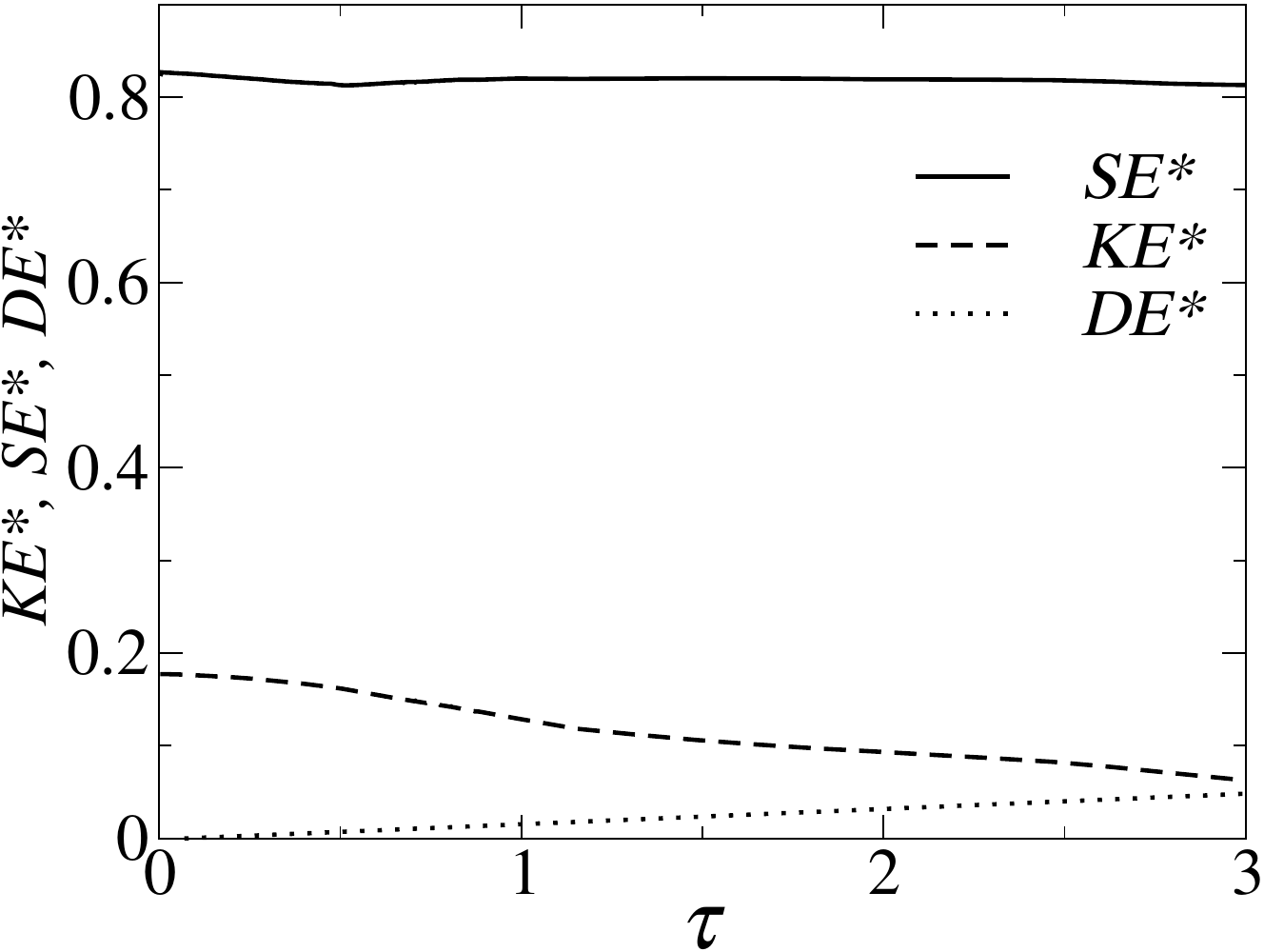} \hspace{0.2cm} \includegraphics[scale=0.25]{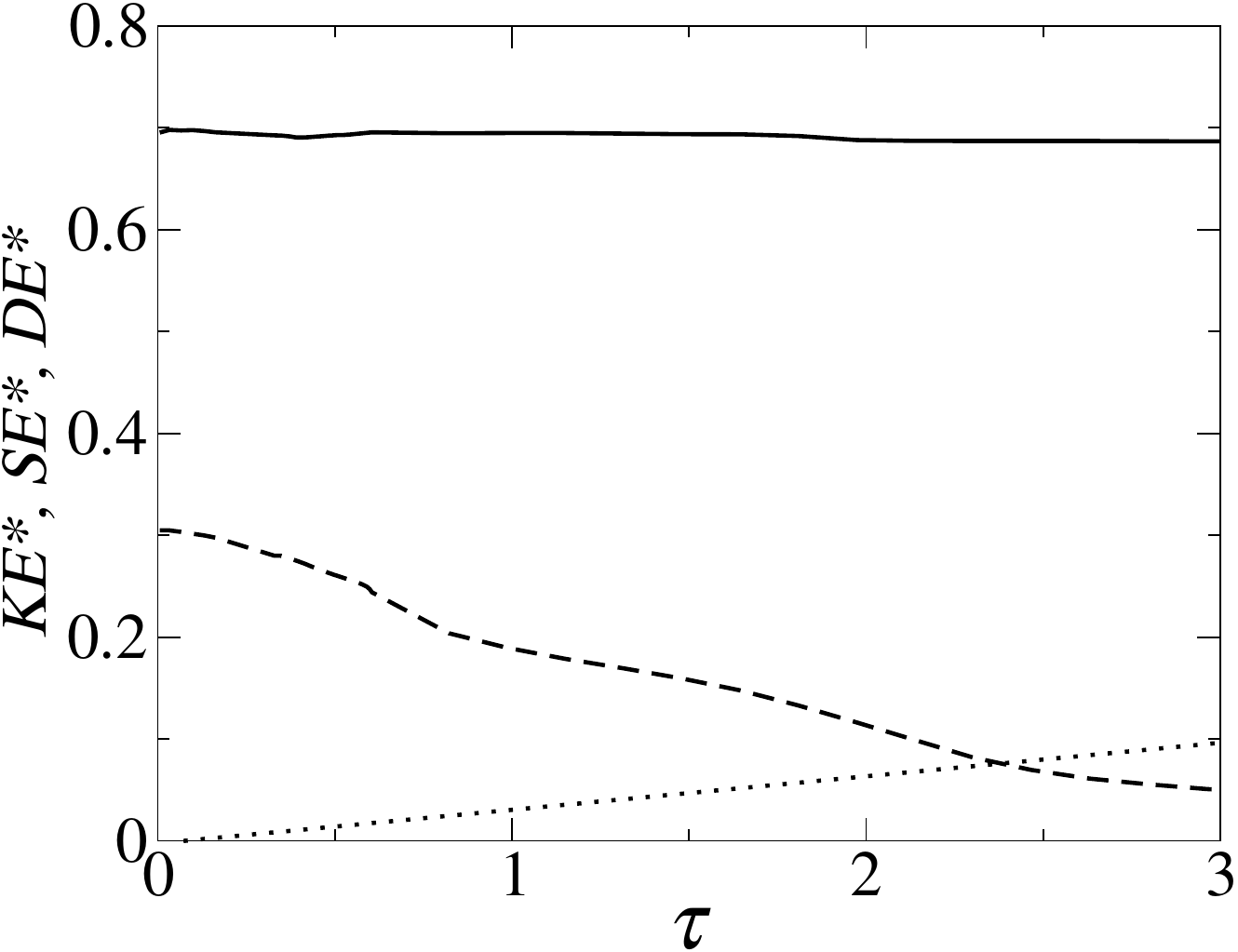} \\
\vspace{0.2cm}
\hspace{0.35cm} {($c$)} \hspace{5.5cm} {($d$)} \\
\includegraphics[scale=0.25]{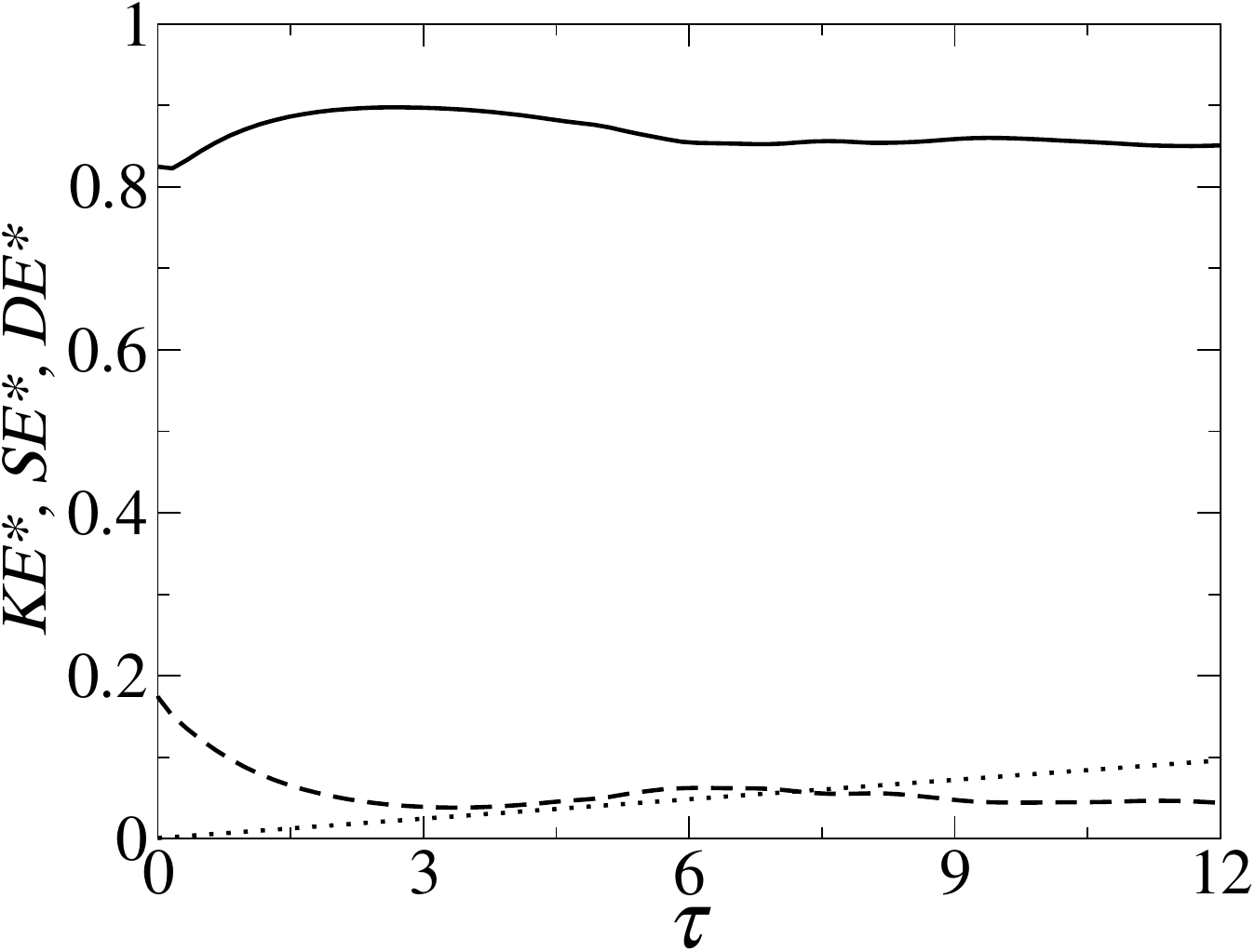} \hspace{0.25cm} \includegraphics[scale=0.25]{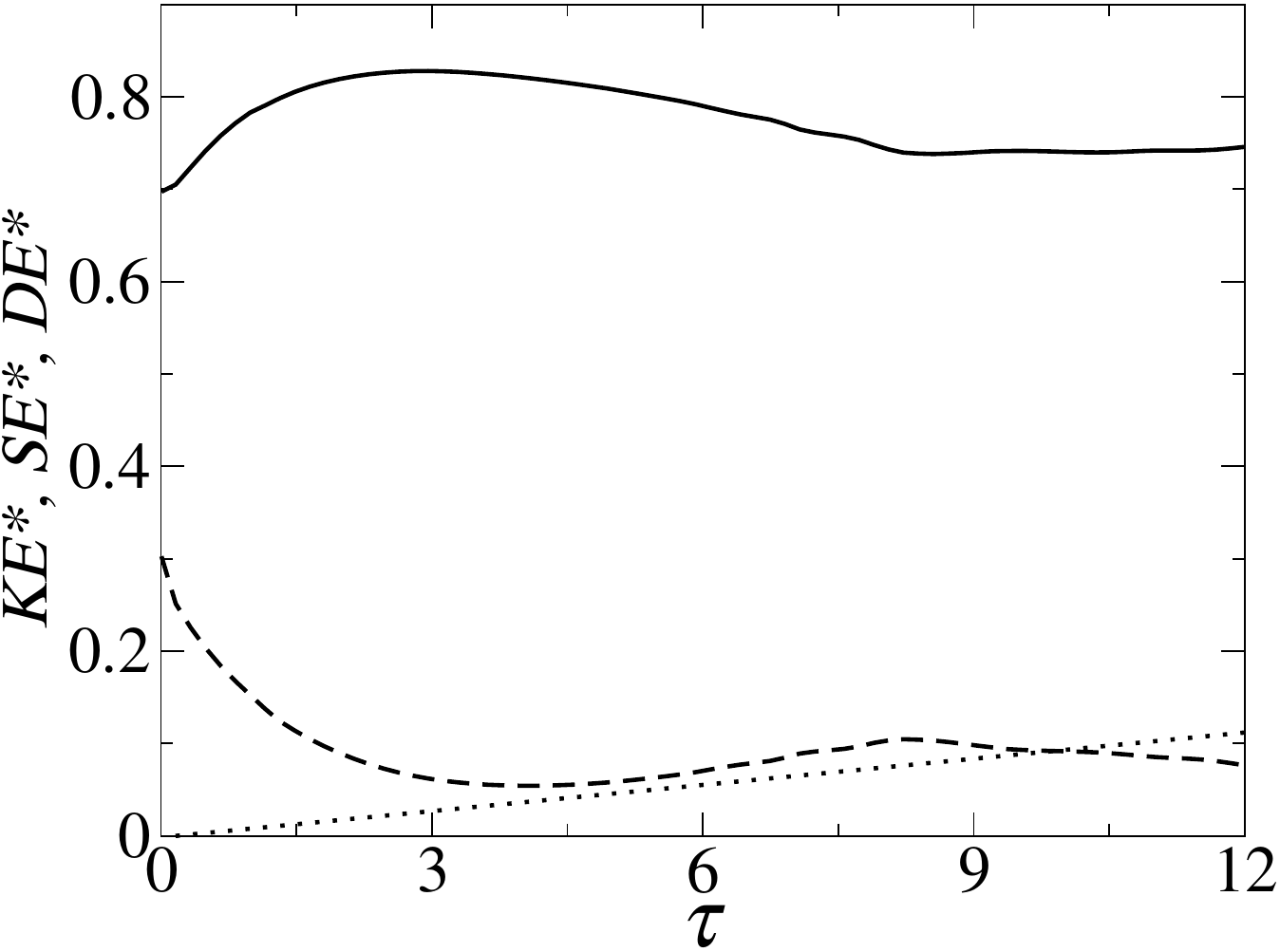} \\
\vspace{0.35cm}
\hspace{0.3cm} {\large ($e$)} \hspace{5.5cm} {\large ($f$)} \\
\includegraphics[scale=0.25]{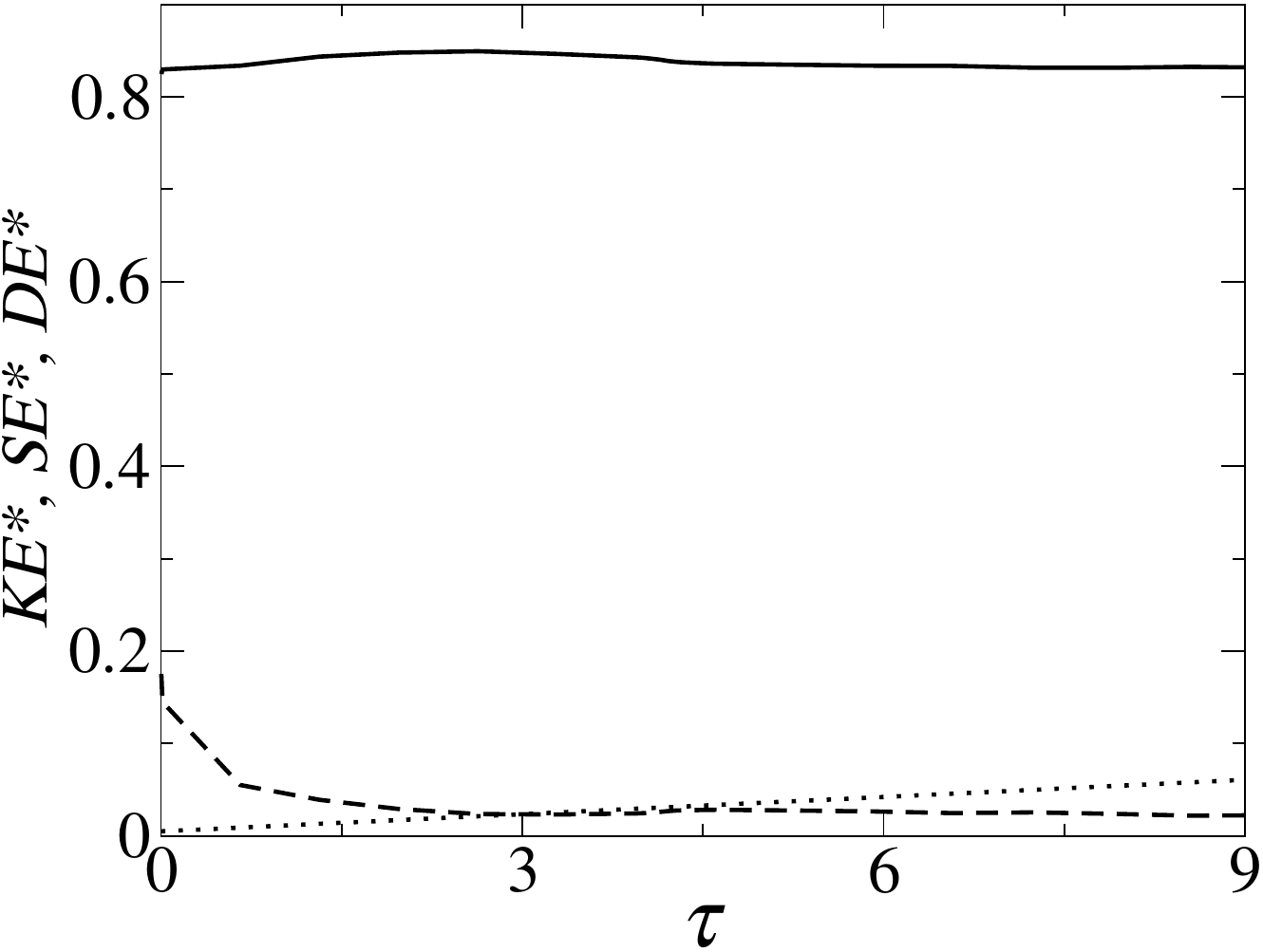} \hspace{0.35cm} \includegraphics[scale=0.25]{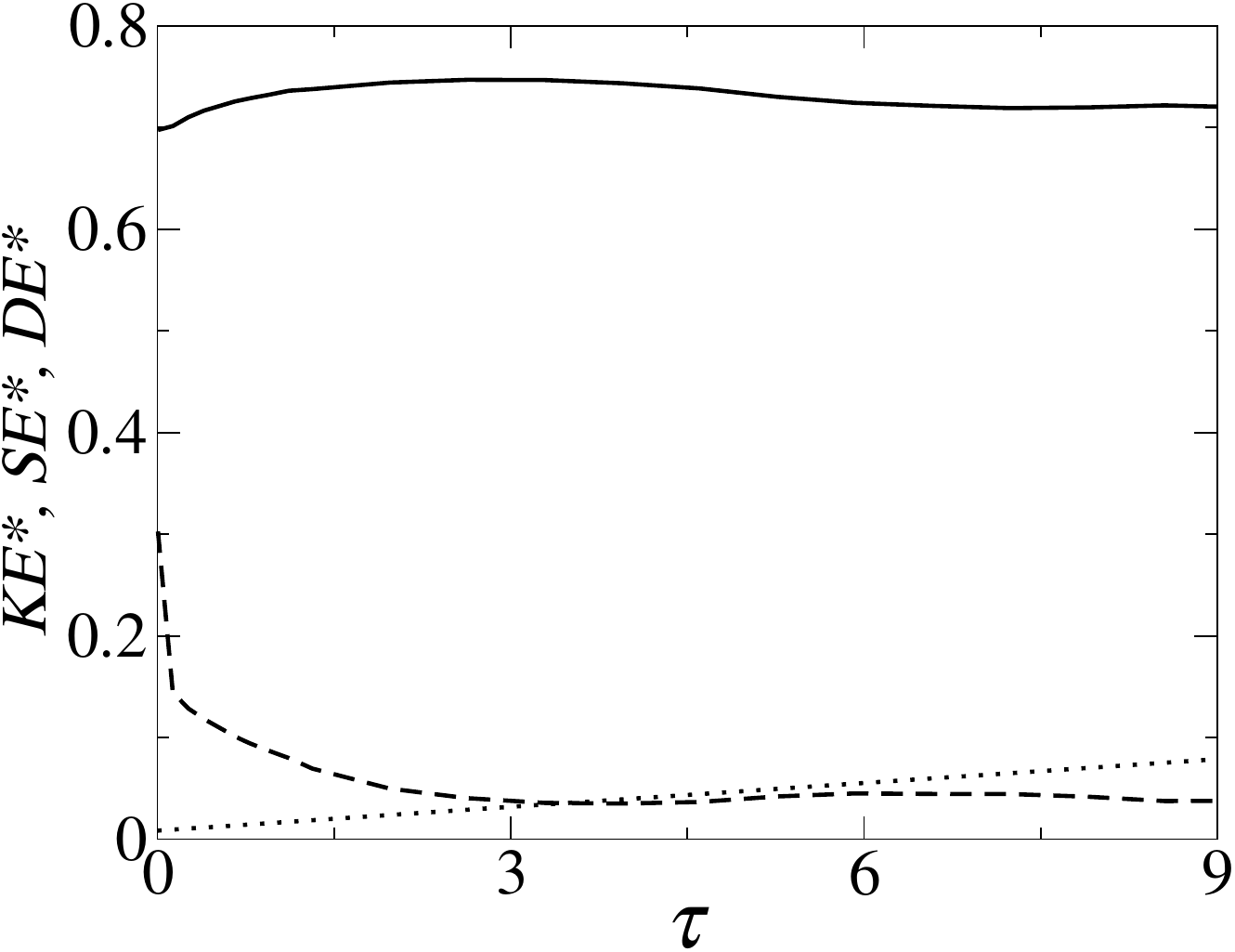}
\caption{\ks{The temporal variations of the normalized kinetic energy $(KE)$, surface energy $(SE)$, and dissipation energy $(DE)$ with the total energy $(TE)$. Here, $SE^* = SE/TE$, $KE^* = KE/TE$, $DE^* = DE/TE$. Panels $(a, b)$, $(c, d)$ and $(e,f)$ are for $A_r=0.2$, $A_r=1.0$ and $A_r=5.0$, respectively. The solid and dashed lines represent the results for $We = 93.4$ and $We = 190.6$, respectively. The rest of the parameters are $\Oh = 7.82 \times 10^{-3}$ and $\Bo = 0.04$.}}
\label{Energy_budget}
\end{figure}

Figure \ref{fig:Energy_splashing}$(a-f)$ depicts the temporal variation of normalized kinetic energy $(KE/KE_0)$ and surface energy $(SE/SE_0)$ for different values of $A_r$ and $\We$. Here, $KE_0$ and $SE_0$ represent the kinetic and surface energies of the system at $\tau = 0$. Initially, for $A_r = 0.2$, the smaller contact area of the droplet on the free surface leads to the creation of a narrower crater (Figure \ref{fig:Splash_3d}). Consequently, there is minimal change in surface energy, while kinetic energy gradually dissipates over time for $\We=93.4$ and 190.6. In contrast, for $A_r = 1.0$ and $5.0$, the impact of the droplet on the water pool results in the formation of a rim-like structure, leading to an increase in surface area. This, in turn, increases surface energy and decreases kinetic energy with time. This trend persists until the wave height recedes, causing the crater to collapse (around $\tau \approx 3$ for $A_r = 1.0$). Notably, the value of $SE$ and the time of its peak are lower for $A_r = 5.0$ due to the oblate drop creating a wider yet shallower crater than $A_r = 1.0$ (Figure \ref{fig:Splash_3d}). An intriguing observation is the crossover in rim diameter for $\We=93.4$ and 190.6 around $\tau \approx 3$ in Figure \ref{fig:crater_diameter}$(b)$ for $A_r=1$. It is interesting to observe that the crossover in rim diameter for $\We=93.4$ and 190.6, as depicted in Figure \ref{fig:crater_diameter}$(b)$ for $A_r=1$, coincides with the instant when $SE$ reaches its maximum. This suggests a close relationship between the dynamics of the rim formation and the variation in surface energy. The absence of the crossover in rim diameter for $A_r = 5.0$ is likely attributed to its significantly lower surface energy than $A_r=1$. However, the apparent features in Figure \ref{fig:crater_diameter}$(c)$ hint at a similar underlying dynamic, even if the crossover itself is not observed. The subsequent collapse of the crater initiates a reduction in the surface area of the air-water interface, leading to a decrease in surface energy, while the kinetic energy increases slightly. Once the crater is collapsed, the catapulting effect induces a whipping motion in the fluid interface, resulting in the formation of a jet that increases the surface area slightly. Subsequently, surface tension attempts to pull the jet downward, causing a slight decrease in $KE$, which stabilizes over time. 

\ks{Figure \ref{Energy_budget} illustrates the temporal variations of the normalized kinetic energy $(KE)$, surface energy $(SE)$, and dissipation energy $(DE)$ with the total energy $(TE)$, such that $SE^* = SE/TE$, $KE^* = KE/TE$, $DE^* = DE/TE$, and $TE = SE_0+KE_0+DE_0$, where $SE_0$, $KE_0$, and $DE_0$ represent the surface energy, kinetic energy, and dissipated energy at $\tau = 0$. The dissipation energy is defined as:
\begin{equation}
\begin{split}
    &DE = \int_{0}^{\tau}\sum \mu_l \Bigg[ 2\mu_r\left( \left( \frac{\partial u}{\partial x} \right)^2+\left( \frac{\partial v}{\partial y}  \right)^2+\left( \frac{\partial w}{\partial z}  \right)^2 \right) \\
    &\quad+ \mu_r\left( \left( \frac{\partial u}{\partial y} +\frac{\partial v}{\partial x} \right)^2 + \left( \frac{\partial v}{\partial z} +\frac{\partial w}{\partial y} \right)^2 + \left( \frac{\partial u}{\partial z} +\frac{\partial w}{\partial x} \right)^2 \right) \Bigg] dV_{\text{cell}} d\tau.
\end{split}
\label{eqn:dissipation}
\end{equation}
In Figure \ref{Energy_budget}, it is evident that the energy dissipation rate (represented by the slope of the dissipation curve), while nearly constant with time, is significantly higher for $A_r$ = 0.2 compared to $A_r$ = 1.0 and 5.0. This difference can be attributed to the shear forces induced by the elongated shape of the droplet, as emphasized by \cite{deka2017regime} in their observations of prolate droplets. Conversely, in the case of oblate and spherical droplets, the broader impact area leads to a greater disturbance in the fluid volume, resulting in a more evenly distributed velocity profile and, subsequently, lower dissipation rates. The disparity in dissipation rate between the prolate shape drop and the non-prolate shape drops (spherical and oblate drops) is further amplified as the dissipation rate is dependent on the square of velocity gradients, which are inherently higher in prolate shapes.  Additionally, higher Weber numbers correspond to elevated dissipation rates, owing to the heightened velocity gradients associated with greater impact velocities. It is worth noting that potential energy, being significantly less compared to the other energies, is disregarded in the energy budget calculations.}

\subsection{Comparison with the freely falling non-spherical drop impacting a liquid pool}

\begin{figure}
\centering
\includegraphics[scale=0.48]{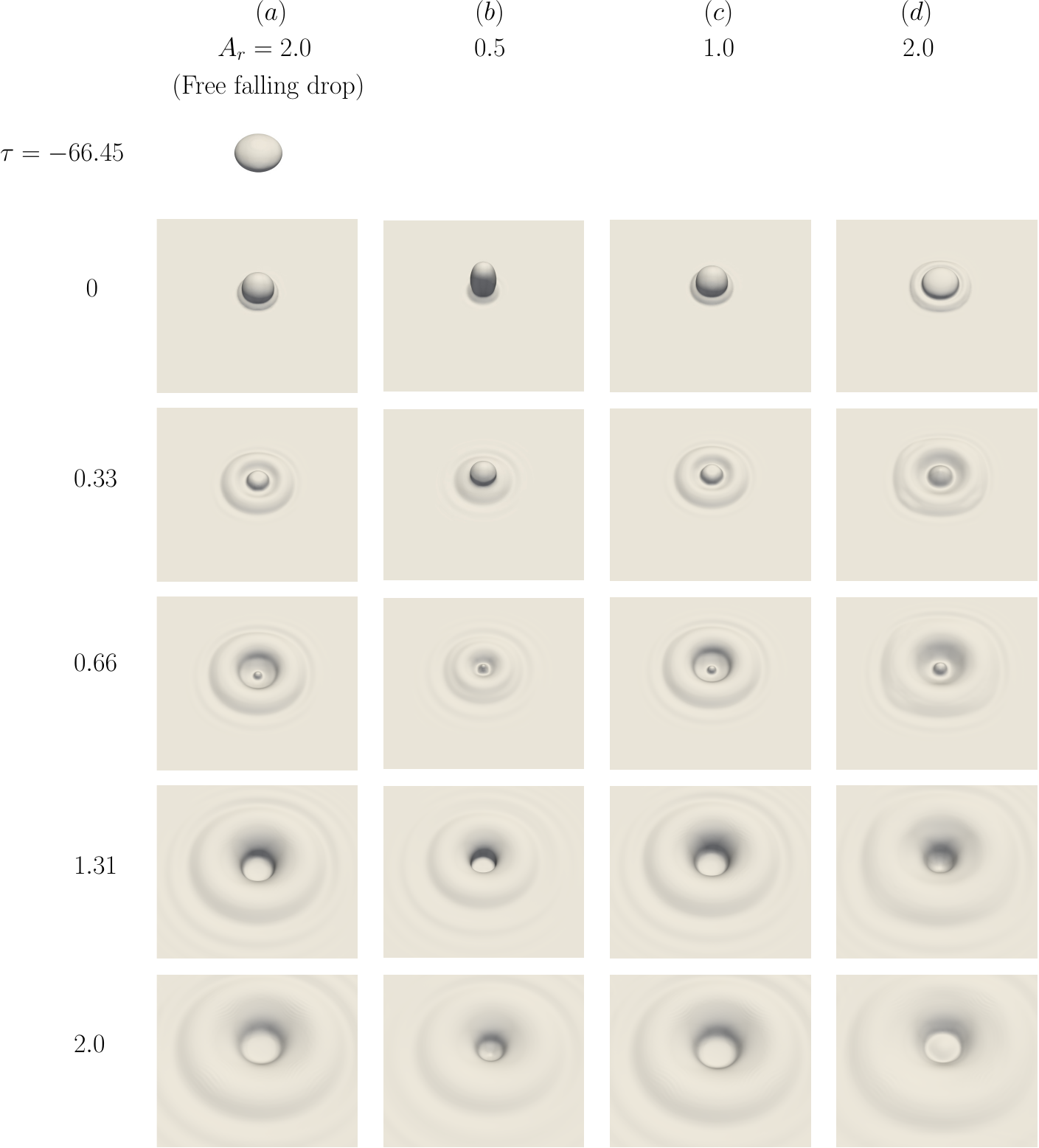}
\caption{The comparison of the impact dynamics of a freely falling droplet from height $L = 92.7 R_{eq}$ from the free surface of the liquid pool with the impact dynamics of parent drops with $A_r=0.5$ (prolate shape), $A_r=1$ (spherical shape) and $A_r=2$ (oblate shape) impacting with an equivalent velocity of $\sqrt{2gL}$. This corresponds to $\We = 7.62$. The rest of the dimensionless parameters are $\Oh = 7.82 \times 10^{-3}$ and $\Bo = 0.04$.}
\label{fig:3D_equivalence}
\end{figure}

Finally, in Figure \ref{fig:3D_equivalence}, we compare the impact dynamics of an oblate drop with an initial aspect ratio of $A_r = 2.0$, descending from a height of $L = 92.7 R_{eq}$ above the free surface of the liquid pool, with the impact dynamics of parent drops of various shapes subjected to impact just above the free surface, with an equivalent velocity of $\sqrt{2gL}$. For instance, a drop falling from a height of $L=5.1$ cm impacts the free surface with a velocity of 1 m/s, which is considered in Figure \ref{fig:3D_equivalence}. For the set of parameters considered (for $\We = 7.62$ defined based on the velocity of the falling drop just above the free surface), it can be seen in Figure \ref{fig:3D_equivalence}$a$ that the oscillations in the free-falling oblate drop mostly dampened out, resulting in its impact with an almost spherical shape. Consequently, the subsequent coalescence dynamics closely resembles that observed in the case of a spherical drop impacting the liquid pool with $\We = 7.62$ (Figure \ref{fig:3D_equivalence}$c$). By conducting an inviscid analysis in the microgravity condition, \cite{tsamopoulos1983nonlinear,lamb1924hydrodynamics} showed that a drop exhibits periodic shape oscillations. Recently, by conducting three-dimensional numerical simulations and experiments, \cite{agrawal2017nonspherical,agrawal2020experimental} demonstrated that a non-spherical drop falling from a height experiences oblate-spherical-prolate shape oscillations due to the interplay between inertia and surface tension forces. However, the viscosity acts to dampen these oscillations. Thus, depending on fluid properties, a freely falling drop can impact the pool with various shapes, as observed in the diverse shapes of raindrops falling on the ocean surface \citep{szakall2009wind}. Figure \ref{fig:3D_equivalence}$(b)$ and $(d)$ depict the coalescence behaviour of a prolate, and an oblate shape parent drop with the same impact velocity ($\We = 7.62$). A closer examination of these results reveals distinct dynamics for the prolate and oblate drops compared to the spherical drop (Figure \ref{fig:3D_equivalence}$c$) and the freely falling oblate drop (Figure \ref{fig:3D_equivalence}$a$) despite their same impact velocity and the same volume (i.e., the same Weber number). Specifically, due to its elongated bullet-like shape, the prolate drop results in a cylindrical crater (Figure \ref{fig:3D_equivalence}$b$). On the other hand, the crater formed by the oblate drop is predominantly hemispherical (Figure \ref{fig:3D_equivalence}$d$). The diameter of the rim formed after impact increases as we transition from prolate to spherical to oblate-shaped drops, as shown in Figure \ref{fig:3D_equivalence}. This analysis offers insights into the distinct coalescence dynamics influenced by drop shape and impact conditions.

\section{Concluding remarks} \label{sec:conc}

We conduct three-dimensional numerical simulations to investigate the impact dynamics of non-spherical drops in a deep liquid pool, varying aspect ratios and Weber numbers. The numerical solver is validated by comparing our results with previous numerical and experimental investigations. Our findings emphasize the significant influence of the shape of the parent droplet (keeping the volume the same) on coalescence dynamics, resulting in distinct morphologies of the liquid-air interface compared to spherical drops studied in earlier literature. First, we investigate the coalescence dynamics of non-spherical parent droplets gently placed on a liquid pool $(\We = 0)$. We observe that a prolate drop makes contact with the liquid pool at almost a single point as it tries to return to a spherical shape under the influence of surface tension. This initiates a competition between the capillary pressure resulting from the negative curvature near the contact point and the Laplace pressure within the parent drop, setting off a velocity field inside the droplet. Due to the substantially higher Laplace pressure than the capillary pressure arising from negative curvature, the drop immediately detaches from the free surface upon contact, showcasing intricate three-dimensional shape oscillations for the high aspect ratio $A_r=0.2$. Subsequently, under the influence of gravity, the drop touches the free surface again, initiating the coalescence process. We found that the parent drop exhibits a partial coalescence phenomenon and the emergence of a daughter droplet for $A_r>0.67$. In contrast to the prolate $(A_r<1)$ and spherical drop $(A_r=1)$, an oblate parent drop with $A_r=5$ encapsulates air in a ring-like bubble within the pool and emerges a liquid column that undergoes Rayleigh-Plateau capillary instability, leading to the formation of two daughter droplets. Further, our observations indicate that the variation of detachment time with aspect ratio exhibits a similar trend to that of the magnitude of the circumference of the droplet (i.e., $\approx \pi(a+b)$).

Then, we explore the collision dynamics of non-spherical parent drops with various aspect ratios, impacting the liquid pool at different velocities (Weber numbers). It is well known that the impact of a spherical drop with a high velocity is characterised by the crater formation, air bubble entrainment inside the pool, the release of a jet and splashing dynamics. Our observations indicate that increasing the Weber number leads to elevated crater heights on the free surface for all shapes of parent drops. As the impact area of a prolate drop is significantly smaller than that of spherical and oblate drops, a prolate drop produces a less pronounced wave swell and a prolonged impact duration. Conversely, an oblate drop generates a much wider wave swell than spherical and prolate drops. For spherical drop ($A_r=1$), we observe an interesting behaviour of the crossover in rim diameter for $\We=93.4$ and 190.6 at around $\tau \approx 3$, which coincides with the instant when surface energy reaches its maximum. This suggests a close relationship between rim formation dynamics and variations in surface energy. Although $A_r = 5.0$ does not show any crossover in rim diameter due to its significantly lower surface energy than $A_r=1$, our results hint at a similar underlying behaviour. 

Finally, we establish an analogy by comparing the dynamics of a freely falling non-spherical drop, experiencing topological oscillations as it descends from a height, with the impact dynamics of parent drops of various shapes striking the liquid surface with an equivalent velocity. This comparative analysis allows us to delve into the impact of shape oscillations, arising from the interplay between inertia and surface tension, on the coalescence phenomenon during the free fall of a drop. This approach contrasts with the extensive studies conducted by various researchers on the impact of a spherical parent drop just above the free surface.\\

\noindent{\bf Acknowledgement:} {K.C.S. thanks the Science \& Engineering Research Board, India, and IIT Hyderabad for their financial support provided through grants CRG/2020/000507 and IITH/CHE/F011/SOCH1. S.B. thanks for the support and resources provided by PARAM Seva under the National Supercomputing Mission, Government of India.}

\section*{Appendix}

 \begin{figure}[!htbp]
\centering
\includegraphics[scale=0.4]{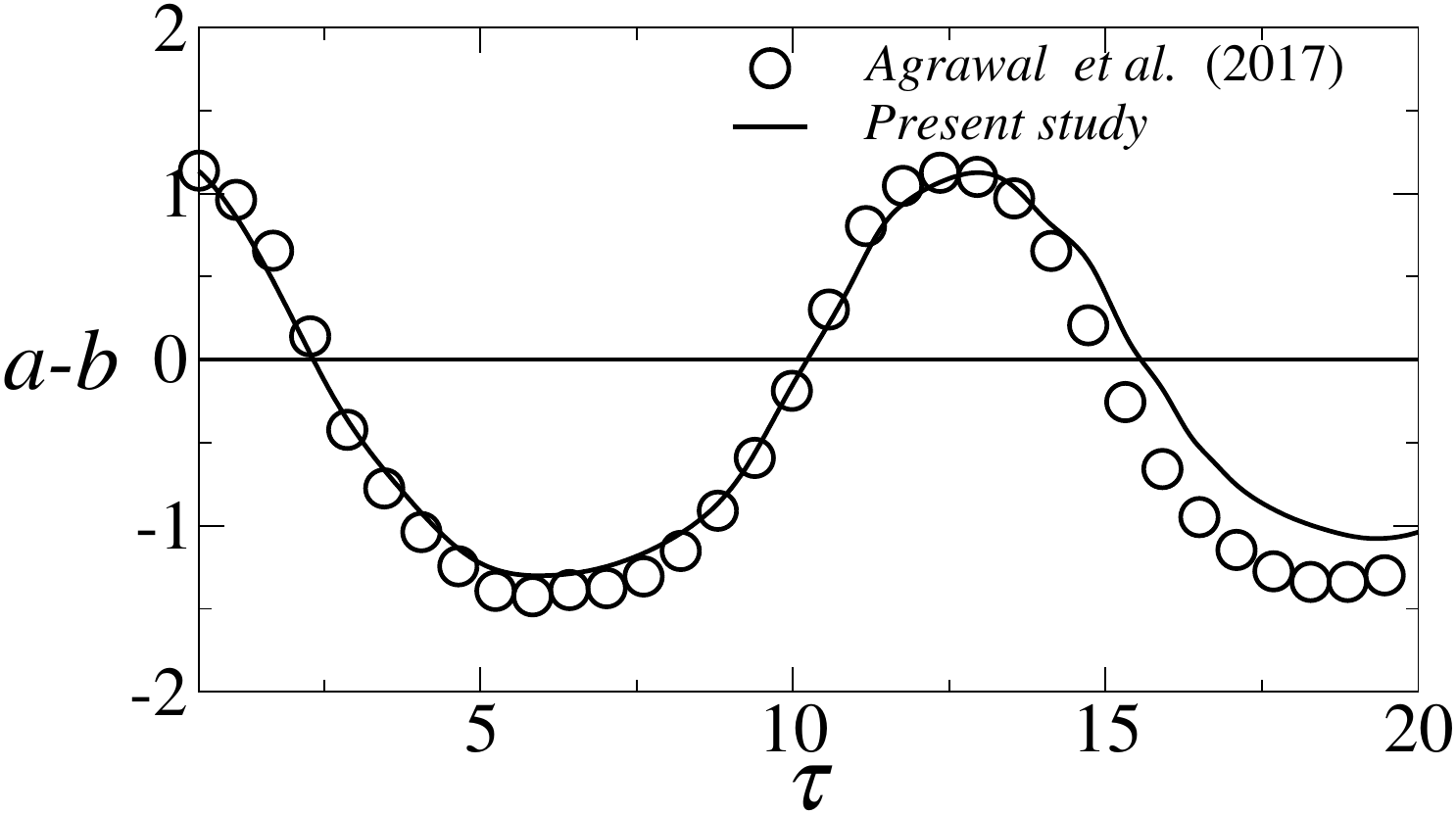}  
\caption{Comparison of the temporal evolution of $(a-b)$ of an oblate droplet of initial $A_r = 1.95$ during its free fall obtained from our numerical simulation with that of \cite{agrawal2017nonspherical}. The values of the non-dimensional numbers considered in our simulations are $Oh = 5.5 \times 10^{-3}$, $We = 0$, $Bo = 1$, $\rho_r = 10^{-3}$ and $\mu_r = 0.018$.}
\label{fig:Agrawal_validation}
\end{figure}


\begin{thebibliography}{55}
\expandafter\ifx\csname natexlab\endcsname\relax\def\natexlab#1{#1}\fi
\providecommand{\url}[1]{\texttt{#1}}
\providecommand{\href}[2]{#2}
\providecommand{\path}[1]{#1}
\providecommand{\DOIprefix}{doi:}
\providecommand{\ArXivprefix}{arXiv:}
\providecommand{\URLprefix}{URL: }
\providecommand{\Pubmedprefix}{pmid:}
\providecommand{\doi}[1]{\href{http://dx.doi.org/#1}{\path{#1}}}
\providecommand{\Pubmed}[1]{\href{pmid:#1}{\path{#1}}}
\providecommand{\bibinfo}[2]{#2}
\ifx\xfnm\relax \def\xfnm[#1]{\unskip,\space#1}\fi
\bibitem[{Agbaglah et~al.(2015)Agbaglah, Thoraval, Thoroddsen, Zhang, Fezzaa
  and Deegan}]{agbaglah2015drop}
\bibinfo{author}{Agbaglah, G.}, \bibinfo{author}{Thoraval, M.J.},
  \bibinfo{author}{Thoroddsen, S.T.}, \bibinfo{author}{Zhang, L.V.},
  \bibinfo{author}{Fezzaa, K.}, \bibinfo{author}{Deegan, R.D.},
  \bibinfo{year}{2015}.
\newblock \bibinfo{title}{Drop impact into a deep pool: vortex shedding and jet
  formation}.
\newblock \bibinfo{journal}{J. Fluid Mech.} \bibinfo{volume}{764},
  \bibinfo{pages}{R1}.
\bibitem[{Agrawal et~al.(2020)Agrawal, Katiyar, Karri and
  Sahu}]{agrawal2020experimental}
\bibinfo{author}{Agrawal, M.}, \bibinfo{author}{Katiyar, R.K.},
  \bibinfo{author}{Karri, B.}, \bibinfo{author}{Sahu, K.C.},
  \bibinfo{year}{2020}.
\newblock \bibinfo{title}{Experimental investigation of a nonspherical water
  droplet falling in air}.
\newblock \bibinfo{journal}{Phys. Fluids} \bibinfo{volume}{32},
  \bibinfo{pages}{112105}.
\bibitem[{Agrawal et~al.(2017)Agrawal, Premlata, Tripathi, Karri and
  Sahu}]{agrawal2017nonspherical}
\bibinfo{author}{Agrawal, M.}, \bibinfo{author}{Premlata, A.R.},
  \bibinfo{author}{Tripathi, M.K.}, \bibinfo{author}{Karri, B.},
  \bibinfo{author}{Sahu, K.C.}, \bibinfo{year}{2017}.
\newblock \bibinfo{title}{Nonspherical liquid droplet falling in air}.
\newblock \bibinfo{journal}{Phys. Rev. E.} \bibinfo{volume}{95},
  \bibinfo{pages}{033111}.
\bibitem[{Balla et~al.(2019)Balla, Tripathi and Sahu}]{balla2019shape}
\bibinfo{author}{Balla, M.}, \bibinfo{author}{Tripathi, M.K.},
  \bibinfo{author}{Sahu, K.C.}, \bibinfo{year}{2019}.
\newblock \bibinfo{title}{Shape oscillations of a nonspherical water droplet}.
\newblock \bibinfo{journal}{Phys. Rev. E.} \bibinfo{volume}{99},
  \bibinfo{pages}{023107}.
\bibitem[{Behera et~al.(2023)Behera, Rao, Dasgupta and
  Chakraborty}]{behera2023investigation}
\bibinfo{author}{Behera, M.R.}, \bibinfo{author}{Rao, A.},
  \bibinfo{author}{Dasgupta, A.}, \bibinfo{author}{Chakraborty, S.},
  \bibinfo{year}{2023}.
\newblock \bibinfo{title}{Investigation of regimes and associated flow
  structures during impingement of a liquid drop on a liquid pool}.
\newblock \bibinfo{journal}{Phys. Rev. Fluids} \bibinfo{volume}{8},
  \bibinfo{pages}{014002}.
\bibitem[{Bisighini et~al.(2010)Bisighini, Cossali, Tropea and
  Roisman}]{bisighini2010crater}
\bibinfo{author}{Bisighini, A.}, \bibinfo{author}{Cossali, G.E.},
  \bibinfo{author}{Tropea, C.}, \bibinfo{author}{Roisman, I.V.},
  \bibinfo{year}{2010}.
\newblock \bibinfo{title}{Crater evolution after the impact of a drop onto a
  semi-infinite liquid target}.
\newblock \bibinfo{journal}{Phys. Rev. E.} \bibinfo{volume}{82},
  \bibinfo{pages}{036319}.
\bibitem[{Blanchette and Bigioni(2006)}]{blanchette2006partial}
\bibinfo{author}{Blanchette, F.}, \bibinfo{author}{Bigioni, T.P.},
  \bibinfo{year}{2006}.
\newblock \bibinfo{title}{Partial coalescence of drops at liquid interfaces}.
\newblock \bibinfo{journal}{Nat. Phys.} \bibinfo{volume}{2},
  \bibinfo{pages}{254}.
\bibitem[{Brackbill et~al.(1992)Brackbill, Kothe and
  Zemach}]{brackbill1992continuum}
\bibinfo{author}{Brackbill, J.U.}, \bibinfo{author}{Kothe, D.B.},
  \bibinfo{author}{Zemach, C.}, \bibinfo{year}{1992}.
\newblock \bibinfo{title}{A continuum method for modeling surface tension}.
\newblock \bibinfo{journal}{J. Comput. Phys.} \bibinfo{volume}{100},
  \bibinfo{pages}{335--354}.
\bibitem[{Charles and Mason(1960)}]{charles1960mechanism}
\bibinfo{author}{Charles, G.E.}, \bibinfo{author}{Mason, S.G.},
  \bibinfo{year}{1960}.
\newblock \bibinfo{title}{The mechanism of partial coalescence of liquid drops
  at liquid/liquid interfaces}.
\newblock \bibinfo{journal}{J. Colloid Sci.} \bibinfo{volume}{15},
  \bibinfo{pages}{105--122}.
\bibitem[{Chen et~al.(2006a)Chen, Mandre and Feng}]{Chen2006b}
\bibinfo{author}{Chen, X.}, \bibinfo{author}{Mandre, S.},
  \bibinfo{author}{Feng, J.J.}, \bibinfo{year}{2006}a.
\newblock \bibinfo{title}{An experimental study of the coalescence between a
  drop and an interface in {N}ewtonian and polymeric liquids}.
\newblock \bibinfo{journal}{Phys. Fluids} \bibinfo{volume}{18},
  \bibinfo{pages}{092103}.
\bibitem[{Chen et~al.(2006b)Chen, Mandre and Feng}]{chen2006partial}
\bibinfo{author}{Chen, X.}, \bibinfo{author}{Mandre, S.},
  \bibinfo{author}{Feng, J.J.}, \bibinfo{year}{2006}b.
\newblock \bibinfo{title}{Partial coalescence between a drop and a
  liquid-liquid interface}.
\newblock \bibinfo{journal}{Phys. Fluids} \bibinfo{volume}{18},
  \bibinfo{pages}{051705}.
\bibitem[{Constante-Amores et~al.(2023)Constante-Amores, Kahouadji, Shin,
  Chergui, Juric, Castrej{\'o}n-Pita, Matar and
  Castrej{\'o}n-Pita}]{constante2023impact}
\bibinfo{author}{Constante-Amores, C.R.}, \bibinfo{author}{Kahouadji, L.},
  \bibinfo{author}{Shin, S.}, \bibinfo{author}{Chergui, J.},
  \bibinfo{author}{Juric, D.}, \bibinfo{author}{Castrej{\'o}n-Pita, J.R.},
  \bibinfo{author}{Matar, O.K.}, \bibinfo{author}{Castrej{\'o}n-Pita, A.A.},
  \bibinfo{year}{2023}.
\newblock \bibinfo{title}{Impact of droplets onto surfactant-laden thin liquid
  films}.
\newblock \bibinfo{journal}{J. Fluid Mech.} \bibinfo{volume}{961},
  \bibinfo{pages}{A8}.
\bibitem[{Cossali et~al.(1997)Cossali, Coghe and Marengo}]{cossali1997impact}
\bibinfo{author}{Cossali, G.E.}, \bibinfo{author}{Coghe, A.},
  \bibinfo{author}{Marengo, M.}, \bibinfo{year}{1997}.
\newblock \bibinfo{title}{The impact of a single drop on a wetted solid
  surface}.
\newblock \bibinfo{journal}{Exp. Fluids} \bibinfo{volume}{22},
  \bibinfo{pages}{463--472}.
\bibitem[{Deka et~al.(2019a)Deka, Biswas, Sahu, Kulkarni and
  Dalal}]{deka2019coalescence}
\bibinfo{author}{Deka, H.}, \bibinfo{author}{Biswas, G.},
  \bibinfo{author}{Sahu, K.C.}, \bibinfo{author}{Kulkarni, Y.},
  \bibinfo{author}{Dalal, A.}, \bibinfo{year}{2019}a.
\newblock \bibinfo{title}{Coalescence dynamics of a compound drop on a deep
  liquid pool}.
\newblock \bibinfo{journal}{J. Fluid Mech.} \bibinfo{volume}{866},
  \bibinfo{pages}{R2}.
\bibitem[{Deka et~al.(2017)Deka, Ray, Biswas, Dalal, Tsai and
  Wang}]{deka2017regime}
\bibinfo{author}{Deka, H.}, \bibinfo{author}{Ray, B.}, \bibinfo{author}{Biswas,
  G.}, \bibinfo{author}{Dalal, A.}, \bibinfo{author}{Tsai, P.H.},
  \bibinfo{author}{Wang, A.B.}, \bibinfo{year}{2017}.
\newblock \bibinfo{title}{The regime of large bubble entrapment during a single
  drop impact on a liquid pool}.
\newblock \bibinfo{journal}{Phys. Fluids} \bibinfo{volume}{29},
  \bibinfo{pages}{092101}.
\bibitem[{Deka et~al.(2019b)Deka, Tsai, Biswas, Dalal, Ray and
  Wang}]{deka2019dynamics}
\bibinfo{author}{Deka, H.}, \bibinfo{author}{Tsai, P.H.},
  \bibinfo{author}{Biswas, G.}, \bibinfo{author}{Dalal, A.},
  \bibinfo{author}{Ray, B.}, \bibinfo{author}{Wang, A.B.},
  \bibinfo{year}{2019}b.
\newblock \bibinfo{title}{Dynamics of formation and oscillation of
  non-spherical drops}.
\newblock \bibinfo{journal}{Chem. Eng. Sci.} \bibinfo{volume}{201},
  \bibinfo{pages}{413--423}.
\bibitem[{Deshpande et~al.(2012)Deshpande, Anumolu and
  Trujillo}]{deshpande2012evaluating}
\bibinfo{author}{Deshpande, S.S.}, \bibinfo{author}{Anumolu, L.},
  \bibinfo{author}{Trujillo, M.F.}, \bibinfo{year}{2012}.
\newblock \bibinfo{title}{Evaluating the performance of the two-phase flow
  solver interfoam}.
\newblock \bibinfo{journal}{Comput. Sci. Discov.} \bibinfo{volume}{5},
  \bibinfo{pages}{014016}.
\bibitem[{Fedorchenko and Wang(2004)}]{fedorchenko2004some}
\bibinfo{author}{Fedorchenko, A.I.}, \bibinfo{author}{Wang, A.},
  \bibinfo{year}{2004}.
\newblock \bibinfo{title}{On some common features of drop impact on liquid
  surfaces}.
\newblock \bibinfo{journal}{Phys. Fluids} \bibinfo{volume}{16},
  \bibinfo{pages}{1349--1365}.
\bibitem[{Gekle and Gordillo(2010)}]{gekle2010}
\bibinfo{author}{Gekle, S.}, \bibinfo{author}{Gordillo, J.M.},
  \bibinfo{year}{2010}.
\newblock \bibinfo{title}{Generation and breakup of worthington jets after
  cavity collapse. part 1. jet formation}.
\newblock \bibinfo{journal}{J. Fluid Mech.} \bibinfo{volume}{663},
  \bibinfo{pages}{293--330}.
\bibitem[{Gielen et~al.(2017)Gielen, Sleutel, Benschop, Riepen, Voronina,
  Visser, Lohse, Snoeijer, Versluis and Gelderblom}]{gielen2017oblique}
\bibinfo{author}{Gielen, M.V.}, \bibinfo{author}{Sleutel, P.},
  \bibinfo{author}{Benschop, J.}, \bibinfo{author}{Riepen, M.},
  \bibinfo{author}{Voronina, V.}, \bibinfo{author}{Visser, C.W.},
  \bibinfo{author}{Lohse, D.}, \bibinfo{author}{Snoeijer, J.H.},
  \bibinfo{author}{Versluis, M.}, \bibinfo{author}{Gelderblom, H.},
  \bibinfo{year}{2017}.
\newblock \bibinfo{title}{Oblique drop impact onto a deep liquid pool}.
\newblock \bibinfo{journal}{Phys. Rev. Fluids} \bibinfo{volume}{2},
  \bibinfo{pages}{083602}.
\bibitem[{Gordillo and Gekle(2010)}]{gordillo2010}
\bibinfo{author}{Gordillo, J.M.}, \bibinfo{author}{Gekle, S.},
  \bibinfo{year}{2010}.
\newblock \bibinfo{title}{Generation and breakup of worthington jets after
  cavity collapse. part 2. tip breakup of stretched jets}.
\newblock \bibinfo{journal}{J. Fluid Mech.} \bibinfo{volume}{663},
  \bibinfo{pages}{331--346}.
\bibitem[{Ivanova and Klyuev(2023)}]{ivanova2023self}
\bibinfo{author}{Ivanova, N.A.}, \bibinfo{author}{Klyuev, D.S.},
  \bibinfo{year}{2023}.
\newblock \bibinfo{title}{Self-sustaining levitation of droplets above a liquid
  pool}.
\newblock \bibinfo{journal}{Appl. Phys. Lett.} \bibinfo{volume}{123}.
\bibitem[{Josserand and Thoroddsen(2016)}]{josserand2016drop}
\bibinfo{author}{Josserand, C.}, \bibinfo{author}{Thoroddsen, S.T.},
  \bibinfo{year}{2016}.
\newblock \bibinfo{title}{Drop impact on a solid surface}.
\newblock \bibinfo{journal}{Ann. Rev. Fluid Mech.} \bibinfo{volume}{48},
  \bibinfo{pages}{365--391}.
\bibitem[{Josserand and Zaleski(2003)}]{josserand2003droplet}
\bibinfo{author}{Josserand, C.}, \bibinfo{author}{Zaleski, S.},
  \bibinfo{year}{2003}.
\newblock \bibinfo{title}{Droplet splashing on a thin liquid film}.
\newblock \bibinfo{journal}{Phys. Fluids} \bibinfo{volume}{15},
  \bibinfo{pages}{1650--1657}.
\bibitem[{Kavehpour(2015)}]{kavehpour2015coalescence}
\bibinfo{author}{Kavehpour, H.P.}, \bibinfo{year}{2015}.
\newblock \bibinfo{title}{Coalescence of drops}.
\newblock \bibinfo{journal}{Ann. Rev. Fluid Mech.} \bibinfo{volume}{47},
  \bibinfo{pages}{245--268}.
\bibitem[{Kirar et~al.(2022)Kirar, Kolhe and Sahu}]{kirar2022coalescence}
\bibinfo{author}{Kirar, P.K.}, \bibinfo{author}{Kolhe, P.S.},
  \bibinfo{author}{Sahu, K.C.}, \bibinfo{year}{2022}.
\newblock \bibinfo{title}{Coalescence and migration of a droplet on a liquid
  pool with an inclined bottom wall}.
\newblock \bibinfo{journal}{Phys. Rev. Fluids} \bibinfo{volume}{7},
  \bibinfo{pages}{094001}.
\bibitem[{Kostinski and Shaw(2009)}]{kostinski2009raindrops}
\bibinfo{author}{Kostinski, A.B.}, \bibinfo{author}{Shaw, R.A.},
  \bibinfo{year}{2009}.
\newblock \bibinfo{title}{Raindrops large and small}.
\newblock \bibinfo{journal}{Nat. Phys.} \bibinfo{volume}{5},
  \bibinfo{pages}{624--625}.
\bibitem[{Lamb(1924)}]{lamb1924hydrodynamics}
\bibinfo{author}{Lamb, H.}, \bibinfo{year}{1924}.
\newblock \bibinfo{title}{Hydrodynamics}.
\newblock \bibinfo{publisher}{University Press}.
\bibitem[{Leneweit et~al.(2005)Leneweit, Koehler, Roesner and
  Sch{\"a}fer}]{leneweit2005regimes}
\bibinfo{author}{Leneweit, G.}, \bibinfo{author}{Koehler, R.},
  \bibinfo{author}{Roesner, K.G.}, \bibinfo{author}{Sch{\"a}fer, G.},
  \bibinfo{year}{2005}.
\newblock \bibinfo{title}{Regimes of drop morphology in oblique impact on deep
  fluids}.
\newblock \bibinfo{journal}{J. Fluid Mech.} \bibinfo{volume}{543},
  \bibinfo{pages}{303--331}.
\bibitem[{Liu and Bothe(2016)}]{liu2016numerical}
\bibinfo{author}{Liu, M.}, \bibinfo{author}{Bothe, D.}, \bibinfo{year}{2016}.
\newblock \bibinfo{title}{Numerical study of head-on droplet collisions at high
  weber numbers}.
\newblock \bibinfo{journal}{J. Fluid Mech.} \bibinfo{volume}{789},
  \bibinfo{pages}{785--805}.
\bibitem[{Liu et~al.(2021)Liu, Lo, Li, Liu, Zhao and Xu}]{liu2021role}
\bibinfo{author}{Liu, Q.}, \bibinfo{author}{Lo, J.H.Y.}, \bibinfo{author}{Li,
  Y.}, \bibinfo{author}{Liu, Y.}, \bibinfo{author}{Zhao, J.},
  \bibinfo{author}{Xu, L.}, \bibinfo{year}{2021}.
\newblock \bibinfo{title}{The role of drop shape in impact and splash}.
\newblock \bibinfo{journal}{Nat. Commun.} \bibinfo{volume}{12},
  \bibinfo{pages}{3068}.
\bibitem[{Liu(2018)}]{liu2018experimental}
\bibinfo{author}{Liu, X.}, \bibinfo{year}{2018}.
\newblock \bibinfo{title}{Experimental study of drop impact on deep-water
  surface in the presence of wind}.
\newblock \bibinfo{journal}{J. Phys. Oceanogr.} \bibinfo{volume}{48},
  \bibinfo{pages}{329--341}.
\bibitem[{Low and List(1982)}]{low1982collision}
\bibinfo{author}{Low, T.B.}, \bibinfo{author}{List, R.}, \bibinfo{year}{1982}.
\newblock \bibinfo{title}{Collision, coalescence and breakup of raindrops. part
  i: Experimentally established coalescence efficiencies and fragment size
  distributions in breakup}.
\newblock \bibinfo{journal}{J. Atmos. Sci.} \bibinfo{volume}{39},
  \bibinfo{pages}{1591--1606}.
\bibitem[{Okawa et~al.(2008)Okawa, Shiraishi and Mori}]{okawa2008effect}
\bibinfo{author}{Okawa, T.}, \bibinfo{author}{Shiraishi, T.},
  \bibinfo{author}{Mori, T.}, \bibinfo{year}{2008}.
\newblock \bibinfo{title}{Effect of impingement angle on the outcome of single
  water drop impact onto a plane water surface}.
\newblock \bibinfo{journal}{Exp. Fluids} \bibinfo{volume}{44},
  \bibinfo{pages}{331--339}.
\bibitem[{Paul et~al.(2023)Paul, Ray, Sahu and Biswas}]{paul2023investigation}
\bibinfo{author}{Paul, A.}, \bibinfo{author}{Ray, B.}, \bibinfo{author}{Sahu,
  K.C.}, \bibinfo{author}{Biswas, G.}, \bibinfo{year}{2023}.
\newblock \bibinfo{title}{An investigation on the impact of two vertically
  aligned drops on a liquid surface}.
\newblock \bibinfo{journal}{Int. J. Multiphase Flow} \bibinfo{volume}{168},
  \bibinfo{pages}{104588}.
\bibitem[{Pumphrey et~al.(1989)Pumphrey, Crum and
  Bjo/Rno/}]{pumphrey1989underwater}
\bibinfo{author}{Pumphrey, H.C.}, \bibinfo{author}{Crum, L.A.},
  \bibinfo{author}{Bjo/Rno/, L.}, \bibinfo{year}{1989}.
\newblock \bibinfo{title}{Underwater sound produced by individual drop impacts
  and rainfall}.
\newblock \bibinfo{journal}{J. Acoust. Soc. Am.} \bibinfo{volume}{85},
  \bibinfo{pages}{1518--1526}.
\bibitem[{Ray et~al.(2010)Ray, Biswas and Sharma}]{ray2010generation}
\bibinfo{author}{Ray, B.}, \bibinfo{author}{Biswas, G.},
  \bibinfo{author}{Sharma, A.}, \bibinfo{year}{2010}.
\newblock \bibinfo{title}{Generation of secondary droplets in coalescence of a
  drop at a liquid--liquid interface}.
\newblock \bibinfo{journal}{J. Fluid Mech.} \bibinfo{volume}{655},
  \bibinfo{pages}{72--104}.
\bibitem[{Ray et~al.(2012)Ray, Biswas and Sharma}]{ray2012oblique}
\bibinfo{author}{Ray, B.}, \bibinfo{author}{Biswas, G.},
  \bibinfo{author}{Sharma, A.}, \bibinfo{year}{2012}.
\newblock \bibinfo{title}{Oblique drop impact on deep and shallow liquid}.
\newblock \bibinfo{journal}{Commun. Comput. Phys.} \bibinfo{volume}{11},
  \bibinfo{pages}{1386--1396}.
\bibitem[{Ray et~al.(2015)Ray, Biswas and Sharma}]{ray2015regimes}
\bibinfo{author}{Ray, B.}, \bibinfo{author}{Biswas, G.},
  \bibinfo{author}{Sharma, A.}, \bibinfo{year}{2015}.
\newblock \bibinfo{title}{Regimes during liquid drop impact on a liquid pool}.
\newblock \bibinfo{journal}{J. Fluid Mech.} \bibinfo{volume}{768},
  \bibinfo{pages}{492--523}.
\bibitem[{Ristenpart et~al.(2009)Ristenpart, Bird, Belmonte, Dollar and
  Stone}]{ristenpart2009non}
\bibinfo{author}{Ristenpart, W.D.}, \bibinfo{author}{Bird, J.C.},
  \bibinfo{author}{Belmonte, A.}, \bibinfo{author}{Dollar, F.},
  \bibinfo{author}{Stone, H.A.}, \bibinfo{year}{2009}.
\newblock \bibinfo{title}{Non-coalescence of oppositely charged drops}.
\newblock \bibinfo{journal}{Nature} \bibinfo{volume}{461},
  \bibinfo{pages}{377--380}.
\bibitem[{Schneiders(2000)}]{schneiders2000octree}
\bibinfo{author}{Schneiders, R.}, \bibinfo{year}{2000}.
\newblock \bibinfo{title}{Octree-based hexahedral mesh generation}.
\newblock \bibinfo{journal}{Int. J. Comput. Geom. Appl.} \bibinfo{volume}{10},
  \bibinfo{pages}{383--398}.
\bibitem[{Sprittles(2024)}]{sprittles2023gas}
\bibinfo{author}{Sprittles, J.E.}, \bibinfo{year}{2024}.
\newblock \bibinfo{title}{Gas microfilms in droplet dynamics: When do drops
  bounce?}
\newblock \bibinfo{journal}{Ann. Rev. Fluid Mech.} \bibinfo{volume}{56}.
\bibitem[{Stone et~al.(2004)Stone, Stroock and Ajdari}]{stone2004engineering}
\bibinfo{author}{Stone, H.A.}, \bibinfo{author}{Stroock, A.D.},
  \bibinfo{author}{Ajdari, A.}, \bibinfo{year}{2004}.
\newblock \bibinfo{title}{Engineering flows in small devices: microfluidics
  toward a lab-on-a-chip}.
\newblock \bibinfo{journal}{Annu. Rev. Fluid Mech.} \bibinfo{volume}{36},
  \bibinfo{pages}{381--411}.
\bibitem[{Szakall et~al.(2009)Szakall, Diehl, Mitra and
  Borrmann}]{szakall2009wind}
\bibinfo{author}{Szakall, M.}, \bibinfo{author}{Diehl, K.},
  \bibinfo{author}{Mitra, S.K.}, \bibinfo{author}{Borrmann, S.},
  \bibinfo{year}{2009}.
\newblock \bibinfo{title}{A wind tunnel study on the shape, oscillation, and
  internal circulation of large raindrops with sizes between 2.5 and 7.5 mm}.
\newblock \bibinfo{journal}{J. Atmos. Sci.} \bibinfo{volume}{66},
  \bibinfo{pages}{755--765}.
\bibitem[{Thomson and Newall(1886)}]{thomson1886v}
\bibinfo{author}{Thomson, J.J.}, \bibinfo{author}{Newall, H.F.},
  \bibinfo{year}{1886}.
\newblock \bibinfo{title}{On the formation of vortex rings by drops falling
  into liquids, and some allied phenomena}.
\newblock \bibinfo{journal}{Proc. Royal Soc. Lond.} \bibinfo{volume}{39},
  \bibinfo{pages}{417--436}.
\bibitem[{Thoroddsen(2002)}]{thoroddsen2002ejecta}
\bibinfo{author}{Thoroddsen, S.T.}, \bibinfo{year}{2002}.
\newblock \bibinfo{title}{The ejecta sheet generated by the impact of a drop}.
\newblock \bibinfo{journal}{J. Fluid Mech.} \bibinfo{volume}{451},
  \bibinfo{pages}{373--381}.
\bibitem[{Thoroddsen et~al.(2008)Thoroddsen, Etoh and
  Takehara}]{thoroddsen2008high}
\bibinfo{author}{Thoroddsen, S.T.}, \bibinfo{author}{Etoh, T.G.},
  \bibinfo{author}{Takehara, K.}, \bibinfo{year}{2008}.
\newblock \bibinfo{title}{High-speed imaging of drops and bubbles}.
\newblock \bibinfo{journal}{Ann. Rev. Fluid Mech.} \bibinfo{volume}{40},
  \bibinfo{pages}{257--285}.
\bibitem[{Thoroddsen and Takehara(2000)}]{thoroddsen2000coalescence}
\bibinfo{author}{Thoroddsen, S.T.}, \bibinfo{author}{Takehara, K.},
  \bibinfo{year}{2000}.
\newblock \bibinfo{title}{The coalescence cascade of a drop}.
\newblock \bibinfo{journal}{Phys. Fluids} \bibinfo{volume}{12},
  \bibinfo{pages}{1265--1267}.
\bibitem[{Tsamopoulos and Brown(1983)}]{tsamopoulos1983nonlinear}
\bibinfo{author}{Tsamopoulos, J.A.}, \bibinfo{author}{Brown, R.A.},
  \bibinfo{year}{1983}.
\newblock \bibinfo{title}{Nonlinear oscillations of inviscid drops and
  bubbles}.
\newblock \bibinfo{journal}{J. Fluid Mech.} \bibinfo{volume}{127},
  \bibinfo{pages}{519--537}.
\bibitem[{Veron(2015)}]{veron2015ocean}
\bibinfo{author}{Veron, F.}, \bibinfo{year}{2015}.
\newblock \bibinfo{title}{Ocean spray}.
\newblock \bibinfo{journal}{Ann. Rev. Fluid Mech.} \bibinfo{volume}{47},
  \bibinfo{pages}{507--538}.
\bibitem[{Wang and Chen(2000)}]{wang2000splashing}
\bibinfo{author}{Wang, A.}, \bibinfo{author}{Chen, C.}, \bibinfo{year}{2000}.
\newblock \bibinfo{title}{Splashing impact of a single drop onto very thin
  liquid films}.
\newblock \bibinfo{journal}{Phys. Fluids} \bibinfo{volume}{12},
  \bibinfo{pages}{2155--2158}.
\bibitem[{Wang et~al.(2013)Wang, Kuan and Tsai}]{wang2013we}
\bibinfo{author}{Wang, A.B.}, \bibinfo{author}{Kuan, C.C.},
  \bibinfo{author}{Tsai, P.H.}, \bibinfo{year}{2013}.
\newblock \bibinfo{title}{Do we understand the bubble formation by a single
  drop impacting upon liquid surface?}
\newblock \bibinfo{journal}{Phys. Fluids} \bibinfo{volume}{25}.
\bibitem[{Weiss and Yarin(1999)}]{weiss1999single}
\bibinfo{author}{Weiss, D.A.}, \bibinfo{author}{Yarin, A.L.},
  \bibinfo{year}{1999}.
\newblock \bibinfo{title}{Single drop impact onto liquid films: neck
  distortion, jetting, tiny bubble entrainment, and crown formation}.
\newblock \bibinfo{journal}{J. Fluid Mech.} \bibinfo{volume}{385},
  \bibinfo{pages}{229--254}.
\bibitem[{Worthington(1877)}]{worthington1877xxviii}
\bibinfo{author}{Worthington, A.M.}, \bibinfo{year}{1877}.
\newblock \bibinfo{title}{Xxviii. on the forms assumed by drops of liquids
  falling vertically on a horizontal plate}.
\newblock \bibinfo{journal}{Proc. R. Soc. Lond.} \bibinfo{volume}{25},
  \bibinfo{pages}{261--272}.
\bibitem[{Zhang et~al.(2019)Zhang, Ling, Tsai, Wang, Popinet and
  Zaleski}]{zhang2019short}
\bibinfo{author}{Zhang, B.}, \bibinfo{author}{Ling, Y.}, \bibinfo{author}{Tsai,
  P.H.}, \bibinfo{author}{Wang, A.B.}, \bibinfo{author}{Popinet, S.},
  \bibinfo{author}{Zaleski, S.}, \bibinfo{year}{2019}.
\newblock \bibinfo{title}{Short-term oscillation and falling dynamics for a
  water drop dripping in quiescent air}.
\newblock \bibinfo{journal}{Phys. Rev. Fluids} \bibinfo{volume}{4},
  \bibinfo{pages}{123604}.

\end{thebibliography}

\end{document}